\documentclass[journal,svgnames]{vgtc}                

\onlineid{1275}

\ifpdf
  \pdfoutput=1\relax                   
  \usepackage{graphicx}                
  \DeclareGraphicsExtensions{.pdf,.png,.jpg,.jpeg} 
\else
  \usepackage{graphicx}                
  \DeclareGraphicsExtensions{.eps}     
\fi%

\graphicspath{{figures/}{pictures/}{images/}{./}} 

\usepackage{microtype}                 
\PassOptionsToPackage{warn}{textcomp}  
\usepackage{textcomp}                  
\usepackage{times}                     
\usepackage{mathptmx}                  
\usepackage{cite}                      
\usepackage{tabu}                      
\usepackage{booktabs}                  
\usepackage[normalem]{ulem}
\usepackage{enumitem}
\usepackage{blindtext}
\usepackage{multirow}
\usepackage{textgreek} 
\usepackage{eurosym}
\usepackage{subcaption}
\usepackage{ccicons}
\usepackage{dingbat}
\usepackage{xspace}
\usepackage{cuted}
\usepackage{flushend}
\usepackage{microtype}
\usepackage[svgnames]{xcolor}
\usepackage{expdlist}
\usepackage{float}
\usepackage{afterpage} 

\useunder{\uline}{\ul}{}

\newcommand{\eg}{e.\,g.}
\newcommand{\ie}{i.\,e.}
\newcommand{\etal}{\textit{et al.}\xspace}
\newcommand{\code}[1]{\texttt{#1}} 

\newcommand{\revis}[1]{\textcolor{FireBrick}{#1}}
\renewcommand{\revis}[1]{\textcolor{Black}{#1}}

\definecolor{discreet-gray}{gray}{0.6}
\newcommand{\discreet}[1]{\textcolor{discreet-gray}{#1}}

\newcommand{\prereg}[2]{\texttt{\href{https://osf.io/#2}{#1}}}

\newcommand{\suppmat}[2]{\href{https://osf.io/#2}{#1}\xspace}

\newcommand{\researchLog}{\href{https://osf.io/7tq9m}{OSF Research log}\xspace}

\newcommand{\scalename}{PREVis\xspace}

\newcommand{\PREVisColors}{\mbox{\raisebox{0.05\baselineskip}{\includegraphics[height=0.15cm,keepaspectratio]{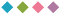}}}\xspace}

\newcommand{\makeSubscaleName}[2]{\mbox{\raisebox{-0.05\baselineskip}{\includegraphics[height=0.25cm,keepaspectratio]{figures/#2.pdf}}} \hspace{-0.4em} \textbf{\textsc{\fontfamily{phv}\fontsize{8}{1.2\baselineskip}\selectfont#1\xspace}}}

\newcommand{\makeSubscaleColorDot}[1]{\mbox{\raisebox{-0.05\baselineskip}{\includegraphics[height=0.25cm,keepaspectratio]{figures/#1.pdf}}} \hspace{-0.4em}\xspace}

\newcommand{\SUn}{\makeSubscaleName{Understand}{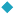}\xspace}
\newcommand{\SUnDot}{\makeSubscaleColorDot{subscale_understand}}

\newcommand{\SLa}{\makeSubscaleName{Layout}{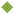}\xspace}
\newcommand{\SLaDot}{\makeSubscaleColorDot{subscale_layout}}

\newcommand{\SDR}{\makeSubscaleName{DataRead}{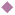}\xspace}
\newcommand{\SDRDot}{\makeSubscaleColorDot{subscale_dataRead}}

\newcommand{\SDF}{\makeSubscaleName{DataFeat}{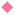}\xspace}
\newcommand{\SDFDot}{\makeSubscaleColorDot{subscale_dataFeat}}

\newcommand{\MTMMoneText}{
\mbox{\raisebox{-.15\baselineskip}{\includegraphics[height=0.3cm,keepaspectratio]{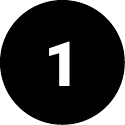}}}\xspace}
\newcommand{\MTMMtwoText}{\mbox{\raisebox{-.15\baselineskip}{\includegraphics[height=0.3cm,keepaspectratio]{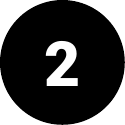}}}\xspace}
\newcommand{\MTMMthreeText}{\mbox{\raisebox{-.15\baselineskip}{\includegraphics[height=0.3cm,keepaspectratio]{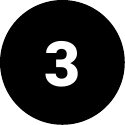}}}\xspace}

\newcommand{\MTMMoneCaption}{\mbox{\raisebox{-.15\baselineskip}{
\hspace{-0.5em}
\includegraphics[height=0.3cm,keepaspectratio]{figures/MTMM_legend_circle1.pdf}}}
\hspace{-0.4em}
}

\newcommand{\MTMMtwoCaption}{\mbox{\raisebox{-.15\baselineskip}{
\hspace{-0.5em}
\includegraphics[height=0.3cm,keepaspectratio]{figures/MTMM_legend_circle2.pdf}}}
\hspace{-0.4em}\xspace
}

\newcommand{\MTMMthreeCaption}{\mbox{\raisebox{-.15\baselineskip}{
\hspace{-0.5em}
\includegraphics[height=0.3cm,keepaspectratio]{figures/MTMM_legend_circle3.pdf}}}
\hspace{-0.4em}\xspace
}

\newcommand{\makeStimLegend}[1]{\mbox{\raisebox{-.15\baselineskip}{
\hspace{-0.2em}{\includegraphics[height=0.3cm,keepaspectratio]{figures/appendix/stimuli_exp_survey/Stimulus#1.pdf}}} \hspace{-0.3em}}}
\newcommand{\stimA}{\makeStimLegend{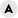}\xspace}

\newcommand{\stimB}{\makeStimLegend{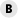}\xspace}

\newcommand{\stimC}{\makeStimLegend{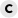}\xspace}

\newcommand{\stimD}{\makeStimLegend{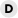}\xspace}

\newcommand{\stimE}{\makeStimLegend{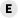}\xspace}

\newcommand{\stimF}{\makeStimLegend{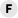}\xspace}

\newcommand{\makeStimVLegend}[1]{\mbox{\raisebox{-.15\baselineskip}{
\hspace{-0.2em}{\includegraphics[height=0.3cm,keepaspectratio]{figures/appendix/stimuli_valid_survey/Stimulus#1.pdf}}} \hspace{-0.3em}}}
\newcommand{\stimVA}{\makeStimVLegend{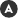}\xspace}
\newcommand{\stimVB}{\makeStimVLegend{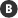}\xspace}
\newcommand{\stimVC}{\makeStimVLegend{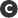}\xspace}

\newlength{\picturewidth}

\vgtccategory{Research}

\vgtcpapertype{please specify}

\newcommand{\mytitle}{\scalename: Perceived Readability Evaluation for Visualizations}
\title{\mytitle}

\author{%
  \authororcid{Anne-Flore\ Cabouat}{0000-0002-3327-2729},
  \authororcid{Tingying\ He}{0000-0002-0500-7995},
  \authororcid{Petra\ Isenberg}{0000-0002-2948-6417},
  \authororcid{Tobias\ Isenberg}{0000-0001-7953-8644}%
}
\authorfooter{
\item
 Anne-Flore Cabouat, Tingying He (\raisebox{-.5pt}{\includegraphics[height=6pt]{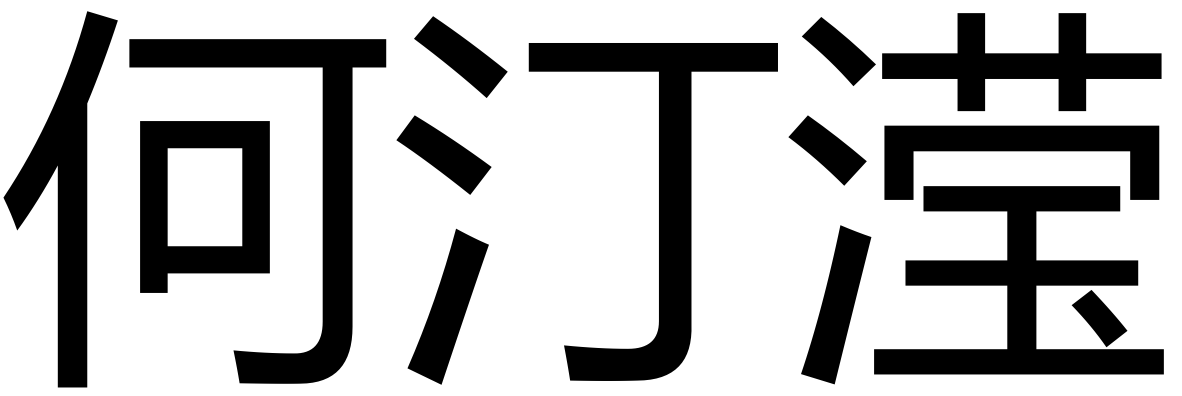}}), Petra Isenberg, and Tobias Isenberg are with Université Paris-Saclay, CNRS, Inria, LISN, France. E-mail: given\_name.family\_name@inria.fr
}

\abstract{%
 We developed and validated an instrument to measure the perceived readability in data visualization: \scalename.
 Researchers and practitioners can easily use this instrument as part of their evaluations to compare the perceived readability of different visual data representations.
 Our instrument can complement results from controlled experiments on user task performance or provide additional data during in-depth qualitative work such as design iterations when developing a new technique.
 Although readability is recognized as an essential quality of data visualizations, so far there has not been a unified definition of the construct 
 in the context of visual representations.
 As a result, researchers often 
 lack guidance for 
 determining how to ask people to rate  their perceived readability of a visualization.
 To address this issue, we engaged in a rigorous process to develop the first \emph{validated} instrument targeted at the subjective readability of visual data representations.
 Our final instrument consists of 11 items across 4 dimensions: understandability, 
 layout clarity, readability of 
 data values, and readability of data patterns. We provide the questionnaire as a document with implementation guidelines 
 on \href{https://osf.io/9cg8j}{\texttt{osf.io/9cg8j}}. 
 Beyond this instrument, we contribute a discussion of how researchers have previously assessed visualization readability, and an analysis of the factors underlying perceived readability in visual data representations.
 
}


\teaser{
  \centering
    \includegraphics[width=\columnwidth]{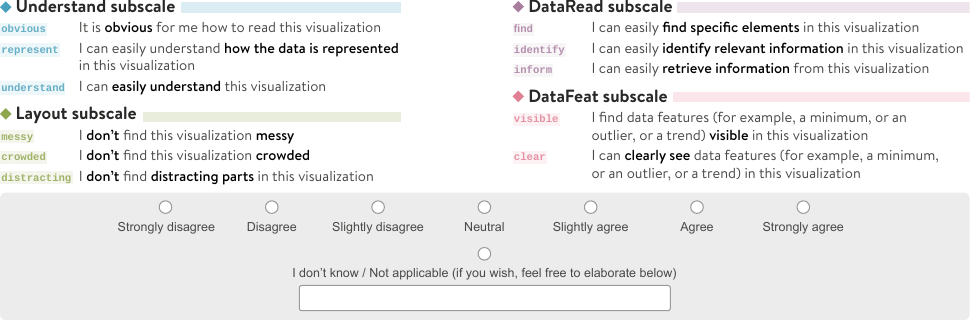}\vspace{-1pt}
  \caption{\protect\revis{\scalename subscales, items, and recommended presentation of answer options for a computer-supported questionnaire.}}
  \label{fig:teaser}
}

\graphicspath{{figs/}{figures/}{pictures/}{images/}{./}} 

\usepackage{tabu}                      
\usepackage{booktabs}                  
\usepackage{lipsum}                    
\usepackage{mwe}                       

\usepackage{mathptmx}                  

\DeclareRobustCommand{\gobblefour}[5]{}
\newcommand*{\SkipTocEntry}{\addtocontents{toc}{\gobblefour}}

\begin{document}



\firstsection{Introduction}

{\renewcommand{\addtocontents}[2]{}\maketitle}


\noindent When looking at examples of data visualizations, it is intuitively clear that some are easier to read than others.
For many data analysis use cases, poor readability will drastically reduce the usefulness of a visual representation of data for the viewer.
As such, readability is a basic quality criterion in data visualization \cite{Kosara:2007:VisualizationCriticism}. 
One of the fundamental challenges in studying the readability of data visualizations, however, is that the concepts of \emph{reading} and \emph{readability} are held as tacit knowledge.
The terms are often used in scientific writing 
without clear definitions of what they specifically mean in the context of data visualization---recalling Kosara's ``empire built on sand'' \cite{Kosara:2016}.

Readability of text is broadly defined as ``the quality of being easy and enjoyable to read'' \cite{Cambridge:Readability:Definition}.
It applies to letters and words as well as entire books. Linguists have developed hundreds of 
formulas to analyze the readability of texts \cite{Benjamin:2012:ReconstructingReadability}, but this approach fails to take into account characteristics of readers.
Since readability is better explained as a function of the interaction between the properties of texts and the characteristics of readers \cite{Bailin:2001:ReadabilityFormulaeCritique}, researchers now seek to analyze text difficulty based on cognitive theories. Such an approach may also be suitable to explore the readability of visual representations of data.

As we discuss in more detail below, a few definitions of ``readability'' exist in the visualization domain, yet they do not fully overlap. As a result, it is unclear to what extent different approaches to measuring readability can thoroughly capture the concept. In addition, we do not have a definition of what ``reading'' a data visualization is as a cognitive activity. Cognitive processes in visualization range from
low-level visual perception \cite{Quadri:2022} to high-level activities such as data exploration \cite{Yalcin:2016}, insight and knowledge generation \cite{Stasko:2014:value-driven-vis, Sacha:2014:Knowledge}, sense-making of unfamiliar visualizations \cite{Lee:2015:sensemaking}, or decision-making \cite{Padilla:2018:decision_model}.
Current cognitive models of visualization comprehension \cite{Hegarty:2011:CogVisualRepresentations, Padilla:2018:decision_model, Fox:2023:theories_models} provide important theoretical grounding to explain how people process information from visual data representations; the models, however, do not specify the boundaries of ``reading'' within the cognition continuum.

\begin{figure*} [t]
    \centering
    \includegraphics[width=\textwidth]{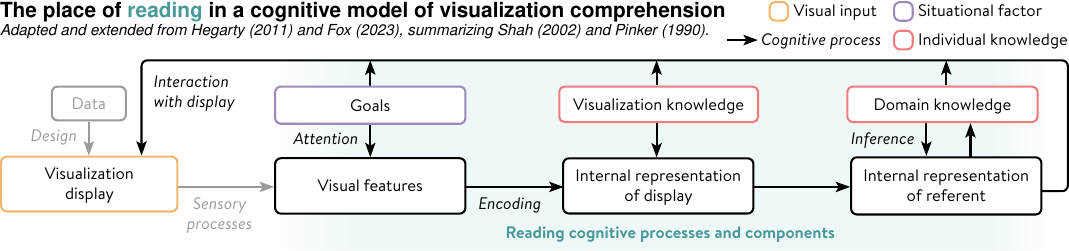}\vspace{-1ex}
    \caption{Our proposition to \revis{place \emph{reading} processes}
    within the model proposed by Hegarty \cite{Hegarty:2011:CogVisualRepresentations}, summarizing Shah \etal \cite{Shah:2005:ComprehensionQuantitativeInformation} and Pinker \cite{Pinker:1990:GraphComprehension}.}\vspace{-1ex}
    \label{fig:model-reading}
\end{figure*}

Our work is based on the fundamental premise that 
readability is a crucially important quality criterion in data visualization. As such, it requires formal definition and \textbf{\emph{empirically verified}} methods to study it.
In this paper we present the development and validation of our \scalename questionnaire. \scalename is a reliable instrument that allows respondents to rate how readable they find a static data visualization across 4 dimensions: layout clarity, ease of understanding, ease of reading data features, and ease of reading data values. 
During the development process, we also had to take first steps in clarifying what readability means in data visualization. This clarification is important because discrepancies in the use of terminology pose issues of comparability and reliability of empirical findings. In particular, we observed that researchers who asked participants to rate the readability of visualization did so using a wide variety of terms and answer options. Our \scalename tool addresses this problem because we followed well-established methodologies in scale development \cite{Devellis:2021:ScaleDevelopment, Boateng:2018:BestPractices}.
Developing a valid scale requires researchers to review existing literature and collect experts' suggestions on how to phrase the scale's items. In that way, it has allowed us to better frame the concept of readability in data visualization. While we do not promise to define readability entirely---yet---, we posit that it is not only a characteristic of the visualization; it is also highly individual and situational.
As such, in our work, we talk about the concept of \emph{perceived readability}, and we emphasize that measures of readability should take the reader's abilities and goals into account, in addition to measurable characteristics of the visualization display---for example, the density of dots in a scatterplot, or the number of edge crossings in a node-link graph.

In summary, with our work we make the following contributions:
\begin{itemize}[nosep,left=0pt .. \parindent]
\item a validated instrument to study perceived visualization readability that can be deployed in empirical research on data visualization,
\item an analysis of the factors underlying perceived readability of visual data representations,
\item an in-depth discussion and analysis of the concept of readability and how it has been used in the visualization field, and
\item an example of careful application of statistical methods from the psychology research field, and how it can contribute to advance theoretical models of data visualization. 
\end{itemize}

\SkipTocEntry\section{Related work}
\label{sec:related}

There are three major related fields that are important to our work: definitions of readability, ways of measuring readability, and scale development. We review these fields next.

\SkipTocEntry\subsection{Defining readability}
\label{subsec:defining-readability}
Readability broadly refers to ``the ease of reading'' in the context of written words. Reading is a complex behavior \cite{elleman:2019:ReadingComprehension}, making it difficult to measure and research.
Reading theorists describe how readers decode letters and words as well as how they integrate them into text comprehension \cite{Perfetti:2014:ReadingSystems}. 
Understanding a text further affords the reader with the ability to produce inferences and complex reasoning.  \revis{While \textit{legibility} essentially focuses on the ease of distinguishing letters, }readability can relate to individual words as well as entire books.

In the context of data visualization, researchers have proposed several models of information processing for graph comprehension \cite{Kosslyn:1989:UnderstandingCharts, Pinker:1990:GraphComprehension, Freedman-Shah:2002:ComprehensionIntegration}.
In these publications, the authors commonly refer to the viewer of a visualization as a ``(graph) reader'' but do not define the boundaries of the reading behavior itself.
Conversely, Curcio \cite{Curcio:1987:MathRelationships} explicitly refers to three levels of reading---reading the data, reading between the data, and reading beyond the data---but does not propose a cognitive model of the processes involved.
Drawing from the literature on reading texts, we posit that the \emph{reading} of visual representations encompasses all processes that allow the reader to transform the visual features retrieved from early visual processes into a meaningful internal representation of the information displayed in the visualization \cite{Cabouat:2023:PonderingReading}. In \autoref{fig:model-reading} we \revis{propose a first attempt at situating reading processes in a visual display comprehension model \cite{Hegarty:2011:CogVisualRepresentations}}.

Yet readability still lacks a formal definition in the data visualization context. Next, we thus review the different components that influence the ease of reading visual representations \cite{Fox:2023:theories_models}: the display, the individual, and the task. We then review existing definitions of readability.

\SkipTocEntry\subsubsection{The influence of display on readability}
Visualization researchers view readability primarily as a quality of the visual object, and have extensively studied the effectiveness of visual encodings for data visualization since early work on the classification of visual variables \cite{Bertin:1983, Cleveland:1984:GraphPerception, Mackinlay:1986:DesignGraph}.
Recent perception studies suggest a need to refine these important foundations because visual variables may interact with each other \cite{Smart:2019:Separability} and with other cognitive processes such as attention \cite{Healey:2012:AttentionVisual}.
Another related challenge in visualization design is to avoid the delivery of overwhelming amounts of information by managing data complexity \cite{Brath:1997:MetricsEffectiveVis}. For example, Henry et al. \cite{Henry:2008:SocialNetworks} refer to readability as \emph{visual complexity} and propose a solution to reduce this complexity in social network representations. To account for the influence of display on readability, in our work we tested our scale items on visualizations showing different amounts of data points and visual variables during the scale development phase.

\SkipTocEntry\subsubsection{The influence of individuals on readability}
Reading is not just seeing: it involves semantic understanding of the content to construct a coherent mental representation \cite{Freedman-Shah:2002:ComprehensionIntegration, Hegarty:2011:CogVisualRepresentations}.
Accordingly, readers need domain knowledge about what is represented---the \emph{referent}---and skills to build an internal representation of the display.
The ability to construct and use mental representations across various types of data visualizations is called \emph{visualization literacy} \cite{Boy:2014:Literacy}, or \emph{graphicacy} in psychology \cite{Postigo:2004:Graphicacy}.
A growing stream of research focuses on testing and understanding factors of visualization literacy \cite{Firat:2022:VLStateOfTheArt, Lee:2017:VLAT, Borner:2016:VLMuseum}.
Yet many questions remain open regarding this skill \cite{Solen:2022:VLFuture}, including how visualization literacy interacts with other literacies (\eg, numeracy) and how it affects the reader's judgment of a data visualization's trustworthiness.
Several individual traits may also play a role in reading data visualizations \cite{Liu:2020:SurveyIndividualDiff, Peck:2012:3DModelIndividual}, such as spatial abilities \cite{Hall:2022:Spatial, Trickett:2006:SpatialCognition}, or verbal working memory \cite{Steichen:2013:AdaptiveVis}---a measure of the ability to store and manipulate verbal information. While we do not know yet how these factors integrate with existing models of visualization comprehension, we tested our candidate scale items on \revis{common and less common visualizations}. 

\SkipTocEntry\subsubsection{The influence of task on readability}
Finally, a reader's goals also have an impact on how they approach a visual representation.
There is always a reason why people engage with data visualizations \cite{Brehmer:2013:MultiLevelTypology}: it can be as diverse as the need to acquire knowledge, the desire to communicate ideas to others, or sheer aesthetic pleasure. 
Researchers have long noticed that the task at hand interacts with visual encodings when a reader processes information from a visualization \cite{Simkin:1987:GraphInformationProcessing}, and can influence perceptual processes \cite{Friel:2001:MakingSense}. Recently, Quadri and Rosen \cite{Quadri:2022} surveyed 132 perception studies and distributed them across 11 low-level tasks from Amar \etal \cite{Amar:2005:LowLevelTasks}.
In a study using visualizations from the MASSVIS data set, Polatsek \etal \cite{Polatsek:2018:SaliencyTask} found that readers' eye gaze patterns were more closely related to the task at hand than to the visual saliency of objects in the visualization.
Wang \etal \cite{Wang:2022:ScatterplotHardComprehend} conducted a systematic study on scatterplots and noticed that the influence of the visual design and the dataset's size on participants' accuracy and response time were different depending on the comprehension task at hand.
In line with existing knowledge on text readability \cite{Meyer:2003:TextReadability}, good readability of a visualization will entail different 
design requirements depending on the reader’s goal.
Since readability is thus dependent on tasks, in our work we ensured during the scale development and validation tests that participants perform reading tasks before they rated the perceived readability of a visualization.

\SkipTocEntry\subsubsection{Definitions of readability}

Despite all of this past work, only few formal definitions of the concept of readability in visualization exist. We are aware of the following:
\begin{itemize}[nosep,left=0pt .. \parindent]
\item \emph{``The relative ease with which the user finds the information he is looking for''} by Ghoniem \etal \cite{Ghoniem:2005:ReadabilityGraphs}, who proposed this definition in the context of comparing user reading performance for matrices with node-link representations of graphs. Similarly, Tu and Shen \cite{Tu:2007:TreemapsChanges} based a definition on eye movement in their work on treemaps: \emph{``how easy to visually scan a layout to find a particular item, based on how many times viewers' eyes have to change scan direction when traversing a layout.''} Although both definitions em\-pha\-size the central role of the reader\revis{'s goal}, they focus on the visual query and do not include the ability to make sense of retrieved visual objects.

\item \emph{``The ability to make direct observations from the visualizations''} \cite{Ruchikachorn:2015:LearningVisualizations}. Ruchikachorn and Mueller proposed this definition to characterize reading tasks in their work on the use of analogies for teaching novices how to read unfamiliar visualizations.
Its minimalist phrasing applies to a wide variety of visual representations, and with the word ``direct'' they seem to remove interactive features from their scope. It describes, however, a learner's ability rather than a property of the interaction between the reader and the visual object. In that sense, it appears to be closer to a definition of visualization literacy coupled with domain knowledge than to a definition of readability.

\item \emph{``The extent to which a visualization supports the graphical perception of the information it contains''} \cite{Thudt:2016:ReadabilityStacked}. In the context of assessing readability of stacked steamgraphs, Thudt \etal proposed this definition centered on the visual object, and added the important notion of \emph{information}. Yet, this definition fails to capture the individual nature of readability as there is no \revis{explicit} mention of a reader. 
We \revis{also} note that this definition relates to a previous definition of \emph{graphical perception}: \emph{``the visual decoding of information encoded on graphs''} \cite{Cleveland:1984:GraphPerception}\revis{, which does not encompass further understanding}. 
\end{itemize}
\revis{While none of these definitions fully captures the \emph{perceived readability} construct we study in this work, each contributes to its broad scope.}

\SkipTocEntry\subsection{Measuring readability}
\label{subsec:measuring_readability}
User studies in data visualization research frequently involve measuring the readability of visual designs.
The most commonly reported \revis{considerations} 
are participant task performance, layout metrics, \revis{subjective} feedback, and, to a lesser extent, eye-tracking recordings.


\textbf{Task performance} assessment consists of measuring task completion time and answer accuracy for data visualization tasks---such as retrieving a value, detecting a trend, or comparing two visual components in the visualization.
For example, Bu \etal \cite{Bu:2021:Sinestream} assessed the readability of stacked area graphs based on accuracy and completion time for three tasks designed by Thudt \etal \cite{Thudt:2016:ReadabilityStacked}: read the thickness of an individual layer, read the variation of an individual layer's thickness, and read the overall thickness of aggregated layers. 
A few studies evaluated readability based on accuracy alone \cite{Skau:2017:Readability}, but more often researchers collected and analyzed both time and accuracy (\eg, \cite{Bu:2021:Sinestream, Wallinger:2021:readability, Henry:2008:SocialNetworks, Ware:2002:CognitiveMeasurements}).

There are some limitations to using task performance as a proxy for readability: the usability of interactive features can arguably affect completion speed, a participant’s prior beliefs can influence the accuracy of their response \cite{Xiong:2023:Beliefs}, and the task at hand might extend beyond the scope of reading (\ie, \revis{mental calculations}).
To put it briefly, task performance on a data visualization may be \emph{influenced by}---but is not a \emph{\revis{targeted} measure of}---readability, as many co-factors play a role.

\revis{\textbf{Layout drawing metrics} are also used to assess readability by approximating representations’ desired visual properties. In node-link visualizations (\eg, \cite{Bachmaier:2007:RadialHierarchical, Haleem:2019:ReadabilityGraphsDeepLearning}), these are called aesthetics metrics. They include} edge crossings, node overlapping, neighborhood preservation, or global symmetry \cite{Purchase:2002:MetricsGraphDrawing, Bennett:2007:Aesthetics}. 
\revis{Beyond node-link layouts,} 
Giovannangeli \etal \cite{Giovannangeli:2023:ScatterplotsVisibility} proposed a \revis{drawing} algorithm to improve the readability of scatterplots, expanding from previous work on overplotting reduction \cite{Li:2023:OverlapFree}, and Goffin \etal \cite{Goffin:2014:WordScalePlacement} computed metrics to quantify how different placements of word-scale visualizations affected the readability of documents.
Still, the range of such metrics is necessarily limited, and \revis{existing} evaluation instruments
cannot be applied to novel types of representations.
This estimative approach also does not \revis{consider individual factors influencing readers' visual perception and understanding.}

\textbf{Eye-tracking recordings} provide spatio-temporal and physiological data that allow researchers to explore the visual behavior of visualization readers \cite{Kurzhals:2014:EyeTrackingVisAnalytics,Koch:2023:VisPsyEyeTracking}, and to evaluate the usefulness of visual representations \cite{Matzen:2016:ImageUtility}.
This technique is already an established method in cognitive science to study reading strategies on visual displays \cite{Matzen:2017:AttentionPatterns, Netzel:2017:ReadingMaps}.
Eye-tracking studies contribute to building datasets, which can then be used to test cognitive \revis{saliency} models \cite{Livingston:2020:PerceptualCognitiveModels,Polatsek:2018:SaliencyTask} or to train AI models for human gaze prediction \cite{shin_2023_ScannerDeeply}.
However, this method requires heavy logistics both for conducting a user study and analyzing the resulting data, and the eye-tracking device can feel invasive for the participant.

Finally, visualization researchers commonly use \textbf{\revis{subjective} assessments} in experiments \cite{Lam:2012:7scenar,Isenberg:2013:SRP} as \revis{such} data contributes to building a more holistic understanding of how people engage with data visualizations \cite{Saket:2016:User_eval}. \revis{Subjective assessment methods include sketching observations \cite{chan_2013_GeneralizedSensitivity}, semi-structured interviews \cite{wen_2023_QuantivineVisualization}, think-aloud protocols \cite{sedlmair_2012_RelExVisualization}, and open-ended comments in surveys \cite{ajani_2022_DeclutterFocus}, often associated with rating questions.}
During the preliminary step of our present work, we reviewed 34 studies with at least one question aiming at collecting \revis{subjective ratings} relevant to the concept of readability (as shown in \autoref{tab:pool_1_sources} in \autoref{sec:app:terms}).
The questionnaires we retrieved not only varied greatly in terms of wording or number of questions asked; we also found a diversity of answer options and numbers of rating categories.
For example, some researchers asked people to agree on several statements using a 7-point Likert scale from ``Totally Disagree (1)'' to ``Totally Agree (7)'' \cite{Bu:2021:Sinestream}; others asked people to rate visualization using 5 unnamed points between polar opposite terms such as ``Well-organized'' and ``Poorly-organized'' \cite{Nusrat:2018:Cartogram}; and even others asked participants to rank 3 visualizations from 1 to 3 based on their ``legibility,'' with the possibility to give them equal rankings \cite{Perin:2016:GapCharts}.
Such discrepancies prevent a comparison of results across studies.
Standardized and validated tools to measure readability of data visualizations from the reader's perspective solve this problem---which is what we aim to provide with the validated \revis{measuring} instrument we develop in this work.

\SkipTocEntry\subsection{Scale development}
\label{subsec:scale_development_related_work}

\begin{figure} [t!]
    \centering%
    \includegraphics[width=.965\linewidth]{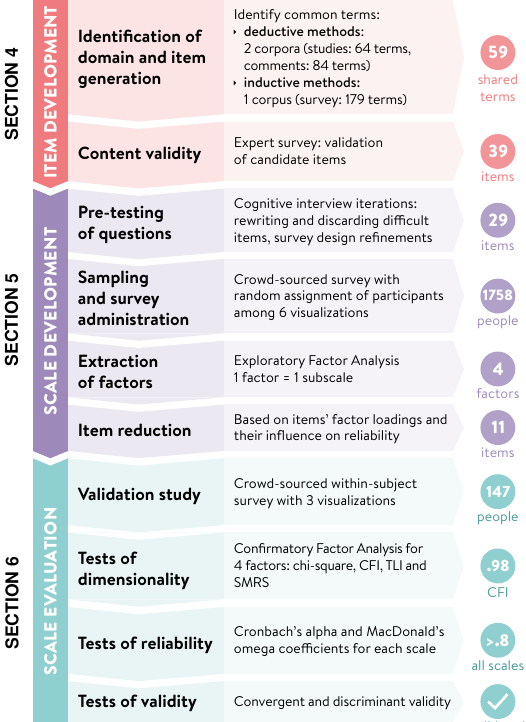}
    \caption{\protect\revis{Summary of our method (adapted from Boateng \etal \cite{Boateng:2018:BestPractices}}). 
    }
    \label{fig:scale-steps}
\end{figure}

\revis{Researchers use psychological scales to} measure a specific, identified construct among the respondents (the ``latent variable'') \cite{Devellis:2021:ScaleDevelopment}. \revis{Scales involve a set of items, each item typically consisting of a statement or a question with matching Likert scale response options.}
Scales can increase statistical power, granting reliable results for smaller sample sizes, which is a particularly relevant issue for usability evaluation studies \cite{Alroobaea:2014:HowManyParticipants}.
As scales provide standardized results, they also facilitate the replicability of studies and comparison of results in a meta-analysis.
But a scale can only provide the aforementioned benefits if researchers follow a rigorous approach to develop and validate its content and its form.
To ensure a high validity and reliability of the final instrument, we rely on recent scale development work in our field \cite{He:2022:Beauvis} and follow best practices in scale development methodology \cite{Boateng:2018:BestPractices, Devellis:2021:ScaleDevelopment} at every step.

\SkipTocEntry\section{What is PREVis and how to use it}

\revis{Following established methodologies for scale development\cite{Boateng:2018:BestPractices, Devellis:2021:ScaleDevelopment}, we contribute \scalename, a validated instrument for comparing the perceived readability of different visualizations. As we show in \autoref{fig:teaser}, it comprises 11 items across four subscales that cover the following concepts (\ie, ``dimensions'': see our glossary in \autoref{sec:app:glossary}):}\vspace{-1.5ex}
\revis{\begin{description}[\compact\setlabelphantom{\SUn:}]
\item[\SUn:] the intelligibility of the encodings for the reader
\item[\SLa:] the visual clarity of the layout
\item[\SDR:] how easily people feel they can read data values
\item[\SDF:] how easily people feel they can read data patterns
\end{description}}
\vspace{-1ex}
\revis{\scalename can be used in experiments, after participants completed reading tasks, to measure their perception of a data visualization's readability in a standardized way. It can also complement qualitative work to evaluate user experience when developing a new technique.
PREVis is not meant to replace existing empirical tools in data visualization research. Instead, our instrument can be used to complement other measuring approaches such as task performance or user interviews and help researchers build a more holistic view of their empirical findings.}

\revis{Together, the four concepts in \scalename capture perceived readability more holistically, but researchers can also use each subscale independently to evaluate the corresponding concept.  
Each \scalename item should be presented using a fully-labeled 7-point Likert scale. We recommend to include an ``I don't know'' option with a text field to collect qualitative feedback when participants feel they cannot answer, and to treat such answers as missing data (N/A) for quantitative analysis purposes.
For each subscale, researchers can obtain a single score by averaging individual items' ratings.
This single score should be viewed comparatively and not interpreted as an absolute measure of the related concept. We do not recommend to calculate an average score between scales as these capture separate dimensions of readability.}

\revis{To create and validate \scalename, we followed a three-phase, nine-step process outlined in \autoref{fig:scale-steps}. We introduce each step in detail in the following sections. We provide details of our methodology and results in our \researchLog, definitions for the technical terms we use in \autoref{sec:app:glossary}, as well as detailed usage guidelines at \href{https://osf.io/9cg8j}{\texttt{osf.io\discretionary{/}{}{/}9cg8j}}.}
\SkipTocEntry\section{Item development}
\label{sec:phase1_items}
In this first phase of the process, we established the boundaries of the readability concept, identified 59 relevant terms and generated a set of 39 candidate items.
To start, we sought to delineate the domain of our scale.
We collected definitions and models of \emph{reading} and \emph{readability} from the dictionary, the psychology literature, and the data visualization literature.
Beyond the summary of our findings from \autoref{subsec:defining-readability}, we offer more details on our explorations in the \emph{Domain definition} section of our \researchLog.
In the absence of an existing theoretical framework for our question, methodological guidelines on scale development recommend that the domain be specified \emph{a posteriori} \cite{Boateng:2018:BestPractices}. 
Accordingly, we moved on to the next step: to identify a pool of suitable terms.

\SkipTocEntry\subsection{Relevant terms identification}
\label{subsec:terms_identification}

To generate a list of rating items relevant to assess readability in data visualization, we first needed to establish a pool of relevant words to use. There are two general ways to identify appropriate terms for scale items: deductive and inductive methods \cite{Boateng:2018:BestPractices}. We used both and constitute 3 different pools of terms, as we show in \autoref{tab:terms_collection}. We provide details and data files in our \suppmat{OSF supplemental material folder}{9cg8j}.

For each corpus, we followed a similar selection process: the first author began by reviewing the corpus' content and retrieving broadly relevant keywords or expressions.
All authors then collectively reviewed this work and selected words according to the inclusion criteria described further below. We refer to the selected words as the collected \revis{\textit{terms}}.
At this stage, we intentionally used over-inclusive criteria because the scale development methodology later allows us to detect weakly related items and it is important not to miss any variables that should have been included \cite{Clark:1995:ConstructingValidity}. After each item collection, we applied \emph{stemming} \cite{Porter:1980:Stemming} to reduce the different variations of a word to its root. For example, the words ``distracting'' and ``distracted'' both yield the ``distract'' root (called \emph{stem}).
We refer to the unique stems as \revis{\textit{unique}} terms in the following paragraphs for the sake of reading ease.

\SkipTocEntry\subsubsection{Deductive method}
\label{subsubsec:deductive_method}
We conducted a literature review on IEEE VIS papers (1990–2022) and TVCG and CG\&A journal papers presented at IEEE VIS (2011–2021).
With the Boolean query \code{"likert" AND ("readab" OR "legib" OR "deciph")} we found 128 papers, from which we derived two sets of terms: terms used in study questionnaires and terms from reported comments of participants. 13 publications provided words for both pools of terms.
We included terms from questions and comments related to the easiness of reading, perception, and understanding, as well as the effectiveness or efficiency of visual elements for a given task.
We excluded questions and comments related to the visualization system (e.g., interactivity, general preference) or to aesthetic judgment.

\textbf{Terms from study questionnaires (Pool 1).} For this pool we focused on collecting words used in items of rating questions in user studies.
In addition to the inclusion criteria above, we included terms from questions that authors related to readability in the method, results, or discussion sections.
We expected the term ``visualization'' to be over-represented and we thus excluded it (see \autoref{sec:app:terms} for more details on the process and results).
We collected 135 terms (64 unique ones) from 34 publications (\autoref{tab:pool_1_sources} in \autoref{sec:app:terms}).

\textbf{Terms from participant comments (Pool 2).} For this pool, we focused on reported comments from participants.
Texts in this corpus were sometimes direct quotations from participants comments, and sometimes we reworded them to summarize comments from multiple participants.
Beyond the exclusion criteria we listed above, we also excluded terms such as ``usable'' as it would be difficult to assess whether it was related to the visual representation alone or to the whole visualization system, including interactive features.
We collected 165 terms (84 unique) from 34 publications (\autoref{tab:pool_2_sources} in \autoref{sec:app:terms}).

\SkipTocEntry\subsubsection{Inductive method}
\label{subsubsec:inductive_method}
To complement our literature review with input from experts, we conducted a survey, which we pre-registered (\prereg{osf.io/4dcav}{4dcav}) and for which we received IRB approval (Inria COERLE, avis \textnumero~2023-17).

\textbf{Participants.} We invited 106 visualization experts by direct e-mail to participate in our survey. We selected them based on our knowledge of their work and their reputation in the visualization community, ensuring that their expertise covers a wide range of topics. We did not compensate participants for taking part in the study. After sending the invitations, we waited for 10 days and, during this time, received 29 complete responses (experience in the field: mean 17.5 years
; 6 women (cis or trans), 21 men (cis or trans), 2 rather not answer). All responses were valid and we included them in our analysis.

\textbf{Procedure.}
We first asked participants to complete the informed consent form and to answer background questions about their gender and years of experience. We then explained the study scenario, which involved wanting to study people's perception about the readability of a visualization they created, using a 7-point Likert scale with the question: ``To what extent do you agree or disagree with the following statement: [\dots].'' We asked each participant to give us at least three statements they would use to fill in the blank in the question. We also gave them the opportunity to leave additional comments after providing us with their items suggestions, if they chose to do so.

\textbf{Results.}
From the 29 completed surveys we collected 132 items and 13 additional comments. We found 3 items aimed at assessing whether the visual object is a data visualization or not; as this was not in the scope of our work, we discarded these 3 items. Then we split the remaining 128 items into 147 statements (\eg, ``I can process the data elements quickly to form an overview of the result.'' was separated in two statements: ``I can process the data elements quickly'' and ``(I can) form an overview of the result''). We used this corpus of statements to establish our third pool of relevant terms, and to find conceptual and phrasing patterns that would inform the writing of our scale's items.

\textbf{Terms from expert survey (Pool 3).}
For this pool, we worked on the assumption that experts proposed content highly relevant to our work. Thus, we only had a few exclusion criteria: we did not include terms describing the visualization (\eg, ``picture'', ``chart''), common verbs (\eg, ``can'' or ``have''), and terms from secondary prepositional phrases (\eg, in the statement ``When I look at this image, I immediately understand how to recover the data values'' we did not retrieve any word from the phrase ``when looking at the visualization'').
We extracted 381 keywords or key expressions (\eg, from the previous example we extracted ``immediately understand; how to; recover; data values''). With this process we collected 447 terms (176 unique terms). We provide this table as a separate file in our \suppmat{supplemental material}{5e3xc}, as it is too long for comfortable reading in a paginated document.

\begin{table}[t]
  \caption{Summary of our terms collection.}\vspace{-1ex}
  \label{tab:terms_collection}%
  \scriptsize%
  \centering%
  \tabulinesep=0.5mm
  \begin{tabu} {@{\extracolsep {0pt minus 1fil}}X[1,l,m]X[1.3,l,m]X[0.7,c,m]X[1.0,c,m]@{\extracolsep {0pt minus 1fil}}} 
    \toprule
    \textbf{method} & \textbf{corpus} & \textbf{sources} & \textbf{terms (unique)} \\
    \bottomrule
    \multirow{2}{2cm}{\rule{0pt}{2em}\textbf{deductive} \emph{literature review: 128 publications}} & questionnaires in studies (Pool 1) & 34 studies & 135 (64) \\
    & \rule{0pt}{1.1em}reported participants' comments (Pool 2)& 34 studies & 165 (84) \\
    \midrule
    \textbf{inductive} \newline \emph{expert survey: 132 statements}  & collected propositions of items (Pool 3)& 29 experts
    & 447 (179) \\
    \bottomrule
    \multicolumn{3}{r}{\rule{0pt}{1em}\textbf{Sum of collected terms (overall unique terms)}} & \textbf{747 (249)}\\
    \bottomrule
  \end{tabu}
\end{table}

\SkipTocEntry\subsection{Item generation}
\label{subsec:item_generation}
We aligned all unique terms from our 3 pools and filtered out terms present in only one pool. We obtained a list of 59 unique terms common to at least 2 pools (\autoref{tab:all_terms_collected} in \autoref{sec:app:terms}).
We excluded 12 irrelevant or overly redundant terms (\eg, we excluded ``hard'' as it was too redundant with ``difficult''). From the remaining 47 terms, we identified 35 primary terms (\eg, ``clear'', ``understand'') and 12 auxiliary ones (\eg, ``data'', ``easy'') as we describe in our \researchLog.

\textbf{Phrasing and conceptual patterns in expert propositions.} Creating scale items does not only require a set of relevant terms; researchers also need to choose how to combine the terms and write sentences that can be used as appropriate rating items \cite{Devellis:2021:ScaleDevelopment}. 
To inform such choices in the generation of our candidate items, we manually analyzed how experts phrased their statements through syntactic roles and conceptual families of words.
We provide an example of this work in \autoref{sec:app:items}, and a more detailed description 
in our \researchLog. Based on this process, we obtained a summary of the experts' statements in the form of hierarchical data representing the flow of sentences, aggregated by conceptual families of terms.
We used this data to help us answer questions such as:
\begin{itemize}[nosep,left=0pt .. \parindent]
\item What kind of objects are referred to as being ``clear''?
\item Are there frequent descriptions of how people should be able to “understand” readable visualizations (\eg, quickly)?
\item Should we write one or two items with the term ``inform'', which stems from two words: ``information'' and ``informative''?
\item Should we write ``This visualization is understandable'' or ``I can understand this visualization''?
\end{itemize}
Referring both to the statements summary and to the original sources of terms, the first author generated a first draft of possible items, with at least one possible item being based on each of the 35 primary terms. When relevant according to the statements summary, we also included items based on auxiliary terms. The team then discussed all items and selected 39.
In particular, we discussed the drawbacks of using reverted items such as ``I find this visualization complex to read.''
\revis{We decided against rewriting these items to keep them as easy to read as possible \revis{until the pre-testing user study (see \autoref{subsec:pretest}).}}
Finally, we harmonized the wording of all 39 items as we show in \autoref{tab:all_items_generated} in \autoref{sec:app:items}.

\SkipTocEntry\subsection{Item validation}
\label{subsection:candidate_items_validation}
Experts are highly knowledgeable in the domain, and are also potential users of our final measuring instrument.
It is thus recommended \cite{Boateng:2018:BestPractices} to seek validation of the generated items among expert judges. To that end, we conducted a study, again pre-registered (\prereg{osf.io/d9nmu}{osf\discretionary{}{.}{.}io\discretionary{/}{}{/}d9nmu})
and IRB-approved (Inria COERLE, avis \textnumero~2023-17).

\textbf{Participants.} We invited the same 106 visualization experts as before to participate in our survey by direct e-mail. Again, we did not compensate participants for taking part in the study. After sending the invitations, we waited for 30 days and, during this time, received 31 complete responses (experience in the field: mean 18.9 years; 8 women (cis or trans), 20 men (cis or trans), 1 non-binary, 2 rather not answer).

\textbf{Procedure.}
We first asked participants to agree to a consent form and to provide background information about their gender and experience in data visualization. We then displayed the list of 39 items and asked them to rate each item according to its relevance for describing the perceived readability of a data visualization, using a 1--5 point Likert scale (1 $=$ ``not at all relevant,'' 5 $=$ ``very relevant''). Participants could leave additional comments after providing us with their ratings. A few comments pointed out that the item ``I find parts of the visualization distracting'' should have been marked as a reverted item, which was an oversight on our part when finalizing the survey.

\textbf{Results.}
We calculated the mean, the mode, and the median score for the 39 items. All means were above 3, and all modes and medians were above or equal to 3 (see \autoref{tab:candidate_items_expert_validation} in \autoref{app:sec:Expert validation survey results}). As a result, we kept all 39 items for the next stage of scale development.

\SkipTocEntry\section{Instrument development: exploratory phase}
\label{sec:phase2_development}
When creating a scale, the goal of this phase is to establish the underlying dimensions of the construct---in our case, ``perceived readability''---and to select the most appropriate items to measure it.
As we found from our related work readings, multiple factors are likely to influence the perception of a visualization's readability, \eg, the clarity of a visual design, or the reader's ability to understand how the data is represented and its topic, or the difficulty of the reading task at hand.
To make this issue even more delicate, such factors may also be correlated among themselves: a reader may, for example, \revis{feel that a visual design is unclear because they are not familiar with the type of representation, or they might find it difficult to observe data patterns because of visual clutter.}
In addition, 39 items would be too many for the easy-to-administer measuring instrument we envisioned.
We thus needed to reduce our pool and to select items that can best reflect how readable respondents find a given data visualization.
Exploratory Factor Analysis (EFA) \cite{Watkins:2018:EFABestPractices} was specifically developed to help researchers determine the underlying factors in a measured construct and to help them identify the most relevant variables---in our case, scale items---to measure it. In scale development, measures of reliability such as Cronbach’s alpha (α) and McDonald’s omega (ω) complement the EFA approach to select items which will best contribute to the instrument's reliability \cite{Devellis:2021:ScaleDevelopment}.
To collect data for conducting such types of analysis we ran a third experiment, again pre-registered (\href{https://osf.io/4dcav}{\texttt{osf.io/4dcav}}) and IRB-approved (Inria COERLE, avis \textnumero 2023-17).

\SkipTocEntry\subsection{Experiment design}
\label{subsec:exploratory_study_design}
The goal of this study was to collect participants' ratings of perceived readability for several data visualizations for conducting EFA and reliability analyses. As our goal was to develop an instrument that could be used to measure perceived readability (1) across multiple populations, (2) with diverse visualization idioms, and (3) across a wide range of readability levels, we now describe our choices regarding these 3 elements in our study design.

\textbf{Target population.} Candidate scale items should be tested on a heterogeneous sample of the target population \cite{Boateng:2018:BestPractices}.
We focused on the general population as a baseline because we wanted our instrument to be useful across multiple populations in research.
In particular, we decided not to apply any exclusion criteria other than English language fluency.
We decided to recruit participants from 
Prolific and to conduct our study online, as we describe in more detail in \autoref{subsec:exploratory_survey}. 

\textbf{Visualization stimuli.}
Here, we refer to the visualizations we asked participants to rate with candidate scale items as ``stimuli.''
As our envisioned instrument should capture information about perceived readability across a broad spectrum of data visualizations, we needed to test the items among various stimuli---\ie, across visualizations presenting variability in aspects that are likely to impact readability. In the absence of an existing objective instrument to evaluate readability for a diverse set of visualizations, we focused on variations in the underlying data (number of data entities and attributes represented) and on the visual encoding (number and appropriateness of visual variables used to encode the data attributes and expected familiarity of visualization idiom). As a result, we used the 6 visualizations in \autoref{fig:teaser} as stimuli for this study; for their characteristics and our design rationale see \autoref{sec:app:exploratory_survey_design}.

\textbf{Reading tasks.} We needed to ensure that participants would at least attempt to read the data visualization, before asking them to rate their perception of its readability.
For an online survey, it meant that we would have to give them reading tasks to perform, before showing the rating items.
As this reading experience would shape their opinion on the readability of the visualization, we needed to ascertain that we would only use tasks that are within the scope of ``reading.''
For instance, a task that requires additional mental calculation such as evaluating an average or a sum would not be appropriate.
We reviewed three main taxonomies of visualization tasks \cite{Amar:2005:LowLevelTasks, Brehmer:2013:MultiLevelTypology, Curcio:1987:MathRelationships} to identify the following list of possible reading tasks: \emph{retrieve value}, \emph{find extremum}, \emph{determine range}, \emph{characterize distribution}, \emph{find anomalies}, \emph{cluster}, \emph{find trend or correlation}, and \emph{make comparisons}.
As an additional criterion, we wanted the reading tasks to be relatively simple and quick to complete for participants, regardless of their experience or skill in reading visualizations.
The quality of the collected ratings would be crucial at this stage of our work, and careful reading and answering of 39 rating questions would require sustained attention from the participants.
As fatigue has been documented to appear after 10 minutes in crowdsourced studies \cite{zhang_2018_UnderstandingFatigue}, our goal was to allow respondents to complete the survey under this threshold. 
While acknowledging that the difficulty of the reading task at hand might have an impact on how readable participants find a visualization, our main concern was to allow respondents to stay focused throughout the entire survey.
Easiness of a visualization task is not absolute; instead, it relates to how appropriate a visualization is for a task. As such, we referred to the work from Lee \etal \cite{Lee:2017:VLAT} on building a Visualization Literacy Assessment Test (VLAT), to select two easy reading tasks for each stimulus, as we describe in more detail in \autoref{sec:app:exploratory_survey_design}.
To sum up, we retained two criteria when designing our tasks: easiness and adequacy to the scope of reading.

\SkipTocEntry\subsection{Pre-testing}
\label{subsec:pretest}
Pre-testing items before administrating a survey is a crucially important step: it helps to ensure that items are actually meaningful to the target population \cite{Boateng:2018:BestPractices}. The goal of pre-testing is to revise the phrasing of items to maximise their clarity, eliminate items that cannot be improved, and examine the extent to which people are able to use the answer options to produce ratings. As such, it is also a way to integrate insights from members of the target population in the scale development process. For this study we recruited 11 participants and conducted cognitive interviews \cite{beatty_2007_ResearchSynthesis}, a form of think-aloud protocol dedicated to evaluating questionnaires. As a result, we made changes to the items in-between rounds of interviews. In particular, we reworded reverted items to negative phrasing, we dropped 11 items, and we created two items to replace the ``cluttered'' term with two related and clearer terms: ``crowded'' and ``messy.'' Participants' feedback also allowed us to refine the presentation of stimuli, the survey's user interface, and the Likert-scale options. For reasons of space, we include the details of this study's design, procedure, and outcomes in \autoref{sec:app:pretest}. The study received approval from our IRB (Inria COERLE, avis \textnumero 2023-17).

\SkipTocEntry\subsection{Survey administration}
\label{subsec:exploratory_survey}
From the pre-test study we obtained a final set of 29 items, which we used to conduct our exploratory survey. We ran the survey in two separate rounds with 6 stimuli described in \autoref{subsec:exploratory_study_design} and \autoref{sec:app:exploratory_survey_design}.

\textbf{Participants.} It is difficult to find a consensus on how large a sample should be for an exploratory study. A general rule is that the more items one wants to test, the more participants are required. In line with suggestions from our methodological references \cite{Boateng:2018:BestPractices, Devellis:2021:ScaleDevelopment}, we targeted a sample size of 300 participants per visualization. We recruited participants from Prolific, who had to be fluent English speakers and of legal age. Participants received a compensation of \euro{}\,11.52 per hour.

\textbf{Procedure.}
After answering a consent form and a question about color-vision deficiency, each participant was randomly assigned to one of the 6 possible stimuli. A short contextual description and a title complemented each stimulus image. We asked participants to answer 2 reading questions and 1 comprehension check question about the visualization.
Regardless of their answers to the reading questions, participants had to answer the comprehension check correctly within two attempts to be able to move on to the next part of the survey. In that final section of the survey, we asked participants to rate the visualization using our 29 candidate items, randomized with one attention check item.
Participants answered on a 7-point agreement scale. Each point was labeled, from ``strongly disagree'' to ``strongly agree,'' and there was a separate option labeled ``I don't know / Not applicable'' with a short text field.
We share additional details in \autoref{sec:app:exploratory_survey_design} (\eg, a screenshot in \autoref{app:fig:survey_screenshot}), and printouts of the surveys in our
\suppmat{supplemental materials}{9cg8j}.

\SkipTocEntry\subsection{Survey results}
\label{subsec:exploratory_survey_results}
We recruited a total of 1,801 participants, who all provided their informed consent. Due to inconsistencies between Prolific and our collected data,
we removed data for 10 participants in the first deployment round.
In addition, we excluded 33 participants from the analysis who failed attention check questions. As a result, we included ratings from 1,758 participants (ages: mean 32.2 years, SD 10.9 years; 39\% female, 59\% male, 2\% non-binary, and $<$1\% gender not disclosed; education: $<$1\% no formal education, 5\% secondary education, 20\% high school diploma, 10\% technical or community college, 41\% undergraduate degree, 22\% graduate degree, 2\% doctorate); color-vision deficiency: 3\% yes). Due to our random assignment of participants to our 6 stimuli, each stimulus received 293 valid ratings on average (SD $=$ 8.27).

\textbf{Missing data.} As we offered an option to answer ``I don't know'' to rating questions in the exploratory survey (\autoref{app:fig:survey_screenshot} in \autoref{app:sec:missing_data}), there was missing data in our collected ratings. As pre-registered, we followed guidelines from Mirzaei \etal \cite{mirzaei_2022_MissingData} on handling such cases by calculating the amount of missing data and testing if it was ``Missing Completely at Random'' (MCAR) using the \texttt{misty} package in \texttt{R}. We did so survey-wise and stimulus-wise. Missing data was not MCAR \revis{for three stimuli}, 
but it was always negligible \revis{($<1\%$)}. We report the details of this analysis and subsequent missing data treatment in \autoref{app:sec:missing_data}.

\SkipTocEntry\subsection{Instrument dimensions and item reduction analyses}
\label{subsec:exploratory_analyses}

Item reduction analysis in scale development aims at maximizing the instrument's measuring accuracy---\ie, minimizing the measurement error, while minimizing its length---and, thus, the time required for respondents to answer all items. Two theories provide tools to assist scale development \cite{Boateng:2018:BestPractices}: Classical Test Theory (CTT) and Item Response Theory (IRT). Each theory's framework provides a different, complementary approach for assessing the measuring performance of items, but they both share the fundamental requirement that scales are \emph{unidimensional} instruments \cite{Devellis:2021:ScaleDevelopment}. In other words, if items are to be combined into a scale, they must reflect one---and only one---construct. When the scale's domain is defined in pre-existing theoretical work, researchers can integrate the fundamental unidimensionality pre-requisite from the start of the item development phase. As a result, a common practice for the scale development phase consists of first analyzing item correlations to reduce the pool of items, before conducting tests of dimensionality \cite{Boateng:2018:BestPractices}. 
We, however, did not have a theoretical framework defining readability in data visualization and, based on our preliminary investigations exposed in \autoref{subsec:defining-readability}, we had suspicions that ``perceived readability'' might not be a unidimensional construct.
Therefore, we first checked that all items were loosely correlated (above 0.3, as shown in \autoref{app:fig:corrMatrix_Agg} in \autoref{app:sec:corr_matrices}),
followed by conducting EFA on the collected data. Only then did we proceed with item reduction analysis, based on factor loadings and reliability tests for each individual factor.

\SkipTocEntry\subsubsection{Exploratory Factor Analysis (EFA)}
\label{subsubsec:EFA}
We conducted our main analysis on the full dataset from valid participations in our survey. As an additional exploratory analysis, however, we also conducted EFA on each individual stimulus' dataset. This extra precaution allowed us to confirm that statistical analyses converged towards 3--5 factors. Finally, we conducted a multi-group Confirmatory Factor Analysis (CFA) to determine whether our questionnaire elicited similar response patterns across stimuli, thus serving as a confirmation of our factor structure before we proceeded with the final item reduction analyses.
We followed Watkin's best practices guide \cite{Watkins:2018:EFABestPractices} for conducting EFA. We provide the \texttt{R} and \texttt{Python} notebooks we used in this process as a part of our \suppmat{supplemental material}{9cg8j}.

\textbf{EFA parameters.} Running EFA requires us to make informed choices regarding analysis parameters \cite{Watkins:2018:EFABestPractices}. We selected the following analysis settings: (1) we used a Principal Axis (PA) factoring method, which does not entail distributional assumptions \cite{fabrigar_1999_EvaluatingUsea}, because our data did not meet the normality requirements for Maximum Likelihood methods when we tested for univariate and multivariate normality; (2) an oblique rotation method (Promax) because underlying factors were likely to violate assumptions of independence, and (3) a common factor analysis model because it is better suited to scale development than Principal Component Analysis \cite{Devellis:2021:ScaleDevelopment}. We also confirmed that our data's correlation matrix was factorable before running the EFA. We report results from all tests we conducted prior to EFA in \autoref{app:sec:data_checks_EFA}.

\textbf{Number of factors.} Following our reference literature recommendations \cite{Watkins:2018:EFABestPractices, Devellis:2021:ScaleDevelopment}, we used \emph{parallel analysis} and \emph{scree plots} (\autoref{app:sec:EFA_screeplots}) to assess how many factors were likely to explain covariance patterns of our candidate items.
In EFA, a parallel analysis determines the number of factors that capture more variance than what would be expected by random chance. In our case, 3--5 factors seemed to be necessary to explain the variance in our data, as we show in \autoref{tab:parallel_analyses_h}.

Parallel analysis, though, does not provide a definitive answer as to how many factors should be retained during scale development \cite{Devellis:2021:ScaleDevelopment}. Instead, researchers should analyze the output \emph{factor loadings} tables and assess whether or not groups of items appear to reflect meaningful constructs. Because scree plots showed a distinct slope break after factor 1, and parallel analysis suggested 3--5 factors depending on the data subsets, we examined factor loadings tables for structures ranging from 1 to 5 factors (more detail in \autoref{app:sec:EFA_loadings}).
We also examined \emph{model fit metrics} for each factor structure \cite{finch_2020_UsingFit}, which we detail in our \researchLog. From these combined observations, we concluded that a 4-factors solution produced the most meaningful grouping of items to describe perceived readability. We then removed items with cross-loadings (loadings in more than one factor with a value above a cutoff value of 0.32 \cite{taherdoost_2014_ExploratoryFactor}) and conducted a Multi-Group Confirmatory Factor Analysis (MG-CFA) to assess how appropriate the 4-factor solution was to explain variance in collected rating, across individual stimuli. We present a short description of this analysis in \autoref{app:sec:EFA_CFA_4factors}. The resulting fit indices shown in \autoref{tab:exp_fit_CFA_fullModel} allowed us to confidently proceed with the development of PREVis as an instrument with four subscales. 

\begin{table}[t]
\centering
\fontsize{7pt}{7pt}\selectfont
\caption{Survey-wise and stimulus-wise parallel analyses suggest that 3 to 5 factors are necessary to explain the data better than at random.}\vspace{-1ex}
\label{tab:parallel_analyses_h}
\tabulinesep=0.8mm
\begin{tabu} to \linewidth 
{X[0.5,c]X[.3,c]X[.3,c]X[.3,C]X[.3,c]X[.3,c]X[.3,c]}
\toprule 
\textbf{Full survey} & A & B & C & D & E & F \\
\midrule
\textbf{5} & 5 & 4 & 3 & 4 & 4 & 3 \\
\bottomrule
\end{tabu}\vspace{-1ex}
\end{table}

\textbf{Four factors in PREVis.} 
As we show in \autoref{app:fig:4factors_loadings_commented} in \autoref{app:sec:EFA_CFA_4factors}, we identified four underlying constructs from our EFA:
\SUn, \revis{ which relates to the ease of understanding visual encodings} with 7 items such as ``It is obvious for me how to read this visualization'';
\SLa, with 4 items related to visual clarity such as ``I don't find this visualization crowded'', or ``I don't find this visualization messy''; 
\SDR, in reference to Curcio's first level of ``\textit{reading the data}'' \cite{Curcio:1987:MathRelationships}, with 5 items such as ``I can easily find specific elements in this visualization'', or ``I can easily retrieve information from this visualization''; and
\SDF, which relates to Curcio's ``\textit{reading between the data}'' \cite{Curcio:1987:MathRelationships},  with the 2 items in our questionnaire related to seeing ``data features (for example, a minimum, or an outlier, or a trend)''. For the remainder of our work, we considered each of these groups of items as individual scales, and we refer to them as \emph{subscales}.

\SkipTocEntry\subsubsection{Item reduction for each subscale}
In this step we evaluated the performance of individual items to identify the most appropriate ones to constitute each subscale. The goal of item reduction is to obtain a parsimonious final instrument, while still ensuring reliability of its measurements. 
The main indicator of reliability used in scale development is Cronbach’s alpha, which estimates the scale’s total variance attributable to a common source \cite{Devellis:2021:ScaleDevelopment}.
DeVellis and Thorpe \cite{Devellis:2021:ScaleDevelopment} consider 
alpha values of 0.7--0.8 to be acceptable, and 0.8--0.9 to be very good. Above 0.9, researchers should consider shortening the list of items. With four subscales in our final instrument, our main goal was to obtain good reliability with a minimum length. We decided to first target a number of 3 items per scale, as a construct with fewer than 3 items is generally weak and unstable \cite{costello_2005_BestPractices}, and to add more items if needed until reaching an alpha coefficient >\,0.8. We had only 2 items to integrate in \SDF, but this factor was nonetheless very stable 
for all EFAs we conducted with 3 or more factors.

We refer to \autoref{app:sec:items_reduction} for more details on the item reduction. We obtained the final set of items summarized in \autoref{tab:exp_models_reliability_loadings} and conducted a Multi-Group Confirmatory
Factor Analysis (MG-CFA) to assess how appropriate our scales are to capture information from the exploratory survey across stimuli. We obtained good reliability for each scale (0.88--0.92, \autoref{tab:exp_reliability_model_final}) and fit metrics for the final model (\autoref{tab:item_reduction_models_fit}).

\SkipTocEntry\section{PREVis subscales validation}
Scales need proper validation before they can be used as data collection instruments. Validation means verifying (a) that the scale's \textbf{dimensionality} (\ie, number of factors) remains the same with an independent sample of the population, (b) that the scales' \textbf{reliability} remains high for this independent sample, and (c) that the scales actually measure the intended \textbf{construct} (here, perceived readability via four subscales: \SUn, \SLa, \SDR, and \SDF). 
To validate PREVis, we ran a fourth experiment, again pre-registered (\href{https://osf.io/yex32}{\texttt{osf.io/yex32}}) and IRB-approved (Inria COERLE, avis \textnumero 2023-17).

\SkipTocEntry\subsection{Study design}
\label{subsec:validation_study_design}
Dimensionality and reliability tests for scale validation are run on data from an independent sample, using the final tool in a similar context. We thus designed a second crowd-sourced survey to administer our scale, similar to that of the exploratory study, with the following differences: we used 3 node-link graphs as stimuli (\autoref{fig:valid_ratings}, top), all par\-ti\-ci\-pants rated all stimuli, and we presented rating items for each subscale on a single screen. For details of our study design see \autoref{app:sec:validation_survey_design}.

In addition, construct validity tests require supplemental measures and correlation tests on three criteria:
\MTMMoneText
\textbf{inter-subscales reliability:} our subscales' scores should highly and positively correlate among themselves, indicating the existence of a shared underlying construct in respondents (\ie, perceived readability);
\MTMMtwoText
\textbf{discriminant validity:} our subscales' scores should not correlate with measures of a different, unrelated construct, measured with a similar method (\ie, in our, case, a 7-point Likert scale); and
\MTMMthreeText
\textbf{convergent validity:} our subscales' scores should positively correlate with another \revis{readability-related indicator measured using a different method}.

For discriminant validity, we chose to use measures of the ``extraversion'' personality trait, which we obtained using the corresponding 2 scale items from a validated 10-items version of the Big Five personality Inventory \cite{rammstedt_2007_MeasuringPersonality}. We chose this measure as it is performed on a 7-point Likert scale, and we have no reason to think that extraversion should correlate---positively or negatively---to perceived readability in visualization. For convergent validity, as we stated in \autoref{sec:related}, we do not have a validated instrument to collect objective or subjective measures of readability. Some instruments exist, however, that can produce metrics related to readability: in particular, algorithms can compute graph readability metrics \cite{dunne_2015_ReadabilityMetrica} on node-link layouts.
We generated 3 different node-link visualizations with a \texttt{D3.js} force-directed component to serve as stimuli in this survey. We then used Gove's \texttt{Greadability.js} library \cite{gove_2018_ItPays} (\texttt{\href{https://github.com/rpgove/greadability}{github\discretionary{}{.}{.}com\discretionary{/}{}{/}rpgove\discretionary{/}{}{/}greadability}}) for calculating graph layout metrics on each graph, as we document in \autoref{app:sec:validation_survey_design}.

\SkipTocEntry\subsection{Survey administration}
\textbf{Participants.} We targeted a sample size of $\geq$\,110 participants (10 per item in the final tool). We recruited participants from Prolific, who had to be fluent English speakers of legal age, excluding people who participated in our previous studies. Having noticed that our population in the exploratory survey was skewed towards men---even though women are more represented on Pro\-li\-fic---, we added a gender distribution criterion. Participants received a compensation of \euro{}\,11.52\,/\,h.

\textbf{Procedure.} Participants in this survey first answered a consent form and a question about color vision deficiency. Then, we gave them a brief explanation on how to read a node-link diagram in the context of the study. In the next part of the survey, we presented participants with 3 different node-link graphs, in random order. For each diagram, respondents answered 2 reading questions and rated the visualization using our 11 \scalename items with a labeled 7-point Likert scale. We presented \scalename items grouped by subscale; we randomized the order of subscales, and the order of items within them.
Participants then had an option to leave an additional text comment. Finally, they answered the two extraversion items. We detail the procedure further in \autoref{app:sec:validation_survey_procedure}.

\SkipTocEntry\subsection{Survey results}
\label{subsec:valid_results}
We had provisioned extra Prolific budget in case answering the survey would take people longer than we expected, but participants actually completed the questionnaire a little faster than expected. This allowed us, as preregistered, to extend the study to more participants until budget exhaustion; as a result, we recruited a total of 148 participants. All participants passed our single comprehension check and at least 2 out of our 3 attention checks. As a result, we included ratings from all participants (ages: mean 28.4 years, SD 7 years; 48\% female, 48\% male, 3\% non-binary, and 1\% gender not disclosed; education: 3\% secondary education, 26\% high school diploma, 13\% technical or community college, 38\% undergraduate degree, 18\% graduate degree, 3\% doctorate); color vision deficiency: 4\% yes).

\begin{figure} [b]
    \centering
    \includegraphics[width=.7\linewidth]{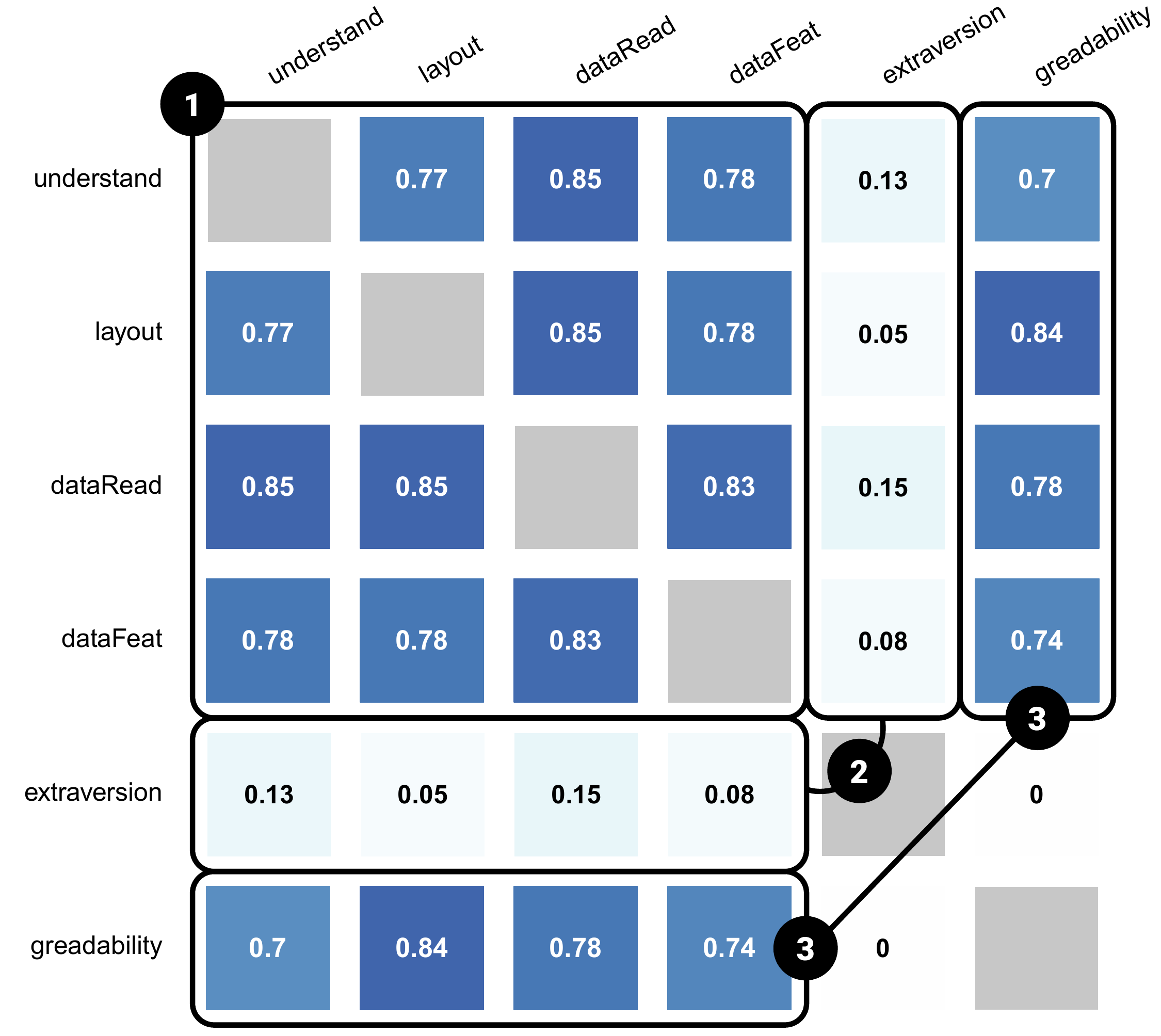}
    \caption{Multi-trait multi-method \protect\revis{composite} correlation matrix \protect\revis{(see details in \protect\hyperlink{appMTMM}{Appx.~\ref{app:sec:construct_validity}})}: \MTMMoneCaption reliability among \scalename subscales, \MTMMtwoCaption discriminant validity from an unrelated personality trait in respondents, and \MTMMthreeCaption convergent validity with graph layout metrics.}\vspace{-1ex}
    \label{fig:MTMM}
\end{figure}

\textbf{Tests of dimensionality.}
To assess the dimensionality of \scalename we conducted a parallel analysis and a Multi-Group Confirmatory Factor Analysis on our collected data, following the same approach as described in \autoref{app:sec:EFA_CFA_4factors} for scale development. We detail in \autoref{app:sec:validation_results} our findings that a 4-factors structure was appropriate, and that the subscales met criteria of good model fit from reference literature \cite{hu_1999_CutoffCriteria}.

\textbf{Tests of reliability.}
We computed alpha and omega coefficients for each subscale. For all scales, all coefficients were in the 0.87--0.96 range (see \autoref{tab:valid_reliabilities} in \autoref{app:sec:validation_results}), which qualifies as very good \cite{Devellis:2021:ScaleDevelopment}.

\textbf{Tests of validity.}
To test the validity of our \scalename instrument, we followed recommendations from our reference literature to create a multi-trait multi-method (MTMM) matrix \cite{Boateng:2018:BestPractices, Devellis:2021:ScaleDevelopment}.

\begin{figure}[b]
    \centering
    \includegraphics[width=1\linewidth]{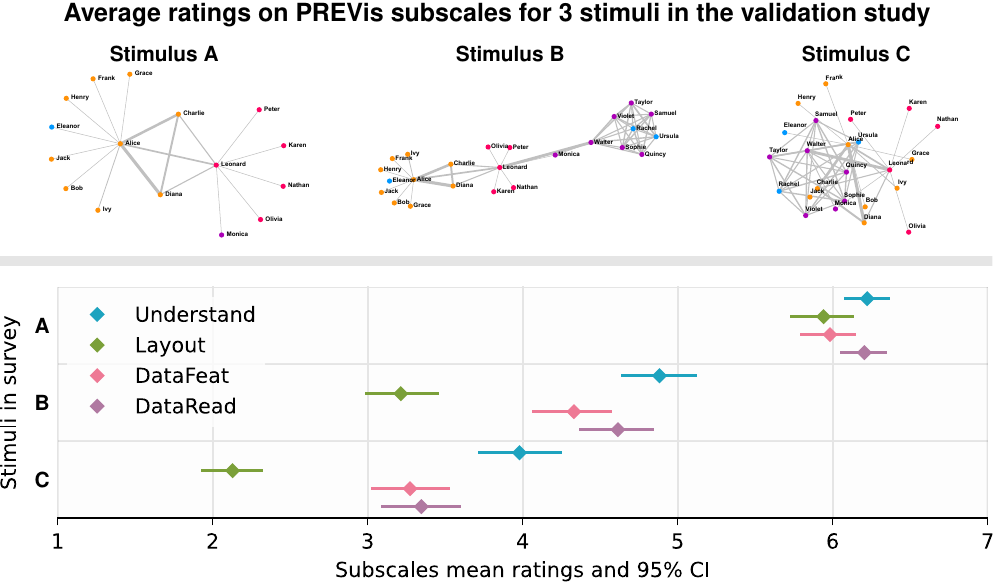}
    \caption{Average ratings (from 1 = ``Strongly disagree'' to 7 = ``Strongly agree'') using the four PREVis \PREVisColors subscales on three node-link visualizations of different readability levels (A $>$ B $>$ C).}\vspace{-1ex}
    \label{fig:valid_ratings}
\end{figure}

A MTMM matrix represents the correlations in data collected from 3 different instruments for assessment of convergent and discriminant validity, as we described in \autoref{subsec:validation_study_design}.
\revis{In \hyperlink{appMTMM}{Appx.~\ref{app:sec:construct_validity}} 
we detail how} we generated the \revis{composite} MTMM matrix in \autoref{fig:MTMM}, which confirmed all 3 criteria:
inter-subscales reliability (positive and high correlations in \MTMMoneText), 
discriminant validity (correlations close to 0 in \MTMMtwoText), 
and convergent validity (positive correlations in \MTMMthreeText). 
Finally, we plotted \scalename \PREVisColors  subscales' average ratings with 95\% CI for each of the 3 stimuli (\autoref{fig:valid_ratings}). For each \PREVisColors subscale, ratings allowed to discriminate clearly between the three stimuli, in the expected order: A $>$ B $>$ C.

\SkipTocEntry\section{Discussion and future work}
\label{sec:discussion}

From our early investigations to define the readability domain, to our exploratory factor analysis in scale development, and our final evaluation of \scalename \PREVisColors performance, we uncovered different components related to how people perceive readability of visualizations. This work allowed us to forge a strong opinion that readability cannot be expressed with a single aggregate measure. For example, when looking at ratings from our validation survey in \autoref{fig:valid_ratings}, or from our exploratory survey in \autoref{app:fig:Exp_ratings_PREVis} in \autoref{app:sec:exp_ratings}, there are clear discrepancies between how easily people estimate they can \SUn a visualization, and how clear they find that visualization's \SLa. 
This is consistent with our view that three families of factors influence readability in data visualization (\autoref{fig:model-reading}): the visualization \textit{display}, the \textit{reading task}, and the \textit{reader}. 

Items from \SUn, in particular, appear to at least partly reflect a \textit{reader's} familiarity with the displayed type of visualization, and by extension to their visualization literacy w.r.t. this specific type of representation \cite{cabouat2024PositionPaperCase}. It is particularly clear in ratings from our exploratory survey (see ratings in \autoref{app:fig:Exp_ratings_PREVis} in \autoref{app:sec:exp_ratings} and visualization stimuli in \autoref{fig:exp_all_stimuli} in \autoref{sec:app:exploratory_survey_design}): the unfamiliar GeneaQuilts visualization (stimulus E) received the lowest \SUn  score among all stimuli, while being rated higher in \SLa  than stimuli D (pie chart with gradient colors encodings) and F (line chart with 18 different categories). 
Future work should investigate the within-participant relationship between visualization literacy tests scores \cite{Lee:2017:VLAT, cui_2024_AdaptiveAssessment} on certain visualizations idioms and \SUn ratings on similar visualizations. \SLa items, conversely, relate more to the clarity of the visualization \textit{display} itself and, without surprise, showed the highest correlation with graph layout metrics among \scalename subscales during our validation study (\textit{r} $=$ 0.74 in \autoref{fig:MTMM}).
Future work should assess to what extent \SLa alone can provide valid measures of
display clarity in visualizations for which we do not have layout metric algorithms. 
The two other dimensions in \scalename relate more to the outcomes of \textit{reading tasks}: visual retrieval of data values in \SDR, and visual saliency of data patterns in \SDF. \revis{These measures, when collected alongside task performance metrics such as speed and accuracy, can provide researchers with a more comprehensive view of how well a given visualization supported participants in their reading tasks.} 
Future work should determine whether\revis{---and under which conditions}---these indicators can predict actual reading task performance in respondents. If such work were to provide positive results, researchers might opt to use \SDR and \textls[-13]{\SDF as proxies for reading task performance measures \revis{in studies where the visualization's objective efficiency is only a peripheral concern.}}

As such, each subscale in \scalename \PREVisColors can be regarded as an individual measure, allowing researchers to choose a subset of the four scales in which they are particularly interested. We draw attention to the fact, however, that the collected data would then be missing the other dimensions we established, thus preventing researchers from drawing conclusions regarding \emph{perceived readability}. It is our opinion that researchers who want to explore higher-level cognitive processes such as decision-making or acceptability of a new visualization system need to take perceived readability into account in its full dimensionality, \revis{among other variables. Future work could also} examine the relationship between \scalename \PREVisColors \revis{measurements} with other aspects of visualization such as user engagement \cite{Saket:2016:User_eval} or visualization trustworthiness \cite{elhamdadi_2024_VistrustMultidimensional}. Findings from such studies would allow us to expand our theoretical understanding of how people process information in visualizations, ultimately providing better guidance for visualization design and recommender systems.

\SkipTocEntry\section{Using \scalename\PREVisColors, limitations, and conclusion}
\label{sec:conclusion}

When implementing \scalename in user studies, we recommend that participants at least attempt a few reading tasks on a visualization, before evaluating it with all items from each subscale \PREVisColors. They should answer \revis{items on} a fully labeled 7-point Likert scale---at \revis{least}, the neutral point and extremities must be labeled, and not numbered. We encourage researchers to provide an ``I don't know / Not applicable'' option or an optional comment field to let participants provide more detailed and qualitative information along with \scalename \PREVisColors ratings. Indeed, in our two crowd-sourced surveys, we received many comments that were very relevant to better understanding \revis{our participants' reading experiences.}
\revis{We highlight that} \SLa items are negatively phrased, which we explain in \autoref{sec:app:pretest}. We advise against reformulating these items---and \scalename items in general. 
However, researchers studying perceived visual clarity alone could choose to use only \SLa, and reformulate those items' statements into affirmative sentences---in which case they should reverse scores before analyzing their data.

A limitation of this work is that we did not attempt to expose participants to more than one visualization before they started to rate images. 
While \scalename \PREVisColors already demonstrated its ability to capture a broad spectrum of readability perceptions, presenting multiple designs prior to rating could potentially enhance the consistency of participants' ratings.
On a related note, during our pre-test interviews we received suggestions from participants to use a readability comparison task (A \textit{vs.} B) rather than a rating scale for a single visualization; similarly, in our validation study we received comments such as ``This one was very messy, compared to the others'', or ``This representation is the best from far !!'' We warn, however, against the temptation to adapt \scalename \PREVisColors items as A/B testing questions because we do not know how reliable such measures would be. Perhaps even more importantly, as researchers have emphasized, generalization of visualization ranking is not without risks \cite{davis_2024_RisksRanking}.
Converting \scalename into a ranking instrument would defeat another purpose of scales, which is to provide standardized results that facilitate cross-study comparisons and meta-analysis.

We developed \scalename as a versatile instrument for diverse visualization types and readers. As the low amount of unanswered questions in our exploratory survey demonstrates (\autoref{subsubsec:EFA}), respondents were able to use \scalename in multiple visualization situations: familiar and unfamilar visualization idioms, categorical, ordinal and continuous numerical data types, a wide range of numbers of data entities, 1--4 encoded data attributes \dots We acknowledge, however, that some visualization contexts may require more specific questions, and that researchers could find that our items are too vaguely phrased to accurately capture perceived readability in their specific research setting. In a network topology reading situation, for instance, one could find it 
useful to replace \SDRDot ``I can easily find specific elements in this visualization'' with ``I can easily find specific nodes in this graph''; or, when working with expert users on cluster detection in graphs, to replace the \SDFDot ``data features'' expression with ``data clusters.''
When doing so, researchers should follow methodological guidelines in scale development and validation---as we extensively documented in this work---to ensure that they produce a reliable and accurate measuring instrument.

Finally, we only focused on static 2D representations with\-out any interaction. \mbox{Readability is already complex to try to} define in such a constrained set of situations; but once it is better established, future research work should expand our understanding of readability to 3D environments, in motion situations \cite{yao_2022_VisualizationMotion}, and with interaction.

In conclusion, our \scalename \PREVisColors instrument is readily deployable in many visualization contexts and, based on our validation, can provide reliable and nuanced measures of perceived readability to researchers. Our work showed that the construct of readability, as visualization readers perceive it, is too complex to be expressed in a single or aggregate measure, and we identified at least four subcomponents. As such, \scalename \PREVisColors also provides multiple opportunities for future research, which can ultimately contribute to better explaining factors and outcomes of readability in data visualization.\vspace{-0.4ex}

\vspace{-.5ex}\acknowledgments{
	We thank Markus Wallinger, who provided their material from \cite{Wallinger:2021:readability}, 
 as well as all Aviz team members for their feedback on the project.\vspace{-0.4ex}}

\vspace{-.5ex}\section*{Supplemental material pointers}
\label{sec:supplemental_materials}

All supplemental materials are available on OSF at
\href{https://osf.io/9cg8j}{\texttt{osf\discretionary{}{.}{.}io\discretionary{/}{}{/}9cg8j}}. We also share our code at \href{https://github.com/AF-Cabouat/PREVis-scales}{\texttt{github.com/AF-Cabouat/PREVis-scales}}.\vspace{-0.4ex}

\vspace{-.5ex}\section*{Figure credits and copyright}
\label{sec:figure_credits}

\autoref{fig:model-reading} was adapted from \cite{Hegarty:2011:CogVisualRepresentations, Fox:2023:theories_models} but is a full re-creation.
All our figures remain under our own copyright and are available under the \href{https://creativecommons.org/licenses/by/4.0/}{Creative Commons \ccLogo\,\ccAttribution\ \mbox{CC BY 4.0}} license; we share them at \href{https://osf.io/9cg8j}{\texttt{osf.io/9cg8j}}.\vspace{-0.4ex}


%

%

\bibliographystyle{abbrv-doi-hyperref-narrow}

\bibliography{abbreviations,template}

\clearpage

\appendix 


\begin{strip} 
\noindent\begin{minipage}{\textwidth}
\makeatletter
\centering%
\sffamily\bfseries\fontsize{15}{16.5}\selectfont
\mytitle\\[.5em]
\large Appendix\\[.75em]
\makeatother
\normalfont\rmfamily\normalsize\noindent\raggedright In this appendix we provide additional explanations, tables, plots, and charts that show data beyond the material that we could include in the main paper due to space limitations or because it was not essential for explaining our approach.
\end{minipage}
\end{strip}

\makeatletter
\renewcommand{\@oddfoot}{\hfil\textrm{\thepage}\hfil}
\renewcommand{\@evenfoot}{\@oddfoot}
\makeatother

\tableofcontents


\section{\revis{Specific glossary}}
\label{sec:app:glossary}
\revis{In the item development section as well as elsewhere in the paper, we use the following terms for which we provide here our contextualized definition to clarify the discussion better:
\begin{itemize}
    \item \textbf{Factor:} we use this word in the context of Exploratory and Confirmatory Factor Analyses 
    where it refers to a latent variable that explains a cluster of covariance among the observed variables (in our case, items). Mathematically, factors are associated with the eigenvectors of a correlation matrix, which we derived from item responses. Each factor is also linked to an eigenvalue, which indicates the proportion of variance in the observed data that the factor accounts for. 
    Within a factor, items have coefficients (\ie, ``loadings'') that represent the degree to which each item is associated with that factor. High loadings on a factor suggest that the items are strongly related to the factor's latent variable. By examining the content and nature of the items, we can interpret the meaning of the factor and propose a name for the underlying latent variable it represents. This how we determined the dimensions in \scalename \PREVisColors as we detail in \autoref{app:sec:EFA_loadings}.
    \item \textbf{Fully-labeled Likert scale:} in the context of items answer options, a fully-labeled Likert scale is a rating scale where each point is associated with a descriptor, also called a \textit{text anchor} \cite{South:2022:EUL}.
    \item \textbf{Instrument:} a measuring tool. In our case, the PREVis instrument consists of a group of 4 related scales.
    \item \textbf{Instrument dimensions:} dimensions are distinct components that contribute to the full construct being assessed (\ie, perceived readability).
    In our case, each dimension is measured using a dedicated \textit{subscale}: \SUn: the intelligibility of the encodings for the reader; \SLa: the visual clarity of the layout; \SDR: how easily people feel they can read data values; and \SDF: how easily people feel they can read data patterns. We derived these dimensions from the four \textit{factors} (see above) we found during our Exploratory Factor Analysis described in \autoref{subsec:exploratory_analyses}.
    \item \textbf{Instrument validity:} validated instruments have been tested to verify that they measure the target construct (\ie, in our case, perceived readability). Validity is established throughout the process of developping an instrument \cite{Boateng:2018:BestPractices}.
    \item \textbf{Item:} in the context of a scale, an item is the combination of a statement or a question with its accompanying answer options. In our work, all items share the same 7-point rating scale answer options.
    \item \textbf{Perceived readability:} how readable a person finds a specific visualization in a given context. It is the construct that PREVis targets in respondents as a measuring \textit{instrument}.
    \item \textbf{Reliability:} a scale's reliability is an indicator of its consistency and stability in measuring what it is intended to measure. Common indices of reliability, such as Cronbach's alpha and McDonald's omega, are based on correlation estimates among the items within the scale. These indices assess the internal consistency, reflecting how well the items correlate with each other and consistently measure the same underlying construct.
    \item \textbf{Psychological scale:} an instrument measuring a single construct in respondents (\ie, how readable people find a visualization). Scales are generally used to help evaluate latent variables, \ie, traits or constructs for which direct observation is not possible. Therefore, scale are considered to be \textit{indicators} of the target variable.
    \item \textbf{Subscale:} in this paper, we call ``subscales'' the four scales that form \scalename \PREVisColors, each measuring a specific dimension of perceived readability. We use the term subscale for the sake of readability in the paper. It is, however, different from the usual acceptance of the term in psychology research, where subscales' scores are usually aggregated to form a higher-level score. For \scalename \PREVisColors, we advise against this practice (see the discussion in \autoref{sec:discussion} as well as in our practical companion PDF on \href{https://osf.io/9cg8j}{\texttt{osf.io/9cg8j}}).
    \item \textbf{Term:} a keyword we collected in one of our source corpora.
\end{itemize}}

\section{Additional Term Collection Details}
\label{sec:app:terms}

\subsection{Deductive method sources}

As described in \autoref{subsubsec:deductive_method}, we extracted terms from study questionnaires (Pool 1) and participant comments (Pool 2). \autoref{tab:polls_1_2_common} shows all 55 sources in this work, and whether they contributed to Pool 1, or Pool 2, or both. 13 sources were common to the two pools of terms.

\begin{table}[h!]
\centering
\fontsize{7pt}{7pt}\selectfont
\caption{55 publications from which we extracted terms for Pool 1 and  Pool 2.}
\label{tab:polls_1_2_common}
\tabulinesep=0.6mm
\begin{tabu} to 0.8\linewidth {X[3,l]X[1,l]X[1,l]}
\toprule
DOI & Pool 1 (studies) & Pool2 (comments)\\ 
\midrule
\href{https://doi.org/10.1109/TVCG.2013.151}{10.1109/TVCG.2013.151} & Yes & Yes\\ 
\href{https://doi.org/10.1109/TVCG.2012.189}{10.1109/TVCG.2012.189} & Yes & Yes\\ 
\href{https://doi.org/10.1109/TVCG.2015.2467872}{10.1109/TVCG.2015.2467872} & Yes & Yes\\ 
\href{https://doi.org/10.1109/TVCG.2021.3068337}{10.1109/TVCG.2021.3068337} & Yes & Yes\\ 
\href{https://doi.org/10.1109/TVCG.2017.2745941}{10.1109/TVCG.2017.2745941} & Yes & Yes\\ 
\href{https://doi.org/10.1109/TVCG.2022.3209475}{10.1109/TVCG.2022.3209475} & Yes & Yes\\ 
\href{https://doi.org/10.1109/TVCG.2020.3030358}{10.1109/TVCG.2020.3030358} & Yes & Yes\\ 
\href{https://doi.org/10.1109/TVCG.2012.255}{10.1109/TVCG.2012.255} & Yes & Yes\\ 
\href{https://doi.org/10.1109/TVCG.2018.2865192}{10.1109/TVCG.2018.2865192} & Yes & Yes\\ 
\href{https://doi.org/10.1109/TVCG.2020.3030388}{10.1109/TVCG.2020.3030388} & Yes & Yes\\ 
\href{https://doi.org/10.1109/TVCG.2011.186}{10.1109/TVCG.2011.186} & Yes & Yes\\ 
\href{https://doi.org/10.1109/TVCG.2017.2744118}{10.1109/TVCG.2017.2744118} & Yes & Yes\\ 
\href{https://doi.org/10.1109/TVCG.2018.2835485}{10.1109/TVCG.2018.2835485} & Yes & Yes\\ 
\href{https://doi.org/10.1109/TVCG.2021.3114789}{10.1109/TVCG.2021.3114789} & Yes & -\\ 
\href{https://doi.org/10.1109/TVCG.2011.183}{10.1109/TVCG.2011.183} & Yes & -\\ 
\href{https://doi.org/10.1109/TVCG.2022.3144975}{10.1109/TVCG.2022.3144975} & Yes & -\\ 
\href{https://doi.org/10.1109/TVCG.2022.3163727}{10.1109/TVCG.2022.3163727} & Yes & -\\ 
\href{https://doi.org/10.1109/TVCG.2020.3030404}{10.1109/TVCG.2020.3030404} & Yes & -\\ 
\href{https://doi.org/10.1109/TVCG.2015.2467035}{10.1109/TVCG.2015.2467035} & Yes & -\\ 
\href{https://doi.org/10.1109/TVCG.2021.3092680}{10.1109/TVCG.2021.3092680} & Yes & -\\ 
\href{https://doi.org/10.1109/TVCG.2010.194}{10.1109/TVCG.2010.194} & Yes & -\\ 
\href{https://doi.org/10.1109/MCG.2016.100}{10.1109/MCG.2016.100} & Yes & -\\ 
\href{https://doi.org/10.1109/TVCG.2014.2346983}{10.1109/TVCG.2014.2346983} & Yes & -\\ 
\href{https://doi.org/10.1109/TVCG.2020.3004137}{10.1109/TVCG.2020.3004137} & Yes & -\\ 
\href{https://doi.org/10.1109/TVCG.2012.225}{10.1109/TVCG.2012.225} & Yes & -\\ 
\href{https://doi.org/10.1109/TVCG.2022.3209354}{10.1109/TVCG.2022.3209354} & Yes & -\\ 
\href{https://doi.org/10.1109/TVCG.2016.2642109}{10.1109/TVCG.2016.2642109} & Yes & -\\ 
\href{https://doi.org/10.1109/TVCG.2020.3030437}{10.1109/TVCG.2020.3030437} & Yes & -\\ 
\href{https://doi.org/10.1109/TVCG.2013.180}{10.1109/TVCG.2013.180} & Yes & -\\ 
\href{https://doi.org/10.1109/TVCG.2018.2865232}{10.1109/TVCG.2018.2865232} & Yes & -\\ 
\href{https://doi.org/10.1109/TVCG.2018.2865049}{10.1109/TVCG.2018.2865049} & Yes & -\\ 
\href{https://doi.org/10.1109/TVCG.2021.3085327}{10.1109/TVCG.2021.3085327} & Yes & -\\ 
\href{https://doi.org/10.1109/TVCG.2019.2941208}{10.1109/TVCG.2019.2941208} & Yes & -\\ 
\href{https://doi.org/10.1109/TVCG.2011.193}{10.1109/TVCG.2011.193} & Yes & -\\ 
\href{https://doi.org/10.1109/TVCG.2021.3114775}{10.1109/TVCG.2021.3114775} & - & Yes\\ 
\href{https://doi.org/10.1109/TVCG.2013.76}{10.1109/TVCG.2013.76} & - & Yes\\ 
\href{https://doi.org/10.1109/TVCG.2022.3209480}{10.1109/TVCG.2022.3209480} & - & Yes\\ 
\href{https://doi.org/10.1109/TVCG.2021.3114822}{10.1109/TVCG.2021.3114822} & - & Yes\\ 
\href{https://doi.org/10.1109/TVCG.2022.3209484}{10.1109/TVCG.2022.3209484} & - & Yes\\ 
\href{https://doi.org/10.1109/TVCG.2014.2346420}{10.1109/TVCG.2014.2346420} & - & Yes\\ 
\href{https://doi.org/10.1109/TVCG.2013.191}{10.1109/TVCG.2013.191} & - & Yes\\ 
\href{https://doi.org/10.1109/TVCG.2022.3209477}{10.1109/TVCG.2022.3209477} & - & Yes\\ 
\href{https://doi.org/10.1109/TVCG.2014.2329308}{10.1109/TVCG.2014.2329308} & - & Yes\\ 
\href{https://doi.org/10.1109/TVCG.2019.2934784}{10.1109/TVCG.2019.2934784} & - & Yes\\ 
\href{https://doi.org/10.1109/TVCG.2009.176}{10.1109/TVCG.2009.176} & - & Yes\\ 
\href{https://doi.org/10.1109/INFVIS.2002.1173148}{10.1109/INFVIS.2002.1173148} & - & Yes\\ 
\href{https://doi.org/10.1109/TVCG.2019.2934557}{10.1109/TVCG.2019.2934557} & - & Yes\\ 
\href{https://doi.org/10.1109/TVCG.2020.2968911}{10.1109/TVCG.2020.2968911} & - & Yes\\ 
\href{https://doi.org/10.1109/TVCG.2014.2337337}{10.1109/TVCG.2014.2337337} & - & Yes\\ 
\href{https://doi.org/10.1109/TVCG.2019.2934337}{10.1109/TVCG.2019.2934337} & - & Yes\\ 
\href{https://doi.org/10.1109/TVCG.2019.2934669}{10.1109/TVCG.2019.2934669} & - & Yes\\ 
\href{https://doi.org/10.1109/TVCG.2013.233}{10.1109/TVCG.2013.233} & - & Yes\\ 
\href{https://doi.org/10.1109/TVCG.2019.2906900}{10.1109/TVCG.2019.2906900} & - & Yes\\ 
\href{https://doi.org/10.1109/VAST.2009.5332595}{10.1109/VAST.2009.5332595} & - & Yes\\ 
\href{https://doi.org/10.1109/TVCG.2018.2864907}{10.1109/TVCG.2018.2864907} & - & Yes\\ 
\href{https://doi.org/10.1109/TVCG.2011.160}{10.1109/TVCG.2011.160} & - & Yes\\ 
\bottomrule
\end{tabu}
\end{table}

\autoref{tab:pool_1_sources} shows the list of studies \revis{and collected questions} from which we extracted terms to form Pool 1 in \autoref{subsubsec:deductive_method}. \revis{We extracted all words from the questions as terms, except for stop words from the \texttt{nltk} package in \texttt{Python} (\eg, ``to'', ``of'', ``for'', ``between'').}

\autoref{tab:pool_2_sources} shows the list of studies with user comments from which we \revis{manually} extracted terms to form Pool 2 in \autoref{subsubsec:deductive_method}\revis{, along with the list of collected terms}.

\begin{table*}[p]
\centering
\fontsize{7pt}{7pt}\selectfont
\caption{34 studies with questionnaires from which we extracted terms in Pool 1 (see \autoref{subsubsec:deductive_method}).}
\label{tab:pool_1_sources}
\tabulinesep=0.8mm
\begin{tabu} to \linewidth {X[0.65,l]X[0.1,l]X[1.2,l]X[1.6,l]}
\toprule DOI & Year & Type of visualization & \revis{Collected questions from which we extracted terms} for item development\\ 
\midrule
\href{https://dx.doi.org/10.1109/TVCG.2022.3175626}{10.1109/TVCG.2022.3175626} & 2010 & trends & readability; ease of use; visually cluttered\\ 
\href{https://dx.doi.org/10.1109/TVCG.2022.3175626}{10.1109/TVCG.2022.3175626} & 2011 & node-links; maps & confidence; effectiveness; frustration; ease; clutter; enjoyment; simplicity; ease to follow\\ 
\href{https://dx.doi.org/10.1109/TVCG.2022.3175626}{10.1109/TVCG.2022.3175626} & 2011 & node-links & intuitive\\ 
\href{https://dx.doi.org/10.1109/TVCG.2022.3175626}{10.1109/TVCG.2022.3175626} & 2011 & paths & clearly visible; visual clutter\\ 
\href{https://dx.doi.org/10.1109/TVCG.2022.3175626}{10.1109/TVCG.2022.3175626} & 2012 & matrix; node-links & clear arrangement\\ 
\href{https://dx.doi.org/10.1109/TVCG.2022.3175626}{10.1109/TVCG.2022.3175626} & 2012 & flows; paths & easy / hard to interpret; easy / hard to understand\\ 
\href{https://dx.doi.org/10.1109/TVCG.2022.3175626}{10.1109/TVCG.2022.3175626} & 2012 & node-links & effectiveness; prefered look\\ 
\href{https://dx.doi.org/10.1109/TVCG.2022.3175626}{10.1109/TVCG.2022.3175626} & 2013 & node-links & easy to learn to read; confident; low clutter\\ 
\href{https://dx.doi.org/10.1109/TVCG.2022.3175626}{10.1109/TVCG.2022.3175626} & 2013 & location; color & easy to gain a good overview of data; easy to understand displayed information; easy to interpret data values (in visualization); confidence in correct answers\\ 
\href{https://dx.doi.org/10.1109/TVCG.2022.3175626}{10.1109/TVCG.2022.3175626} & 2014 & nan & easy / difficult\\ 
\href{https://dx.doi.org/10.1109/TVCG.2022.3175626}{10.1109/TVCG.2022.3175626} & 2015 & diagrams & easy to read text; easy to read symbols; easy to see link between components; good overview of diagram in mind\\ 
\href{https://dx.doi.org/10.1109/TVCG.2022.3175626}{10.1109/TVCG.2022.3175626} & 2015 & parallel plots &  easy to understand\\ 
\href{https://dx.doi.org/10.1109/TVCG.2022.3175626}{10.1109/TVCG.2022.3175626} & 2016 & line charts & easy / difficult to identify\\ 
\href{https://dx.doi.org/10.1109/TVCG.2022.3175626}{10.1109/TVCG.2022.3175626} & 2016 & cartograms & poor / excellent readability\\ 
\href{https://dx.doi.org/10.1109/TVCG.2022.3175626}{10.1109/TVCG.2022.3175626} & 2017 & scatter plots; 3D scatter plots & feature easy to perceive\\ 
\href{https://dx.doi.org/10.1109/TVCG.2022.3175626}{10.1109/TVCG.2022.3175626} & 2017 & flows, paths & able to read\\ 
\href{https://dx.doi.org/10.1109/TVCG.2022.3175626}{10.1109/TVCG.2022.3175626} & 2018 & flows; maps; paths & good visual design; ease of use\\ 
\href{https://dx.doi.org/10.1109/TVCG.2022.3175626}{10.1109/TVCG.2022.3175626} & 2018 & slides with multiple types of vis & easy to read; easy to follow\\ 
\href{https://dx.doi.org/10.1109/TVCG.2022.3175626}{10.1109/TVCG.2022.3175626} & 2018 & vector glyphs & easy / hard to interpret\\ 
\href{https://dx.doi.org/10.1109/TVCG.2022.3175626}{10.1109/TVCG.2022.3175626} & 2018 & multiple types; point-based; text-based & intuitive\\ 
\href{https://dx.doi.org/10.1109/TVCG.2022.3175626}{10.1109/TVCG.2022.3175626} & 2019 & calendar: mutliple charts in analytics view & easy to understand; fast to find the information\\ 
\href{https://dx.doi.org/10.1109/TVCG.2022.3175626}{10.1109/TVCG.2022.3175626} & 2020 & steamgraphs & good readability\\ 
\href{https://dx.doi.org/10.1109/TVCG.2022.3175626}{10.1109/TVCG.2022.3175626} & 2020 & multiple type of vis; surface-based; colormaps; glyph-based; volumes; continuous colors & readable; feature visible; recognizability; enable overview; unambiguous (obvious to understand); at a glance; able to determine; assessable; interpret\\ 
\href{https://dx.doi.org/10.1109/TVCG.2022.3175626}{10.1109/TVCG.2022.3175626} & 2020 & storyline; paths; timelines; node-links & see clearly\\ 
\href{https://dx.doi.org/10.1109/TVCG.2022.3175626}{10.1109/TVCG.2022.3175626} & 2020 & node-links; color-coded text & easy to understand \\ 
\href{https://dx.doi.org/10.1109/TVCG.2022.3175626}{10.1109/TVCG.2022.3175626} & 2020 & maps; VR & ease of use\\ 
\href{https://dx.doi.org/10.1109/TVCG.2022.3175626}{10.1109/TVCG.2022.3175626} & 2021 & bar charts; line charts & I am confused - it makes sense\\ 
\href{https://dx.doi.org/10.1109/TVCG.2022.3175626}{10.1109/TVCG.2022.3175626} & 2021 & infographics & readability\\ 
\href{https://dx.doi.org/10.1109/TVCG.2022.3175626}{10.1109/TVCG.2022.3175626} & 2021 & multiple; text-based; line-based; dot-based; maps; ... & easy to use for understanding\\ 
\href{https://dx.doi.org/10.1109/TVCG.2022.3175626}{10.1109/TVCG.2022.3175626} & 2022 & multiple charts; bar graphs; line charts; pie charts; dot based; areas & facilitate understanding\\ 
\href{https://dx.doi.org/10.1109/TVCG.2022.3175626}{10.1109/TVCG.2022.3175626} & 2022 & bar charts; line charts; VR & easy to learn; easy to understand\\ 
\href{https://dx.doi.org/10.1109/TVCG.2022.3175626}{10.1109/TVCG.2022.3175626} & 2022 & glyph based; surface based; bar charts & quick to read; easy to read\\ 
\href{https://dx.doi.org/10.1109/TVCG.2022.3175626}{10.1109/TVCG.2022.3175626} & 2022 & tables; text; path; node-link & interpretable\\ 
\href{https://dx.doi.org/10.1109/TVCG.2022.3175626}{10.1109/TVCG.2022.3175626} & 2022 & bar charts & clear to understand; easy to read\\ 
\bottomrule
\end{tabu}
\end{table*}

\begin{table*}[t]
\centering
\fontsize{7pt}{7pt}\selectfont
\caption{36 studies with user comments from which we extracted terms in Pool 2 (see \autoref{subsubsec:deductive_method}).}
\label{tab:pool_2_sources}
\tabulinesep=0.8mm
\begin{tabu} to \linewidth {X[0.65,l]X[0.1,l]X[1.2,l]X[1.6,l]}
\toprule DOI & Year & Type of visualization & Terms collected for item development\\ 
\midrule
\href{https://dx.doi.org/10.1109/INFVIS.2002.1173148}{10.1109/INFVIS.2002.1173148} &2002 &node-links &readable; visible \\
\href{https://dx.doi.org/10.1109/TVCG.2009.176}{10.1109/TVCG.2009.176} &2009 &labels; maps &legible \\
\href{https://dx.doi.org/10.1109/VAST.2009.5332595}{10.1109/VAST.2009.5332595} &2009 &timelines &readable; simple \\
\href{https://dx.doi.org/10.1109/TVCG.2011.186}{10.1109/TVCG.2011.186} &2011 &maps; node-links &better; suitable; difficult; follow; lost \\
\href{https://dx.doi.org/10.1109/TVCG.2011.160}{10.1109/TVCG.2011.160} &2011 &charts &decipher; confusing \\
\href{https://dx.doi.org/10.1109/TVCG.2012.255}{10.1109/TVCG.2012.255} &2012 &node-links &rich; usable; comprehensibly; easily; simplicity; speed; understandable \\
\href{https://dx.doi.org/10.1109/TVCG.2012.189}{10.1109/TVCG.2012.189} &2012 &paths &meaningful; view; simple; follow; understand; helpful \\
\href{https://dx.doi.org/10.1109/TVCG.2013.151}{10.1109/TVCG.2013.151} &2013 &node-links &difficult \\
\href{https://dx.doi.org/10.1109/TVCG.2013.191}{10.1109/TVCG.2013.191} &2013 &charts &readability; easy to see; easy to read; legible; ease of visibility; understand; clean; bold; untidy; cluttered \\
\href{https://dx.doi.org/10.1109/TVCG.2013.233}{10.1109/TVCG.2013.233} &2013 &flowcharts &easy; see; difficult; clutter; use \\
\href{https://dx.doi.org/10.1109/TVCG.2014.2346420}{10.1109/TVCG.2014.2346420} &2014 &node-links &occluded; confusing; recognize; simple; guide; intuitive \\
\href{https://dx.doi.org/10.1109/TVCG.2015.2467872}{10.1109/TVCG.2015.2467872} &2016 & parallel coordinates plot &strong; easy \\
\href{https://dx.doi.org/10.1109/TVCG.2017.2745941}{10.1109/TVCG.2017.2745941} &2018 &scatter plots; 3D scatter plots &judge; ease; clear \\
\href{https://dx.doi.org/10.1109/TVCG.2017.2744118}{10.1109/TVCG.2017.2744118} &2018 &strorylines &understand; recognize; disorientating \\
\href{https://dx.doi.org/10.1109/TVCG.2018.2865192}{10.1109/TVCG.2018.2865192} &2019 &flows; maps; paths &easy; difficult; follow; unexpected; clear; distinguish; encoding; intuitive; sparse \\
\href{https://dx.doi.org/10.1109/TVCG.2018.2864907}{10.1109/TVCG.2018.2864907} &2019 &motion patterns &overview; instantly; interpret; hard; cognitive \\
\href{https://dx.doi.org/10.1109/TVCG.2019.2934784}{10.1109/TVCG.2019.2934784} &2020 &fact sheets &easy; understand; meaningful; present (?) \\
\href{https://dx.doi.org/10.1109/TVCG.2019.2934557}{10.1109/TVCG.2019.2934557} &2020 &matrices &easy; perceive; helpful; distinguish; coding; obvious; effective \\
\href{https://dx.doi.org/10.1109/TVCG.2019.2934337}{10.1109/TVCG.2019.2934337} &2020 &dot plots; motion paths; surface-based &discern; interpret; understand \\
\href{https://dx.doi.org/10.1109/TVCG.2019.2934669}{10.1109/TVCG.2019.2934669} &2020 &line-based; bar charts &unclear; hard; meaning \\
\href{https://dx.doi.org/10.1109/TVCG.2020.3030388}{10.1109/TVCG.2020.3030388} &2021 &multiple types: surface-based; colormaps; glyph-based; volumes; continuous colors &clear; intuitive; interpret; correctly; spot; easily \\
\href{https://dx.doi.org/10.1109/TVCG.2020.3030358}{10.1109/TVCG.2020.3030358} &2021 &node-links; color-coded text &difficult; understand; easy; visualize; conceptualize \\
\href{https://dx.doi.org/10.1109/TVCG.2013.76}{10.1109/TVCG.2013.76} &2013 &maps; areas &distracting; confusing; hard; lost; follow; chaos \\
\href{https://dx.doi.org/10.1109/TVCG.2014.2329308}{10.1109/TVCG.2014.2329308} &2015 &maps; motion paths &easier; busy; messy; comfortable \\
\href{https://dx.doi.org/10.1109/TVCG.2014.2337337}{10.1109/TVCG.2014.2337337} &2015 &node-links &cluttered; richer; neat \\
\href{https://dx.doi.org/10.1109/TVCG.2018.2835485}{10.1109/TVCG.2018.2835485} &2019 &multiple types: point-based; text-based &intuitively; easy; hamper; useful; appropriate; confusing; distinguish; efficient; highlight; attract attention; recognize \\
\href{https://dx.doi.org/10.1109/TVCG.2019.2906900}{10.1109/TVCG.2019.2906900} &2020 &node-links; paths &complex \\
\href{https://dx.doi.org/10.1109/TVCG.2020.2968911}{10.1109/TVCG.2020.2968911} &2021 &glyphs &simple; complex \\
\href{https://dx.doi.org/10.1109/TVCG.2021.3068337}{10.1109/TVCG.2021.3068337} &2022 &bar charts; ligne charts &distracting; messy; cluttered; disorganized; unappealing \\
\href{https://dx.doi.org/10.1109/TVCG.2021.3114822}{10.1109/TVCG.2021.3114822} &2022 &multiple types: glyph based; timeline; line-based &useful; insight; easily; understand; prefer; useful; see; relevant; show; attention; obvious \\
\href{https://dx.doi.org/10.1109/TVCG.2021.3114775}{10.1109/TVCG.2021.3114775} &2022 &animation of charts &balance \\
\href{https://dx.doi.org/10.1109/TVCG.2022.3209477}{10.1109/TVCG.2022.3209477} &2023 &multiple types: node-links; matrix; dot based &understand; interpretation \\
\href{https://dx.doi.org/10.1109/TVCG.2022.3209475}{10.1109/TVCG.2022.3209475} &2023 &bar charts; line charts; VR &occlusion; balance \\
\href{https://dx.doi.org/10.1109/TVCG.2022.3209484}{10.1109/TVCG.2022.3209484} &2023 &networks &intuitive; easy, understand \\
\href{https://dx.doi.org/10.1109/TVCG.2022.3209480}{10.1109/TVCG.2022.3209480} &2023 &flows; maps; paths &easy; hard; confusing; visible; see; crowded; convenient; show \\
\href{https://dx.doi.org/10.1109/TVCG.2022.3144975}{10.1109/TVCG.2022.3144975} &2023 &tables; text; flow &intuitive; informative \\
\bottomrule
\end{tabu}
\end{table*}

\subsection{Inductive method sources}
We provide the table of term collection from \autoref{subsubsec:inductive_method} as a separate file in our \suppmat{supplemental material}{5e3xc} because it contains 152 lines, which makes it too long for comfortable reading from a paginated PDF. To form this pool of terms, we extracted individual words from a selection of key expressions. Therefore, the column ``key\_expressions'' hold the collected terms in the reference file.

We also provide a \suppmat{printout of the survey}{4f8va} we used to collect statements from experts as supplemental material.

\subsection{Combining all terms pools}
\autoref{tab:all_terms_collected} shows all collected terms from the 3 different pools of items described in \autoref{subsec:terms_identification}.
\begin{table*}[p]
\centering
\fontsize{7pt}{7pt}\selectfont
\caption{Terms related to readability retrieved from deductive and inductive methods (as described in \autoref{subsec:terms_identification})}
\label{tab:all_terms_collected}
\tabulinesep=0.8mm
\begin{tabu} to \linewidth {X[0.8,l]X[2,l]X[0.5,l]X[1.5,l]X[1.5,l]X[1.5,l]}
\toprule
unique terms & collected terms & overall counts & collected terms from studies & collected terms from comments & collected terms from survey\\ 
\midrule
easi & easy & 40 & easy & easy & easy\\
understand & understand, understanding, understandable & 36 & understand, understanding & understandable, understand & understand, understanding\\ 
interpret & interpret, interpretation & 16 & interpret & interpret, interpretation & interpret\\ 
inform & information, informative & 15 & information & - & information, informative\\ 
clear & clearly, clear & 15 & clearly, clear & clear & clear, clearly\\ 
read & read, reading & 14 & read & read & reading, read\\ 
visual & visually, visual, visualization, visualize & 12 & visually, visual, visualization & visualize & visual, visually\\ 
easili & easily & 12 & - & easily & easily\\ 
clutter & cluttered, clutter & 11 & cluttered & cluttered, clutter & clutter, cluttered\\ 
data & data & 11 & data & - & data\\ 
readabl & readable, readability & 11 & readable, readability & readable, readability & readable\\ 
see & see & 11 & see & see & see\\ 
difficult & difficult & 11 & difficult & difficult & difficult\\ 
confus & confused, confusing & 10 & confused & confusing & confused, confusing\\ 
use & use, useful & 9 & use & use, useful & useful\\ 
hard & hard & 8 & hard & hard & hard\\ 
intuit & intuitive, intuitively & 8 & intuitive & intuitive, intuitively & intuitive\\ 
eas & ease & 6 & ease & ease & -\\ 
confid & confidence, confident & 6 & confidence & - & confident\\ 
overview & overview & 6 & overview & overview & overview\\ 
identifi & identify, identifiable & 6 & identify & - & identifiable, identify\\ 
show & show, shows & 6 & - & show & show, shows\\ 
mean & meaning, means & 6 & - & meaning & meaning, means\\ 
encod & encoding, encoded, encodings & 5 & - & encoding & encoding, encoded, encodings\\ 
visibl & visible, visibility & 5 & visible & visible, visibility & -\\ 
simpl & simple & 5 & simple & simple & -\\ 
relev & relevant & 5 & - & relevant & relevant\\ 
valu & value, values & 4 & value & - & values\\ 
meaning & meaningful & 4 & - & meaningful & meaningful\\ 
help & helpful, help & 4 & - & helpful & help\\ 
present & present, presentation, presented & 4 & - & present & presentation, presented, present\\ 
find & find & 4 & find & - & find\\ 
recogn & recognize & 4 & recognize & recognize & recognize\\ 
learn & learn, learned & 4 & learn & - & learned, learn\\ 
look & look & 4 & look & - & look\\ 
quick & quick, quickly & 4 & quick & - & quickly\\ 
attent & attention & 3 & - & attention & attention\\ 
complex & complex & 3 & - & complex & complex\\ 
answer & answer & 3 & answer & - & answer\\ 
effect & effective & 3 & effective & effective & -\\ 
follow & follow & 3 & follow & follow & -\\ 
distract & distracting, distracted & 3 & - & distracting & distracted\\ 
design & design & 3 & design & - & design\\ 
obvious & obvious & 3 & obvious & obvious & -\\ 
featur & feature, features & 3 & feature & - & features\\ 
arrang & arrangement & 2 & arrangement & - & arrangement\\ 
correct & correct, correctly & 2 & correct & correctly & -\\ 
text & text & 2 & text & - & text\\ 
compon & component, components & 2 & component & - & components\\ 
perceiv & perceive & 2 & perceive & perceive & -\\ 
glanc & glance & 2 & glance & - & glance\\ 
grasp & grasp & 2 & grasp & - & grasp\\ 
discern & discern & 2 & discern & discern & -\\ 
determin & determine & 2 & determine & - & determine\\ 
deciph & decipher & 2 & - & decipher & decipher\\ 
unexpect & unexpected & 2 & - & unexpected & unexpected\\ 
easier & easier & 2 & - & easier & easier\\ 
attract & attract, attracting & 2 & - & attract & attracting\\ 
insight & insight & 2 & - & insight & insight\\
\bottomrule
\end{tabu}
\end{table*}

\textbf{Error log.} In both \autoref{tab:pool_1_sources} and \autoref{tab:all_terms_collected} we noticed a term which should have been excluded: ``visualization''. Its presence is the result of a manual copy error, which we noticed too late in our qualitative analysis process for correcting the data. As all other instances of this word were initially excluded, the associated count do not reflect the actual content of the analyzed questionnaires.

\clearpage

\section{Generated items details}
\label{sec:app:items}
\subsection{Table of 39 generated items}
\autoref{tab:all_items_generated} contains all items generated as described in \autoref{subsec:item_generation}, and with additional details below. 
\begin{table*}[p]
\centering
\footnotesize
\tabulinesep=0.8mm
\caption{Outcome of the item generation step (\autoref{subsec:item_generation}): 39 items generated from primary and secondary relevant terms.}
\label{tab:all_items_generated}
\begin{tabu} to \textwidth{ X[2, l] X[0.5, l] X[1, l] }
\toprule
\textbf{Item statement} & \textbf{Primary term} & \textbf{Auxiliary term(s)}\\
\midrule
I can answer questions about the data after reading this visualization & answer & data, read\\ 
The visual encodings (e.g., color, shapes...) in this visualization make the information clear to me & clear & encod, inform\\ 
This visualization shows information in a clear way for me & clear & inform\\ 
I find this visualization cluttered [reverted item] & clutter &  -\\ 
I find this visualization complex to read [reverted item] & complex & read\\ 
I am confident in my interpretation of this visualization & confid & interpret\\ 
I find this visualization confusing [reverted item] & confus &  -\\ 
This visualization allows me to correctly interpret the data & correct & data\\ 
I find the visual encodings (e.g., color, shapes...) difficult to decipher in this visualization [reverted item] & deciph & difficult, encod\\ 
I can determine key information in this visualization & determin & key, inform\\ 
This visualization allows me to discern elements of the data & discern & data\\ 
I find parts of the visualization distracting [reverted item] & distract &  -\\ 
This visualization effectively shows the data to me & effect & data, show\\ 
I can understand how the data is encoded in this visualization & encod & data, understand\\ 
I can easily find specific elements or information in this visualization & find & easili, inform\\ 
I find this visualization easy to follow & follow & easi\\ 
This visualization helps me understand the data & help & understand, data\\ 
I can identify relevant information in this visualization & identifi & relev, inform\\ 
I can easily retrieve information from this visualization & inform & easili\\ 
I find this visualization informative & inform &  -\\ 
I find this visualization easy to interpret & interpret & easi\\ 
I find this visual design intuitive & intuit & visual, design\\ 
The visual encodings (e.g., color, shapes) in this visualization make sense to me & make sense & encod
\emph{(Note: "make sense" is used as a replacement for "meaningful")}\\ 
I can easily interpret the overall meaning of the data visualization & mean & overall, interpret\\ 
I can understand what the visual components of the visualization mean & mean & understand, visual, compon\\ 
It is obvious for me how to read this visualization & obvious & read\\ 
I find this visualization well organized & organiz & \emph{(Note: "organiz" is used as a replacement for "arrang")}\\ 
This visualization provides me with a good overview of the data & overview & data\\ 
I can easily perceive data features (e.g., trends, minimums, outliers...) in this visualization & perceiv & easili, data, featur\\ 
I can easily read this visualization & read & easili\\ 
I find this visualization readable & readabl &  -\\ 
I can recognize data features (e.g., trends, minimums, outliers...) in this visualization & recogn & data, featur\\ 
I can clearly see data features (e.g., trends, minimums, outliers...) in this visualization & see & data, featur, clear\\ 
This visualization shows the data in an appropriate manner for me & show & data\\ 
I find this visualization simple to read & simpl & read\\ 
I can easily understand this visualization & understand & easili\\ 
I can quickly understand this visualization & understand & quick\\ 
I can extract data values from this visualization & valu & data\\ 
I find data features (e.g., trends, minimums, outliers...) visible in this visualization & visibl & featur\\ 
\bottomrule
\end{tabu}
\end{table*}
\subsection{A detailed example of how we built a candidate item}
\label{app:subsec:item_generation_example}
As mentioned in \autoref{subsec:item_generation}, to combine primary and secondary terms into candidate scale items, we referred to phrasing and conceptual patterns extracted from the statements collected in our expert survey (see \autoref{subsubsec:inductive_method}).
We extracted these patterns by analyzing syntactic roles and conceptual families of words in experts' statements.
\subsubsection{Splitting complex proposed items}
\begin{figure} [b]
    \centering
    \includegraphics[width=\linewidth]{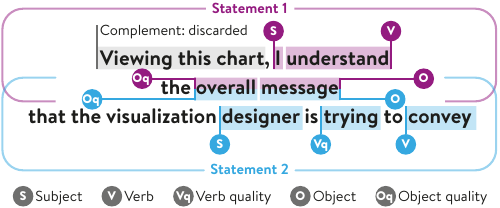}
    \caption{An example of separation of clauses and syntactic role attribution, as described in \autoref{app:subsec:item_generation_example}.}
    \label{app:fig:clause_separation}
\end{figure}
As a first step in this analysis, we split complex items (nested clauses and compound sentences) into individual statements, with a effort to put resulting single-clause sentences in a affirmative, active form. For example, as shown in \autoref{app:fig:clause_separation}, the sentence ``\emph{Viewing this chart, I understand the overall message that the visualization designer is trying to convey}'' was separated in two statements:
\begin{flushleft}
\begin{enumerate}
    \item ``\emph{Viewing this chart, I understand the overall message}'', and
    \item ``\emph{The visualization designer is trying to convey \mbox{an overall message}}''.
\end{enumerate}
\end{flushleft}

In doing this, we transformed 128 proposal items into 147 statements. We then proceeded to the analysis of the structures and concepts in each statement. Here, we will present a detailed example of how we processed the following statement: \newline\newline
\large\centerline{\textbf{\emph{``I can recognize main characteristics of the data''}}}\normalsize\newline

\subsubsection{Categorizing words by role in sentences}
As a second step, we identified the syntactic role and meaningfulness of each term regarding the statement:
\newcommand\itemT{\item[$\rightarrow$]}
\newcommand\itemTq{\item[$\longrightarrow$]}

\begin{itemize}
    \itemT \discreet{(Subject)} \emph{viewer} 
    \begin{itemize}
        \itemTq \discreet{(Subject quality)} \emph{able}
        
    \end{itemize}
    \itemT \discreet{(Verb)}  \emph{recognize}
    \itemT \discreet{(Object)} \emph{data characteristics} 
    \begin{itemize}
        \itemTq \discreet{(Object quality)} \emph{main} 
    \end{itemize}
\end{itemize}
Our work here consisted in capturing the essence of what each statement described and making it consistent. For example, we transformed questions into affirmations, and discarded complements such as ``in the visualization'' or ``on the screen''. We also transformed some words at this stage.
For example, we coded all instances of ``I'' our ``you'' terms into a generic ``viewer'' term, and all terms such as ``chart'', ``visual data representation'', etc. into a generic ``visualization'' term.
We also transformed verbs such as ``can'' and ``am able to'' into a characteristic of the subject labeled ``able'' (as in the example above), which allowed us to focus on the visualization activity as the main predicate.

We refer the reader to our \suppmat{file of annotated statements}{qtjk7} in our supplemental material for a complete account of this process.

\subsubsection{Coding conceptual families}
For each type of role (\discreet{Subject}, \discreet{Subject quality}, \discreet{Verb}, etc.) in our annotated table of expert propositions, we extracted a table of all terms.
Two researchers of the team independently attributed terms to abstract semantic groups: the \textsc{concept} families. They coded a main concept, and an additional one if needed.
For example, the first author coded all following \discreet{Object} terms as \textsc{data feature}: \emph{trends, outliers, pattern, features, trend, patterns, clusters, groups, relationships, communities, activity, level, characteristics}.
Then the two researchers discussed results harmonization on the main concept level, and the remaining uncertainties or unresolved conflicts were taken to the full research team for final decision. Only the main concept was saved for the next step of the work. We then mapped the \textsc{concepts} back to the original structure of each sentence. The example above thus became:
\begin{itemize}
    \itemT \discreet{(S)} \textsc{READER}
    \begin{itemize}
        \itemTq \discreet{(Sq)} \textsc{SUPPORT}
    \end{itemize}
    \itemT \discreet{(V)} \textsc{QUERY}
    \itemT \discreet{(O)} \textsc{DATA}, \textsc{DATA FEATURES}
    \begin{itemize}
        \itemTq \discreet{(Oq)} \textsc{KEY} 
    \end{itemize}
\end{itemize}

With this approach, we produced an set of sentence patterns with 60 \textsc{concepts} from the original 176 unique terms collected in Pool 3 (see \autoref{subsubsec:inductive_method}). 13 additional \textsc{concepts} were found from terms in statements that we did not include in Pool 3.
Our syntactic analysis approach allowed us to synthetize the survey results in sentence patterns while keeping a refined association between original words and their aggregation as \textsc{concepts}.
After having finalized the definition of \textsc{concepts} based on the survey content, the main author expanded the concepts attribution to terms from the other pools from the deductive method described in \autoref{subsubsec:deductive_method}. It was not necessary to create new \textsc{concepts} at this stage. 

\begin{figure} [t]
    \centering
    \includegraphics[width=\linewidth]{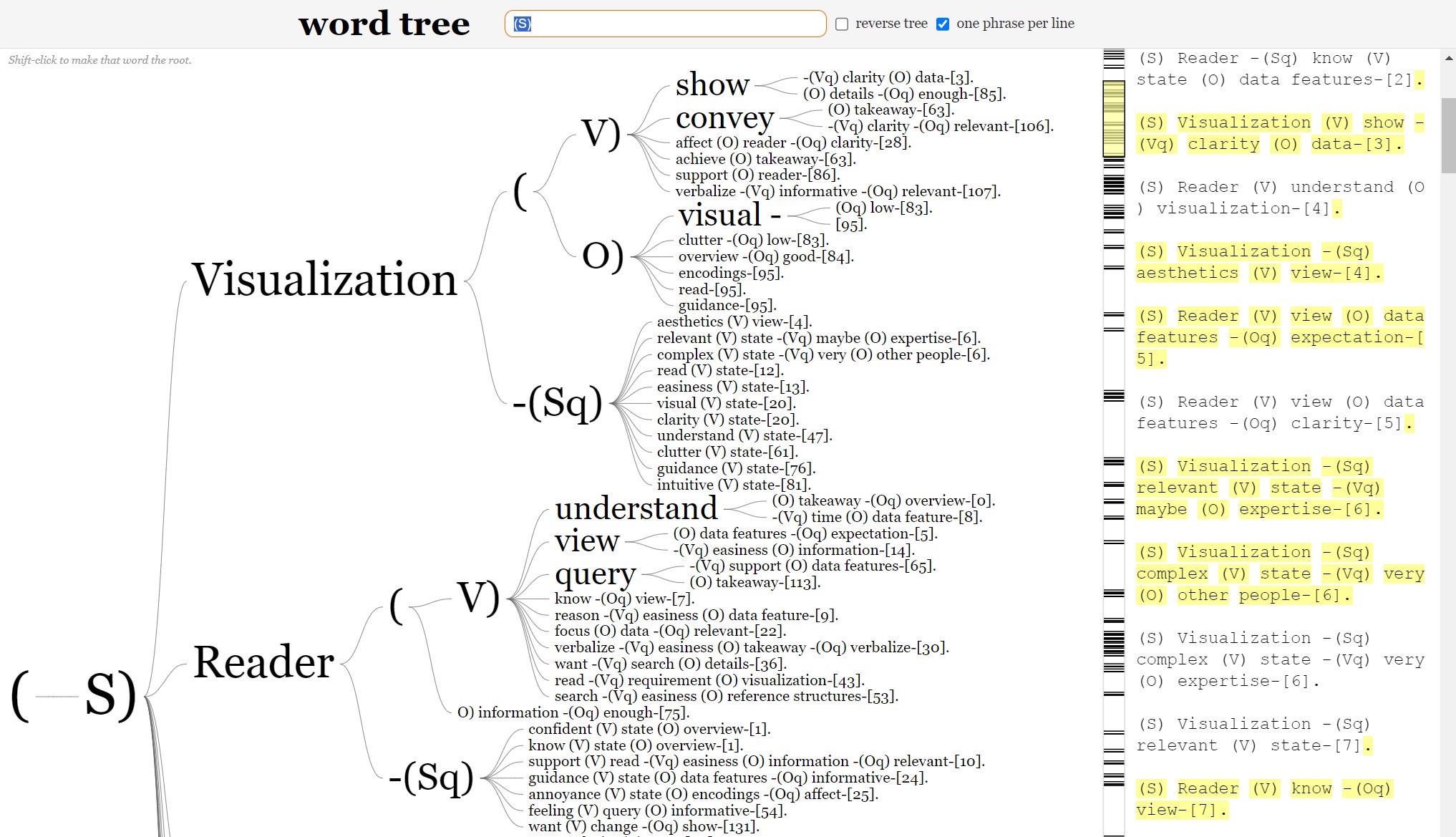}
    \caption{Among other tools, we used Jason Davies's Word Tree online tool at \href{https://www.jasondavies.com/wordtree/}{\texttt{jasondavies.com/wordtree}} to visualize conceptual patterns in collected expert propositions of statements for assessing perceived readability.}
    \label{app:fig:wordTree}
\end{figure}

This work served as guide for item writing: we referred to visual representations of expert statements patterns (for example, using Word Tree, \autoref{app:fig:wordTree}), in combination with the dictionaries of term-\textsc{concept} associations.

These dictionaries (which we called ``lexicons'') were of particular importance because some words can have multiple meanings. For example ‘‘look'' has different meanings between ``it looks awesome'' and ``I don't know where to look''). What's more, at this stage all terms from our 3 pools had been reduced to their root (\ie, their \textit{stem}). Stemming is useful to reduce the number of terms; however it also reduces semantic precision, for example when aggregating ‘‘information'' ‘‘informative'' in a unique ‘‘inform'' \textit{stem}. For these reasons, we decided it was important to refer to the conceptual categorization, which was built from words rather that from stems.

We refer the reader to our \researchLog for a complete account of the writing procedure and a description of the files provided as supplemental material in the OSF repository.

\begin{table}[H]
\centering
\caption{Mean, median, and modes of experts ratings on a scale from 1 to 5 for each candidate item, as explained in \autoref{subsection:candidate_items_validation}}
\label{tab:candidate_items_expert_validation}
\fontsize{7pt}{7pt}\selectfont
\tabulinesep=0.8mm
\begin{tabu} to \linewidth{ X[2, l] X[1, l] X[1, l] X[1, l] X[1, l] X[1, l] X[1, l] }
\toprule
term &median &mean &largest mode &smallest mode &all modes &keep \\\midrule
answer &5 &4.26 &5 &5 &5 &yes \\
clearEncod &5 &4.23 &5 &5 &5 &yes \\
clearInform &4 &4.03 &4 &4 &4 &yes \\
clutter &3 &3.23 &4 &4 &4 &yes \\
complex &4 &3.84 &4 &4 &4 &yes \\
confid &4 &3.81 &5 &5 &5 &yes \\
confus &4 &4.03 &4 &4 &4 &yes \\
correct &4 &3.48 &5 &4 &(4, 5) &yes \\
deciph &4 &3.94 &5 &5 &5 &yes \\
determin &4 &4.00 &4 &4 &4 &yes \\
discern &4 &3.26 &4 &4 &4 &yes \\
distract &3 &3.13 &4 &4 &4 &yes \\
effect &4 &4.06 &4 &4 &4 &yes \\
encod &5 &4.23 &5 &5 &5 &yes \\
find &4 &3.81 &5 &5 &5 &yes \\
follow &4 &3.81 &4 &4 &4 &yes \\
help &4 &3.81 &5 &5 &5 &yes \\
identifi &4 &3.90 &5 &4 &(4, 5) &yes \\
inform1 &4 &4.29 &4 &4 &4 &yes \\
inform2 &3 &3.23 &4 &4 &4 &yes \\
interpret &4 &4.23 &5 &5 &5 &yes \\
intuit &4 &3.77 &4 &4 &4 &yes \\
makeSense &4 &4.00 &5 &5 &5 &yes \\
meanCompon &4 &4.16 &4 &4 &4 &yes \\
meanOveral &4 &3.90 &4 &4 &4 &yes \\
obvious &4 &4.23 &5 &4 &(4, 5) &yes \\
organiz &4 &3.58 &4 &4 &4 &yes \\
overview &3 &3.58 &3 &3 &3 &yes \\
perceiv &4 &3.97 &4 &4 &4 &yes \\
read &4 &4.19 &5 &5 &5 &yes \\
readabl &4 &3.90 &4 &4 &4 &yes \\
recogn &4 &3.81 &4 &4 &4 &yes \\
see &4 &3.90 &4 &4 &4 &yes \\
show &4 &3.77 &5 &5 &5 &yes \\
simpl &4 &3.97 &5 &4 &(4, 5) &yes \\
understandEasi &4 &4.06 &5 &5 &5 &yes \\
understandQuick &4 &3.90 &4 &4 &4 &yes \\
valu &4 &3.65 &4 &4 &4 &yes \\
visibl &4 &3.65 &4 &4 &4 &yes \\
\bottomrule
\end{tabu}
\end{table}

\newpage

\section{Survey results for expert validation of items}
\label{app:sec:Expert validation survey results}

We provide in \autoref{tab:candidate_items_expert_validation} the expert ratings on our 39 candidate items, which we collected in the study described in \autoref{subsection:candidate_items_validation}. All candidate items received a mean, mode and median above 3. Therefore we kept them all for the scale development process, starting with the pre-test study we describe next in \autoref{sec:app:pretest}.

We also provide the \suppmat{questionnaire used for this study}{mx7jy} and the \suppmat{collected data}{jqmc6} in our supplemental materials on OSF.
Experts had the opportunity to comment: we read and integrated those comments in our reflection process.
Three comments pointed out that the statement “I find parts of the visualization distracting” was lacking the “[reversed item]” tag that we put for others statements where a high agreement would indicate lower readability (\eg, ``I find this visualization complex to read'').

\section{Pre-testing study details}
\label{sec:app:pretest}
In this section of the appendix, we provide important details on the pre-testing study we conducted before administrating the exploratory survey on Prolific for the \scalename development phase in \autoref{sec:phase2_development}.

\textbf{Visualization stimuli.} In this survey we used 4 stimuli from the VLAT \cite{Lee:2017:VLAT} (line chart, histogram, choropleth map, and bubble chart) and 2 of their associated VLAT questions, as found on the reVISit webpage \cite{reVISit}. This choice was motivated by the need to test our rating items among a range of low to high readability situations. In the absence of an existing set of readable to non-readable visualizations, we considered two main factors: efficiency of visual design, or difficulty of reading tasks.
In this pre-test, we chose the latter because it offers a principled assessment of the difficulty of design-task pairs.
We provide additional details about the choice of stimuli and associated tasks in our \researchLog.

\textbf{Participants.} We invited participants by sending an email to the staff and students of the first author's university. We did not compensate participants for taking part in this study. We received 12 positive answers but one participant canceled their interview slot. As a result, 11 people participated in one of this study's 3 rounds (4 men, 7 women; age: mean 30, min: 22, max: 47; highest degree completed: 5 Bachelor, 6 Master). 

\textbf{Procedure.} We conducted interviews online \cite{shepperd_2021_GuidelinesConducting} or in-person, depending on the participants' preferences. In both case, participants connected to our online survey platform. We first asked them to read and agree to the informed consent form, and to answer background questions about their gender, education, and English fluency. Participants then completed a training think-aloud task. When they agreed to continue the study after this task, we started the recording of their voice and screen. We then randomly assigned the participants to one of 4 possible stimuli. In round 1 and 2, we ensured that each stimulus was seen at least by one person. Round 3 had only participant and the stimuli was assigned at random. 
Participants answered 2 reading tasks and one comprehension check about the stimulus, after what they rated the visualization using the 39 candidate items and a 7-point agreement scale.
The interviewer observed the participant's behavior to identify hesitations, and occasionally prompted them to elucidate problems if the participants had gone silent. Interviews duration ranged from 45 to 93 minutes.

\textbf{Data analysis.} We used the problem respondent matrix \cite{conrad_1996_ImpressionsData} to analyse the content of the interviews. This tool allows to break down the questionnaire answering process into 3 stages: understanding (\ie, making sense of the question), task performance (\eg, forming an opinion, or recalling an experience), and response formatting (\eg, choosing a Likert-scale answer option); and to categorize respondents potential issues into 5 groups of problems: lexical, temporal, logical, computational, and omission/inclusion. Cognitive interviews are qualitative methods to detect problems, the goal is not to perform quantitative analysis. We did not conduct analysis of other data such as comments on the definitions of ``reading'', ``understanding'' and ``interpreting'' or reading tasks.

\begin{table}[H]
\centering
\caption{Our respondent problem matrix for coding interviews.}
\label{}
\fontsize{7pt}{7pt}\selectfont
\tabulinesep=0.8mm
\begin{tabu}to \linewidth{ X[1.6, l]  X[1, l] X[1, l] X[1, l] X[1, c]} 
\toprule
\textbf{} &\multicolumn{3}{c}{\textbf{RESPONSE STAGE}} \\
\cmidrule{2-4}
\textbf{PROBLEM TYPE} &\textbf{Understanding} &\textbf{Task performance} &\textbf{Response Formatting} \\
\midrule
\textbf{Lexical} &ULe &TLe &RLe \\
\textbf{Logical} &ULo &TLo &RLo \\
\textbf{Computational} &UCo &TCo &RCo \\
\textbf{Omission/Inclusion} &UOm &TOm &ROm \\
\bottomrule
\end{tabu}
\end{table}

\textbf{Items changes.}
Over the 3 rounds of interviews we rephrased items that lacked clarity, and dropped items that could not be clarified. We present the final list of items statements in \autoref{tab:all_items_exp}. In our supplemental material, we provide a \suppmat{summary of item changes}{j7xry} over the 3 rounds. We also share the codings from cognitive interviews in \suppmat{round 1}{pgafr}, \suppmat{round 2}{jsnbp}, and \suppmat{round 3}{7j9ks}. Finally, we share our \href{https://osf.io/hqytm}{methodological notes on cognitive interviewing} as separate material \footnote{We do not share this document in the main supplemental material OSF repository as it contains images from previous work which cannot be redistributed under the CC BY licence}.

In particular, we noticed that participants often incorrectly rated reversed items (\eg, ``I find parts of this visualization distracting''). In many cases, they did not notice it until asked to talk more about their choice of answer. We found that, most items being of positive valence (\eg, ``I find this visualization easy to read''), participants tended to associate the right side of the Likert options (towards ``Strongly agree'') to having a ``good opinion'' of the visualization regarding the thematic (\eg, understanding or distraction), rather than an agreement with the statement itself. Therefore, after round 3 we reworded items of negative valence with a negative turn of phrase (\eg, `I don't find distracting parts in this visualization'').

\begin{figure} [h!]
    \centering
    \includegraphics[width=\linewidth]{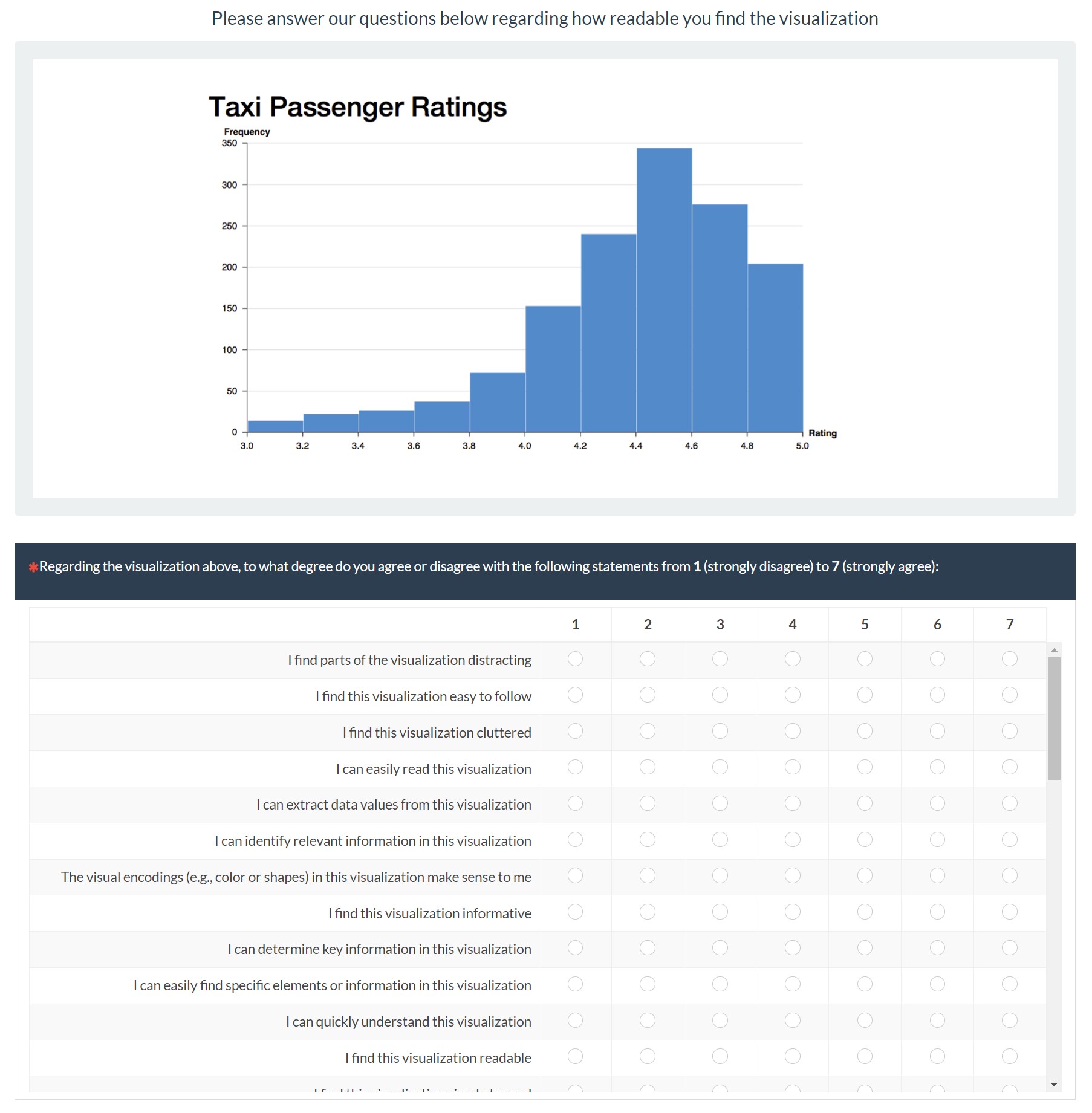}
    \caption{Presentation of a rating item in round 1 of the pre-test study.}
    \label{app:fig:pretestround1}
\end{figure}

\begin{figure} [h!]
    \centering
    \includegraphics[width=\linewidth]{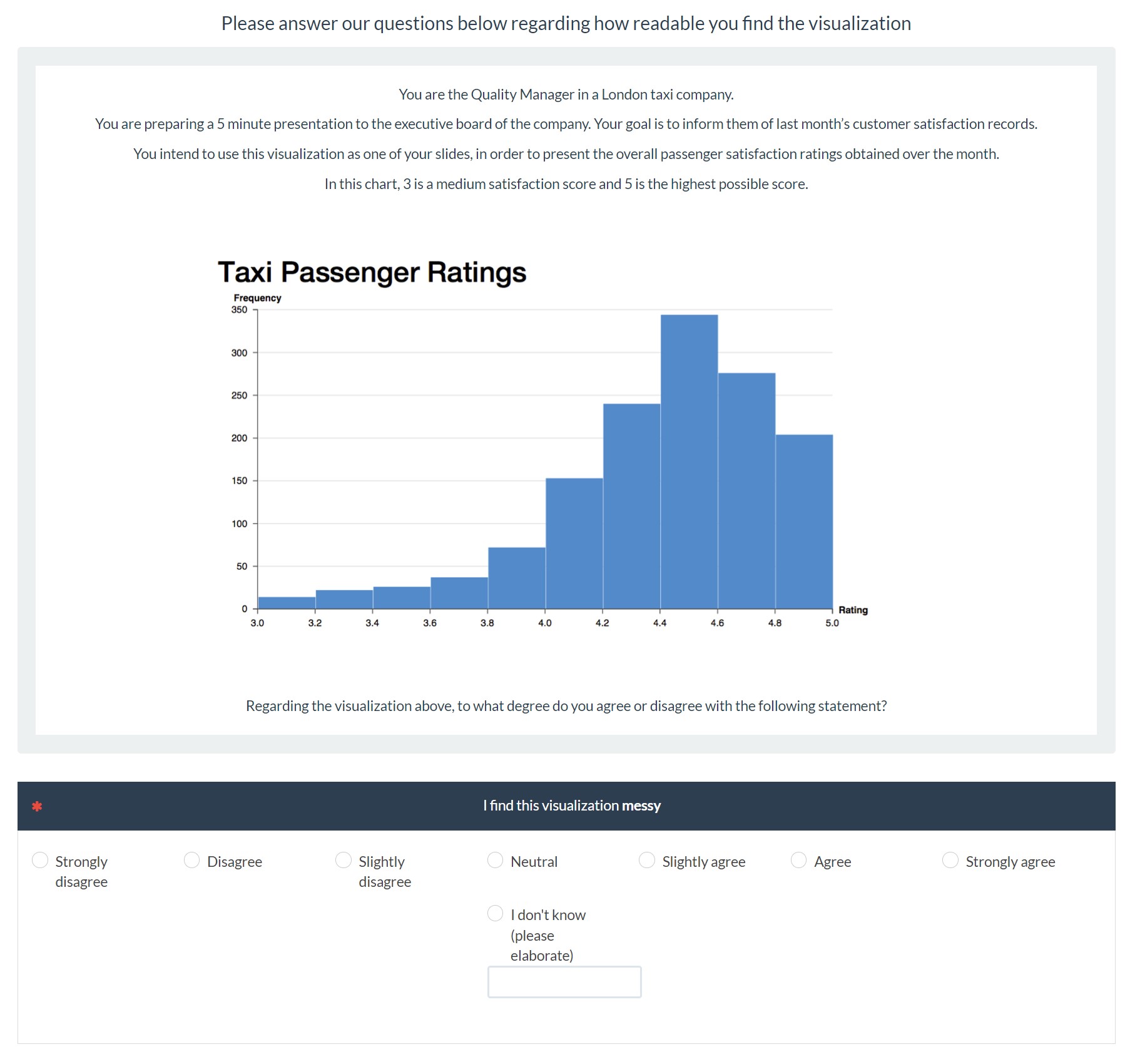}
    \caption{Presentation of a rating item in round 3 of the pre-test study.}
    \label{app:fig:pretestround3}
\end{figure}

\textbf{Survey design changes.}
Comments from participants and observations from the interviewer also allowed us to refine our survey's design. As such, in the first round of our pre-test study, participants answered 39 rating items and one attention check on a single screen (see \autoref{app:fig:pretestround1}). We imagined it would make the task less tedious, but we found that respondents could easily miss an item in this setting. When it was the case, participants then had to scroll back up to spot the empty item, which they found tedious. As a result, we designed the final survey to display each rating item on an individual screen \autoref{app:fig:pretestround3}, and we added an option to use the ``Enter'' key from the keyboard to access the next screen more quickly than with a mouse interaction.
We also refined the presentation of the 7 points Likert-scale over rounds: in the first round, respondents saw numbers from 1 to 7; while in the final survey, they could choose between 7 points, all labeled (from ``Strongly disagree'' to ``Strongly agree''). We also added a separate option labeled ``I don't know (please elaborate)'' with a short text field. Although such an answer option generates missing data in the collected answers, it can also reduce noise and is generally recommended in surveys \cite{chyung_2017_EvidenceBasedSurvey}. 

We provide a visual summary of changes in survey design during and after pre-test as \suppmat{supplemental material}{gbjxv}.

An other important outcome of the pre-test study was our decision not to use the VLAT items as stimuli for our final exploratory survey. This was motivated by our observation that the stimuli were visually very clean, and participants tended to comment a lot about their \emph{interpretability} rather than \emph{readability} in the think-aloud procedure. We thus decided to focus the design of our final stimuli on different criterion, as we describe next in \autoref{sec:app:exploratory_survey_design}. We detail this decision a little more in our \researchLog.

\clearpage

\section{Exploratory survey stimuli visualizations and tasks details}
\label{sec:app:exploratory_survey_design}
This section complements \autoref{subsec:exploratory_study_design}, as it provides additional details regarding the content of the survey used to develop \scalename.

We provide \suppmat{a summary of the stimuli and their reading tasks}{wbghn} as supplemental material.

\subsection{6 stimuli visualizations}

We wanted to develop and validate an instrument that could be used in a variety of visualization and readability situations. To that end, for our exploratory data collection we aimed at creating a sample of visualizations stimuli to test with variety in terms of:
\begin{itemize}
    \item \textbf{underlying data} w.r.t. data domain (what they represent), data structure (\eg, tables and networks), number of data points, and number of data attributes;
    \item \textbf{visualization idioms} used (\eg, glyph-based, bar chart, point-based, line-based, node-link, areas, surfaces and volumes, matrix, text-based, continuous color…); and
    \item \textbf{presumed readability} for the selected tasks and target audience.
\end{itemize}

It would not have been feasible to engage in a systematic exploration of all possible variations for so many characteristics; however we attempted to obtain at least some variability for all of them.

\autoref{tab:exp_survey_stimuli_design} summarizes the design characteristics of the visualizations we used as stimuli in our crowd-sourced experiment (for which the procedure can be found in \autoref{subsec:exploratory_survey}). 
Stimuli \stimA \stimB \stimC (shown in \autoref{fig:exp_all_stimuli}) had presumably rather good readability, while \stimD \stimE \stimF (also in \autoref{fig:exp_all_stimuli}) had lower presumed readability based on: difficult encodings in \stimD, unfamiliarity for \stimE, and visual clutter in \stimF.

\subsubsection{Number of visualization stimuli---Pilot study}

We had a limited Prolific budget which would only allow us to fund 156 hours worth of participation.
We thus ran a pilot study to assess how much time was needed for respondents to answer our rating items. The results allowed us to calculate how many visualizations we would be able to test in the exploratory survey within our budget restrictions.

We recruited 14 Masters students from a visualization class we teach. Participation was voluntary and they did not receive any compensation. After agreeing to a consent form and answering 5 demographic questions (color vision deficiency, age, gender, english fluency, and education), each participant saw 2 visualizations out of a set of 4 possible visualizations and answered all questions of: 2 reading tasks, 1 comprehension check and 29 candidate rating items, and 1 attention check in the form of a rating item.

The mean time required to complete the survey was 20.19 min (std = 3.8). Our results (see \autoref{app:tab:pilot_times}) indicated that Masters students needed about 2 minutes to answer the consent form and demographic questions, leaving 18 min for 2 stimuli. We estimated that crowd-sourced workers from Prolific would be a little faster as they tend to minimize the time spent on tasks. In addition, we would retrieve demographic information from Prolific data (except color vision deficiency). We thus estimated each Prolific participant would need 7--8 minutes to complete our survey with one visualization.

As fatigue has been documented to appear after 10 minutes in crowdsourced studies \cite{zhang_2018_UnderstandingFatigue}, our goal was to allow respondents to complete the survey under this threshold. This fortified our decision to show only one stimulus to each participant, rendering our study design effectively cross-sectional with independent groups assessing each visualization.

Our Prolific budget was 156 hours worth of participation: in the best case scenario of 7 minute for each visualization, we could reach a maximum of 1330 ppeople---1170 for 8 minutes.
To reach our target sample size of 300 participants per visualization (more details on that in our \researchLog), we would be able to use 4 different stimuli.
We later received more funds and could add 2 further stimuli and expand on our stimulus characteristics.

As a result, we designed 6 stimuli for our exploratory survey; and we used stimuli \stimA \stimB \stimC \stimD for our first run of the survey; only later did we run a separate survey with survey \stimE and \stimF.


\begin{table}[h]
    \centering
    \fontsize{7pt}{7pt}\selectfont
    \caption{Time spent by participants in our pilot study. Each participant randomly saw 2 our of 4 visualizations.}
    \label{app:tab:pilot_times}
    \tabulinesep=0.8mm
    \begin{tabu} to \linewidth {X[0.6, r]X[1.5, c]X[1.5, c]}
    \toprule
    \textbf{Participant} &\textbf{Total time spent on survey (min)} &\textbf{Time spent on consent and demopgraphics (min)} \\\midrule
    537729235 &18.9 &1.6 \\
    2097021850 &20.7 &2.4 \\
    2130749281 &19.4 &4.7 \\
    190639113 &27.5 &1.8 \\
    442561266 &18.6 &2.4 \\
    733260304 &17.1 &2.0 \\
    575889510 &19.8 &1.1 \\
    805650123 &25.8 &1.3 \\
    406801568 &14.4 &1.0 \\
    925228469 &23.9 &1.6 \\
    1655604134 &22.6 &2.7 \\
    1020624217 &19.0 &1.5 \\
    293208060 &14.4 &1.4 \\
    383121506 &20.7 &2.2 \\
    \midrule
    \textbf{mean} &\textbf{20.2} &\textbf{2.0} \\
    \textbf{std} &\textbf{3.8} &\textbf{0.9} \\
    \bottomrule
    \end{tabu}
\end{table}

In the following subsections, we briefly address each of the characteristics cited above and how we considered it in our stimuli design.

\subsubsection{Underlying data}

Here we briefly describe our rationales regarding the data we plotted in our stimuli visualizations. The first concern we had was that our instrument was not meant to capture a participant's knowledge of the domain. Although this factor could influence a reader's ability to correctly interpret a visualization, it is unclear whether it is useful to measure at the \textit{reading} level. Therefore, a fundamental requirement for our stimuli was that we did not want readers to require any domain knowledge regarding the represented data. Keeping this in mind, we explored possible data characteristics for which we should consider variations in our stimuli to best capture the range of possible factors affecting readability during the exploratory study.

A first point of attention was the nature of \textit{what} the data represented. In their seminal work, Card \etal \cite{Card:1999} distinguished \textbf{physical data} with a spatial mapping---``\textit{the human body, earth, modecules or others}''---from \textbf{nonphysical information}---``\textit{such as financial data, business information, collections of documents, and abstract conceptions}''. In our survey, stimulus \stimC was a spatial representation (world map) while others were not--- namely, we represented financial data (prices in \stimB, budget in \stimD, profits and losses in \stimF), genealogical information in \stimE, and internet bandwidth speed in \stimA.

A close concept to that of \textit{what} the data represents is the notion of \textit{how} the data is structured: different structures might entail different readability factors in their representations. Munzner \cite{Munzner:2014} identifies four \textbf{dataset types}: tables, networks, fields and geometry (spatial). Stimuli \stimA, \stimB, \stimD \stimF are represented from tables, while \stimE is a representation of a network, and \stimC is a spatial representation.

Another major factor influencing design choices in data visualization is the number and type of \textbf{data attributes} to represent. The data we used for all stimuli has categorical and numerical attributes. \stimE is somewhat different in the sense that it produces countable elements in the visualization from entirely categorical data.

Lastly, the number of represented \textbf{data points} can create visual clutter and make it harder for a reader to retrieve individual values; and, in dense layouts, large amounts of data points can cause overplotting, impeding the visual detection of higher-order patterns in the data. We did not provide extreme cases of data points density as to not overwhelm our participants, but the represented data in our visualizations ranged from 7 bars in \stimA or 3 lines in \stimB, to 36 lines in \stimF. In \stimC, we plotted 176 countries and our dataset contained data values for 122 of them; but we labeled only 10.

\subsubsection{Visualization idioms}

\afterpage{
    \clearpage
    \begin{figure*} []
        \centering
        \includegraphics[width=1\textwidth]{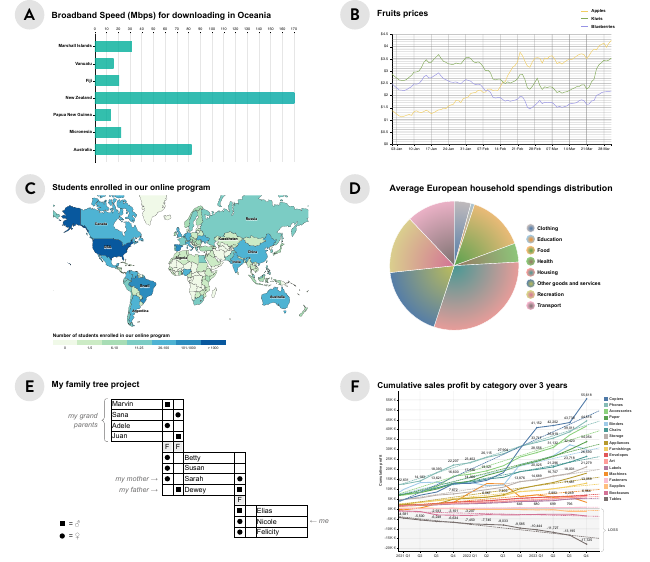}
        \caption{The 6 stimuli visualizations we created for our exploratory survey to develop \scalename.}
        \label{fig:exp_all_stimuli}
        \vspace{1cm}
    \end{figure*}
   \begin{table*}[t]
\centering
\footnotesize
\tabulinesep=0.8mm

\caption{Design characteristics of the 6 stimuli used in our survey and how they related to the 3 families of visualization characteristics possibly influencing readability.}
\label{tab:exp_survey_stimuli_design}
\begin{tabu} to \textwidth{X[0.5] X[1.5] X[0.9] X[0.5]}
\toprule
\textbf{Stimuli} &\textbf{Display: encodings, data attributes} \newline (on a canvas of max 800 x 500px) &\textbf{Individual: idiom familiarity} &\textbf{Task difficulty} \newline in VLAT\cite{Lee:2017:VLAT} \\\midrule
\textbf{A - Bar chart} &2 visual variables, few entities (7 bars), clear layout &Very familiar or intuitive &Easy \\
\textbf{B - Line chart} &3 visual variables, few entities (3 lines), cluttered layout (grid) &Very familiar or intuitive &Easy \\
\textbf{C - Choropleth map} &4 visual variables, many entities (world countries), simple encodings &Very familiar or intuitive &Easy \\
\textbf{D - Pie chart} &2 visual variables, few entities (8 categories), messy encodings (colors, labels, ) &Very familiar or intuitive &Easy \\
\textbf{E - GeneaQuilts} \newline (family tree in a matrix style \cite{bezerianos_2010_GeneaQuiltsSystem}) &3 visual variables, reasonable amount of entities (11 people), simple encodings &Not familiar, possible to infer mapping rules from labels with effort &- (not in VLAT) \\
\textbf{F - Many lines chart} &4 visual variables, many entities (18 categories x 2 types of lines), overplotting (too much information) &Somewhat familiar intuitive (people might no be familiar with general trend dashed lines) &Easy \\
\bottomrule
\end{tabu}
\end{table*}
   \clearpage
} 

Our preliminary work led us to think that readability could be affected by readers' familiarity with a particular idiom \cite{Lee:2015:sensemaking} or their visualization literacy \cite{Boy:2014:Literacy, Borner:2016:VLMuseum, Lee:2017:VLAT}. Therefore, we needed our exploratory data to also encompass this dimensions. That being said, we did not aim at creating an instrument dedicated to measuring such a skill in participants. Therefore we used five idioms that were widely familiar to broad audiences, and one unfamiliar type of representation.

For familiar idioms we used: bar chart in \stimA, line chart in \stimB and \stimF, choropleth map in \stimC, and pie chart in \stimD. It's worth noting that, while the line chart has widespread use as we use it in \stimB, we also plot trend lines in \stimF, possibly adding to the reading difficulty for this last stimulus.

For the unfamiliar visualization, we considered Parallel Coordinate Plots (PCP), which Lee \etal used in the VLAT \cite{Lee:2017:VLAT} validation study. However, it can be very difficult for readers to intuitively guess how to read a PCP---if not impossible. In fact, Lee \etal provided participants with a training in their study design. We, on the other hand, would not provide a training for other stimuli. Deviating from our study parameters for one of our 6 independent groups would endanger the reliability of our findings.

\begin{figure}[H]
    \centering
    \includegraphics[]{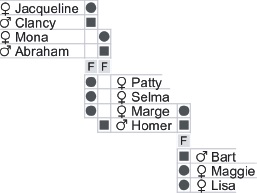}
    \caption{GeneaQuilts Visualization of the Simpson Family in Bezerianos \etal's original work \cite{bezerianos_2010_GeneaQuiltsSystem}. Image is \textcopyright{} 2010 IEEE, used with permission.}
    \label{app:fig:GeneaQuilts_Simpsons}
\end{figure}

Instead, we chose to use a visualization idiom that is both novel to our audience and ``self-teachable'': the GeneaQuilts technique \cite{bezerianos_2010_GeneaQuiltsSystem}, a matrix-based representation of genealogical data. GeneaQuilts are most useful to represent large genealogy datasets; but with small dataset, they provide a very interesting combination of qualities for our current work: they can look cryptic at first sight, but their design is minimalist.
What's more, given basic knowledge of the family links in the represented genealogy, a viewer can autonomously deduce how the GeneaQuilts encodings work (see the Simpson's family GeneaQuilts representation in \autoref{app:fig:GeneaQuilts_Simpsons}). We implement GeneaQuilts in \stimE.

\subsubsection{Visual display design choices to alter readability}
For each visualization we created design variations that presumably would affect readability for respondents. We summarize our choices in the following paragraph a summary. Details can be found in a \suppmat{separate document}{wbghn} from our supplemental material.

Firstly, our goal was to produce an instrument that would be usable in wide variety of \textbf{\textit{likely}} visualization reading situations. Therefore, we did not include completely chaotic encodings (such as a complete mismatch of datatypes with idioms like plotting time trends on a pie chart). The furthest we went into that direction was the use of gradient colors encodings in \stimD without in-chart labels.
Secondly, our goal was to avoid potential influence of aesthetics judgement---for which there is already an existing measuring scale: \cite{He:2022:Beauvis}---on perceived readability. therefore, we attempted to keep a relatively clean an professional look for all images we produced.

To introduce design choices that would presumably increase or decrease perceived readability, we introduced facilitators (\eg, good separability of colors or direct labeling) and difficulties (\eg, absence of data label coupled to an arbitrary sorting of the accompanying legend).







\subsubsection{Color impairment simulations}

\begin{figure}
    \centering
    \includegraphics[width=.7\linewidth]{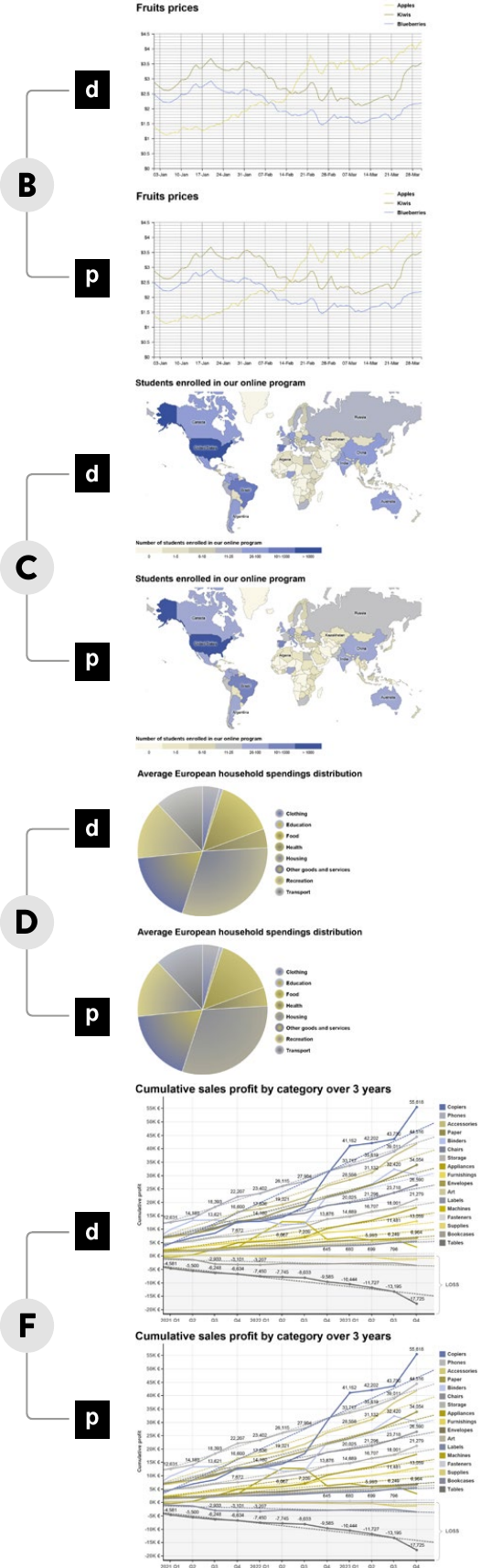}
    \caption{Deuteranipa (d) and protanopia (p) simulations in Adobe Illustrator for all stimuli relying on color encodings for data attributes in our exploratory survey.}
    \label{fig:exp_survey_stimuli_color_vision}
\end{figure}

For stimuli relying on color encodings for data attributes, we chose color that would remain distinguishable for people with color vision deficiencies. We assessed the results in Adobe Illustrator using the View $>$ Proof Setup for deuteranopia and protanopia, as shown in \autoref{fig:exp_survey_stimuli_color_vision}.


\raggedbottom

\subsection{Reading tasks}

Before asking participants to assess their perception of readability in a visualization, we need them to read it. 
Since this reading experience will shape their opinion on readability, we considered them as an integral aspect of our stimuli for this exploratory study. Therefore, we paid a particular attention in selecting select tasks that were in the scope of “reading.” We provide details on our review of three task taxonomies \cite{Curcio:1987:MathRelationships, Amar:2005:LowLevelTasks, Brehmer:2013:MultiLevelTypology} and how we also referred to item difficulty from VLAT items \cite{Lee:2017:VLAT} to form our starting set of reading tasks before selecting stimuli-task pairs.

In contrast with our difficulty-based approach during pre-test (see \autoref{sec:app:pretest}), in our final survey we chose tasks that demonstrated low discriminating power in VLAT and were easy to perform, meaning that they did not require specific levels of visualization reading skills in our target audience.
This could somewhat reduce the ability of our instrument to capture information about the respondent's visualization literacy; but then again, our goal was to focus on multiple factors of readability and we had already accounted for the influence of individual knowledge on answers variance by implementing the unfamiliar GeneaQuilts visualization in \stimE.

\subsubsection{Rating items}
\autoref{tab:all_items_exp} presents the list of rating items we presented to participants after they answered reading questions. This list was refined from the initial 39 candidate statements through the pre-test study described in \autoref{sec:app:pretest}.

\begin{table}[t]
\centering
\footnotesize
\tabulinesep=0.8mm
\caption{29 items and one attention check presented to participants in the exploratory survey to rate the readability of visualization stimuli.}
\label{tab:all_items_exp}
\begin{tabu} to \linewidth{ X[0.5, l] X[2, l]}
\toprule
\textbf{Item code} &\textbf{Item statement} \\\midrule
answer &I can easily \textbf{answer some questions} about the represented data with this visualization \\
clearData &This visualization shows the data in a \textbf{clear way} for me \\
clearRepresent &The representation of the data makes the \textbf{information clear} to me in this visualization \\
complex &I don't find this visualization \textbf{complex to read} \\
confid &I am \textbf{confident} in my understanding of this visualization \\
confus &I \textbf{don't} find this visualization \textbf{confusing} \\
crowd &I \textbf{don't} find this visualization \textbf{crowded} \\
deciph &I \textbf{don't} find the presentation of the data \textbf{difficult to decipher} in this visualization \\
distinguish &I can easily \textbf{distinguish individual elements} of the represented data (for example individual lines, or dots, or areas, or colors...) \\
distract &I \textbf{don't} find parts of the visualization \textbf{distracting} \\
effect &This visualization \textbf{effectively} shows the data to me \\
find &I can easily \textbf{find specific elements} in this visualization \\
identifi &I can easily \textbf{identify relevant information} in this visualization \\
inform &I can easily \textbf{retrieve information} from this visualization \\
lost &I \textbf{don't} feel \textbf{lost} trying to read this visualization \\
meanElem &I can easily understand what the \textbf{different elements} of the visualization \textbf{mean} \\
meanOveral &I can easily understand the \textbf{overall meaning} of this data visualization \\
messi &I \textbf{don't} find this visualization \textbf{messy} \\
obvious &It is \textbf{obvious} for me how to read this visualization \\
organiz &I find this visualization well \textbf{organized} \\
read &I can \textbf{easily read} this visualization \\
readabl &I find this visualization \textbf{readable} \\
represent &I can easily understand \textbf{how the data is represented} in this visualization \\
see &I can \textbf{clearly see} data features (for example, a minimum, or an outlier, or a trend) in this visualization \\
simpl &I find this visualization \textbf{simple to read} \\
understandEasi &I can \textbf{easily understand} this visualization \\
understandQuick &I can \textbf{quickly understand} this visualization \\
valu &I can \textbf{read data values} from this visualization \\
visibl &I find data features (for example, a minimum, or an outlier, or a trend) \textbf{visible} in this visualization \\
\midrule
attentionCheck &For calibration purposes, please select \textbf{slightly agree} with this item \\
\bottomrule
\end{tabu}
\end{table}

\subsection{Two rounds of survey implementation}
We ran the survey in two separate rounds: the first one over the course of one week with stimuli \stimA \stimB \stimC \stimD, and the second with stimuli \stimE and \stimF, over the course of 1 + 2 days, separated by a 2 week delay for transferring additional budget to the Prolific platform. We did not make any change between the two rounds other than adding \stimE \stimF---and removing \stimA \stimB \stimC \stimD for which we had already collected our exploratory data in the previous round. We share printouts of \suppmat{the first round's survey}{3hd84} and \suppmat{the second round's survey}{t75vw} as supplemental material.

We pre-registered (\href{https://osf.io/4dcav}{\texttt{osf.io/4dcav}}) our first round of the study. We updated the pre-registration on on March 9, 2024, after receiving the additional funds and before running our second round on Prolific; however, we encountered a glitch with the update feature, which we only noticed later. We contacted the OSF team and they manually fixed the registration on March 18, 2024.

\begin{figure} [h]
    \centering
    \includegraphics[width=\linewidth]{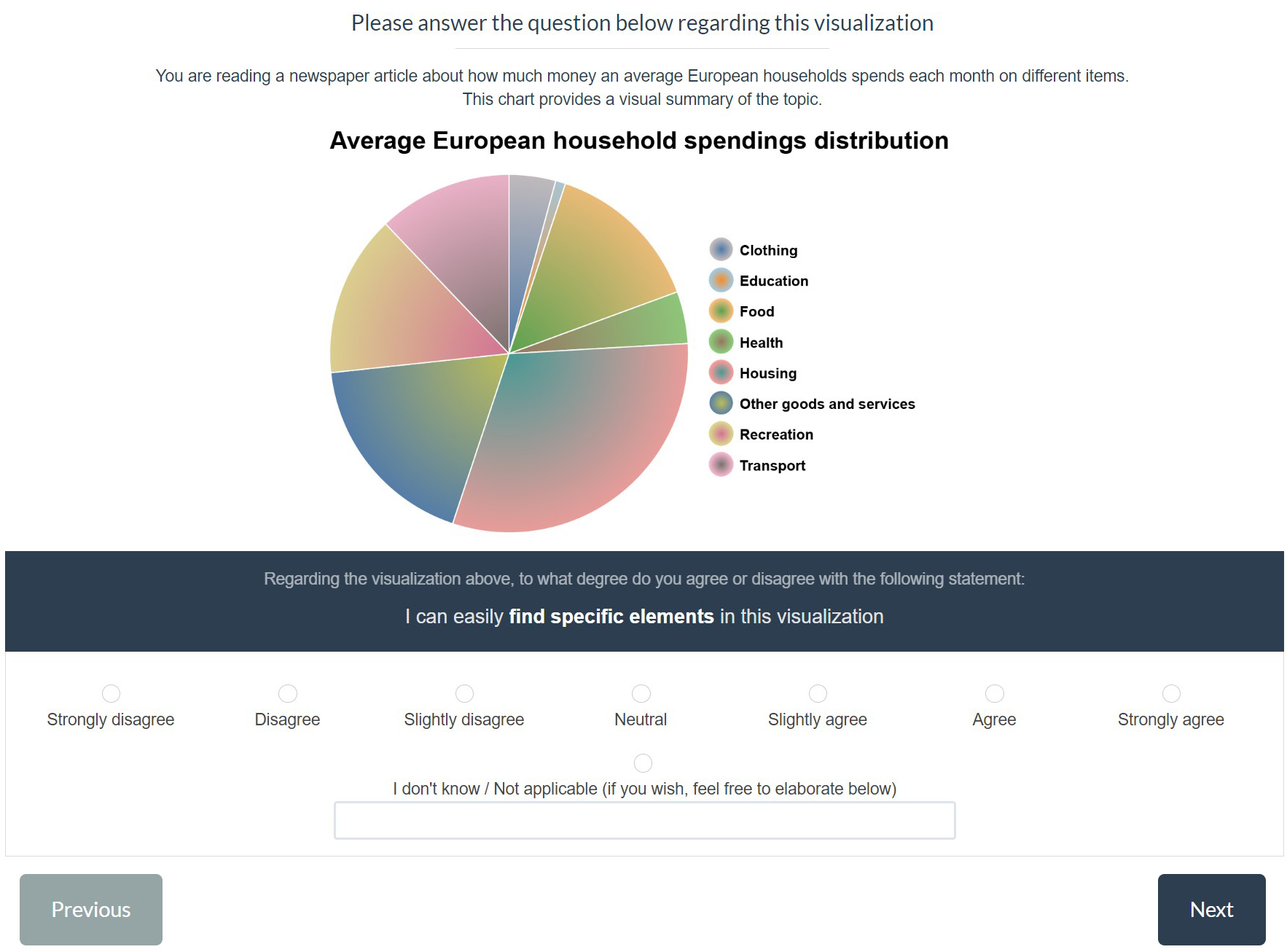}
    \caption{An example screenshot from our exploratory survey, consisting of: (1) a stimulus visualization (D: pie chart) with a title and a short contextual explanation, (2) a candidate rating item (``find'') where keywords have been highlighted, and (3) 8 answer options consisting of a 7 points Likert-scale with individual labels, and one `Ì don't know / Not applicable'' option, allowing the participant to elaborate on their reason for choosing this answer if they wish.}
    \label{app:fig:survey_screenshot}
\end{figure}

\clearpage

\section{Missing data handling for exploratory survey}
\label{app:sec:missing_data}
As we provided an option to answer ``I don't know'' to rating questions in the exploratory survey, there was missing data in our collected ratings.
Missing data, if not carefully handled, can impede the reliability of data analysis results. In particular, missing data can affect factors extraction in EFA \cite{mcneish_2017_ExploratoryFactor}. Mirzaei \etal \cite{mirzaei_2022_MissingData} provide a decision tree on handling missing data in surveys. We implemented the two checks they recommend in our data analysis code in \texttt{R}:

\textbf{Step 1: Calculate the percentage of missing data.} We calculated the amount of missing data and checked it again the thresholds Mirzaei \etal proposed:
\begin{itemize}
    \item \textbf{$<$ 5\% missing data}: missing data is negligible and researchers may choose to handle missing data with deletion or imputation methods without significantly affecting their subsequent analysis;
    \item \textbf{5-10\% missing data}: a grey area, where Mirzaei \etal recommend that the researcher refer to the theory regarding the phenomenon of interest before deciding on the next step;
    \item \textbf{10-40\% missing data}: missing data is not negligible, but researchers may use imputation methods, depending on the results of the second step;
    \item \textbf{$>$ 40\% missing data}: missing data is too high for imputation methods: researcher should conduct a quantitative and qualitative investigation.
\end{itemize}

\textbf{Step 2: Perform Little's test of missingness.} A significant p-value result for this test \cite{little_1988_TestMissing} indicates that the null hypothesis that data is Missing Completely At Random (MCAR) is rejected, and therefore that a pattern exists to the missing data. When data is not MCAR, 

We report the results of the missing data analysis in \autoref{tab:missing_data}.
\begin{table}[h!]
\centering
\fontsize{7pt}{7pt}\selectfont
\caption{Survey-wise and stimulus-wise frequencies of missing data and p-value in Little's test of Missing Completely At Random (MCAR) \cite{little_1988_TestMissing}.}
\label{tab:missing_data}
\tabulinesep=0.8mm
\begin{tabu} to \linewidth {X[1,l]X[1,l]X[1,l]}
\toprule Dataset & Amount of missing data & Little's test p-value \\
\midrule

Full survey & 0.25\% & <0.001 (data is MCAR) \\
Stimulus A & 0.25\% & 0.624 (data is not MCAR) \\
Stimulus B &  0.14\% & <0.001 (data is MCAR) \\
Stimulus C & 0.22\% & <0.001 (data is MCAR) \\
Stimulus D & 0.45\% & 0.109 (data is not MCAR) \\
Stimulus E & 0.84\% & 0.013 (data is MCAR) \\
Stimulus F & 0.28\% & 0.268 (data is not MCAR) \\
\bottomrule
\end{tabu}
\end{table}

The tests on the complete dataset showed that all data was 0.25\% and MCAR.
Each individual stimulus' dataset also had less that 5\% of missing data; however, Little's test of missingness showed that data was not MCAR in the datasets from stimuli A, D and F were not MCAR. As a result, we decided to use imputation methods, which tend to perform better than deletion methods according to the literature \cite{mirzaei_2022_MissingData, mcneish_2017_ExploratoryFactor}.

We used the \texttt{mifa} package in \texttt{R} to generate the correlation matrix on which the EFA would be based. This package was developed specifically to perform multiple imputation for EFA \cite{nassiri_2018_mifa} and relies on the \texttt{mice} package to perform multiple imputation  Mutlivariate Imputation by Chained Equation for the estimation of a covariance matrix of incomplete data.

As participants were able to comment on the reason why they chose the ``I don't know / Not applicable'' answer option, we also report all comments collected this way along with the number of time respondents chose this option (regardless of whether or not they commented on it) as \suppmat{supplemental material}{2qav9}.

\section{Correlation matrices from exploratory survey data}
\label{app:sec:corr_matrices}
\begin{figure*} []
    \centering
    \includegraphics[width=\textwidth]{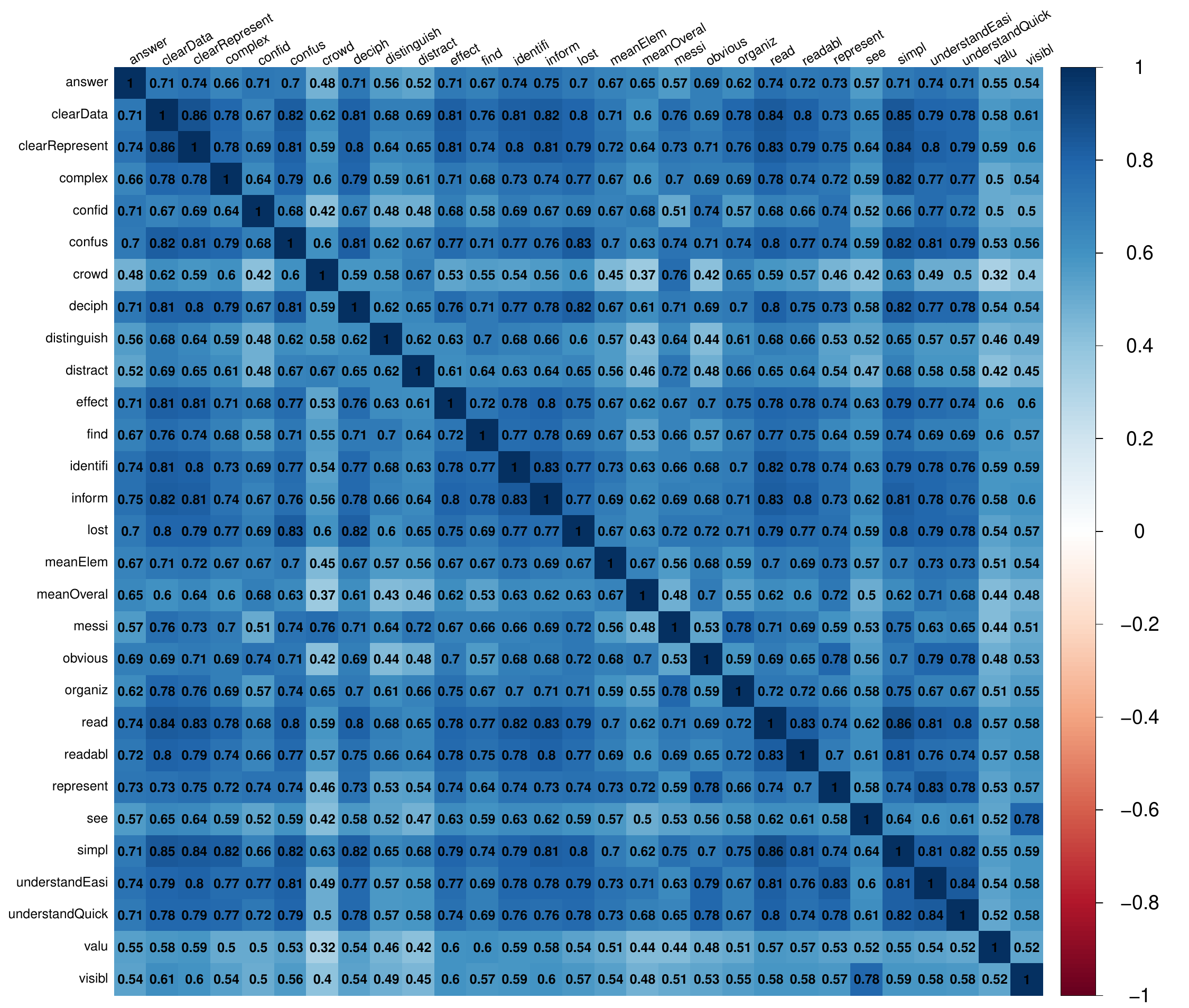}
    \caption{Correlation matrix from our survey data after missing data treatment detailed in \autoref{app:sec:missing_data}. All items have positive correlations, meaning that they tend to co-vary in the same direction. The correlation value indicate to what extent: in here, we observe that all items have a correlation $>$0.3 with all other items, and a large majority exhibit correlations $>$0.5. The correlation matrix serves as input data for Exploratory Factor Analysis.}
    \label{app:fig:corrMatrix_Agg}
\end{figure*}

\begin{figure} []
    \centering
    \includegraphics[width=0.7\linewidth]{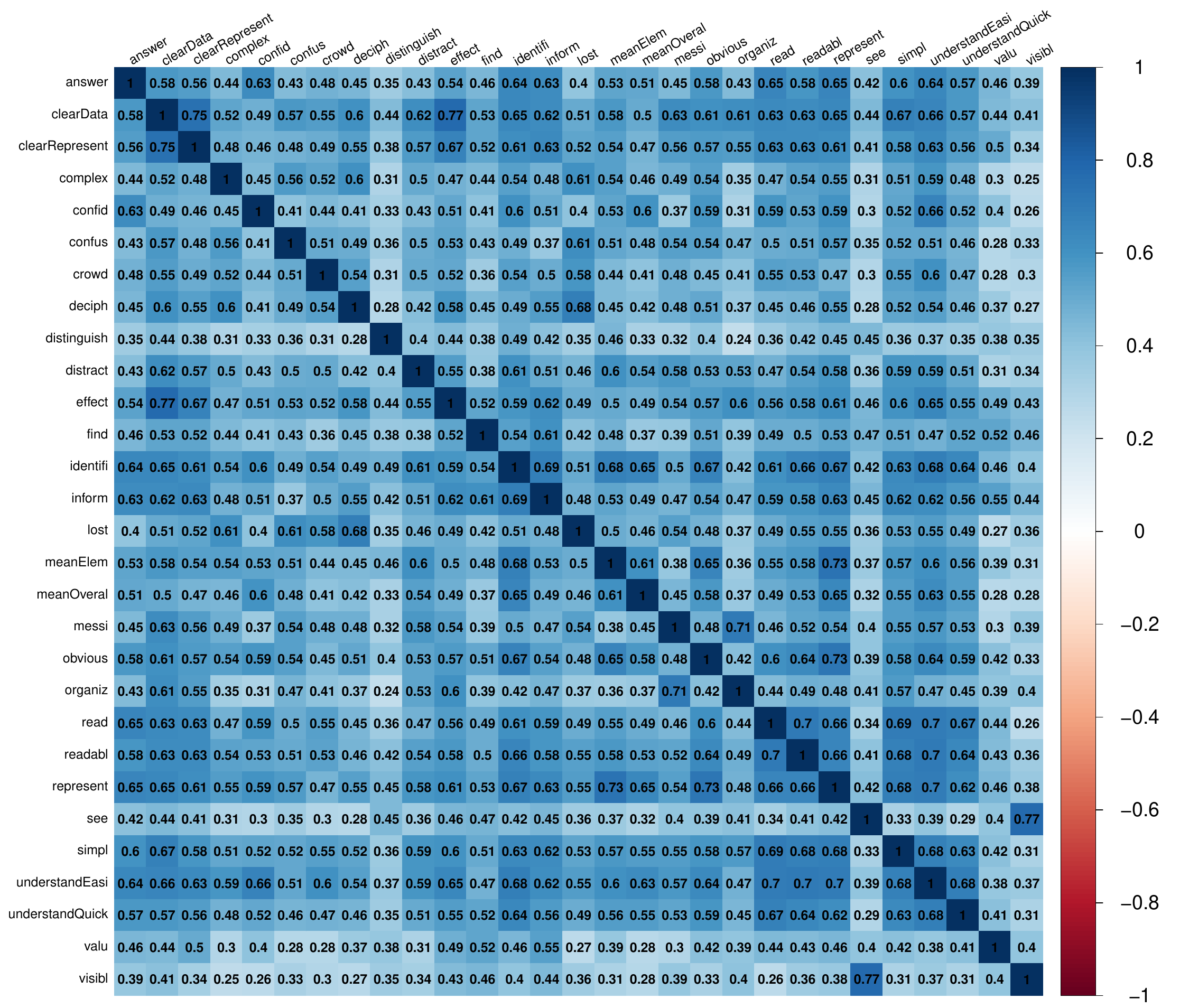}
    \caption{Correlation matrix from our survey data for \textbf{stimulus A} (bar chart) after missing data treatement detailed in \autoref{app:sec:missing_data}.}
    \label{app:fig:corrMatrix_A}
\end{figure}

\begin{figure} []
    \centering
    \includegraphics[width=0.7\linewidth]{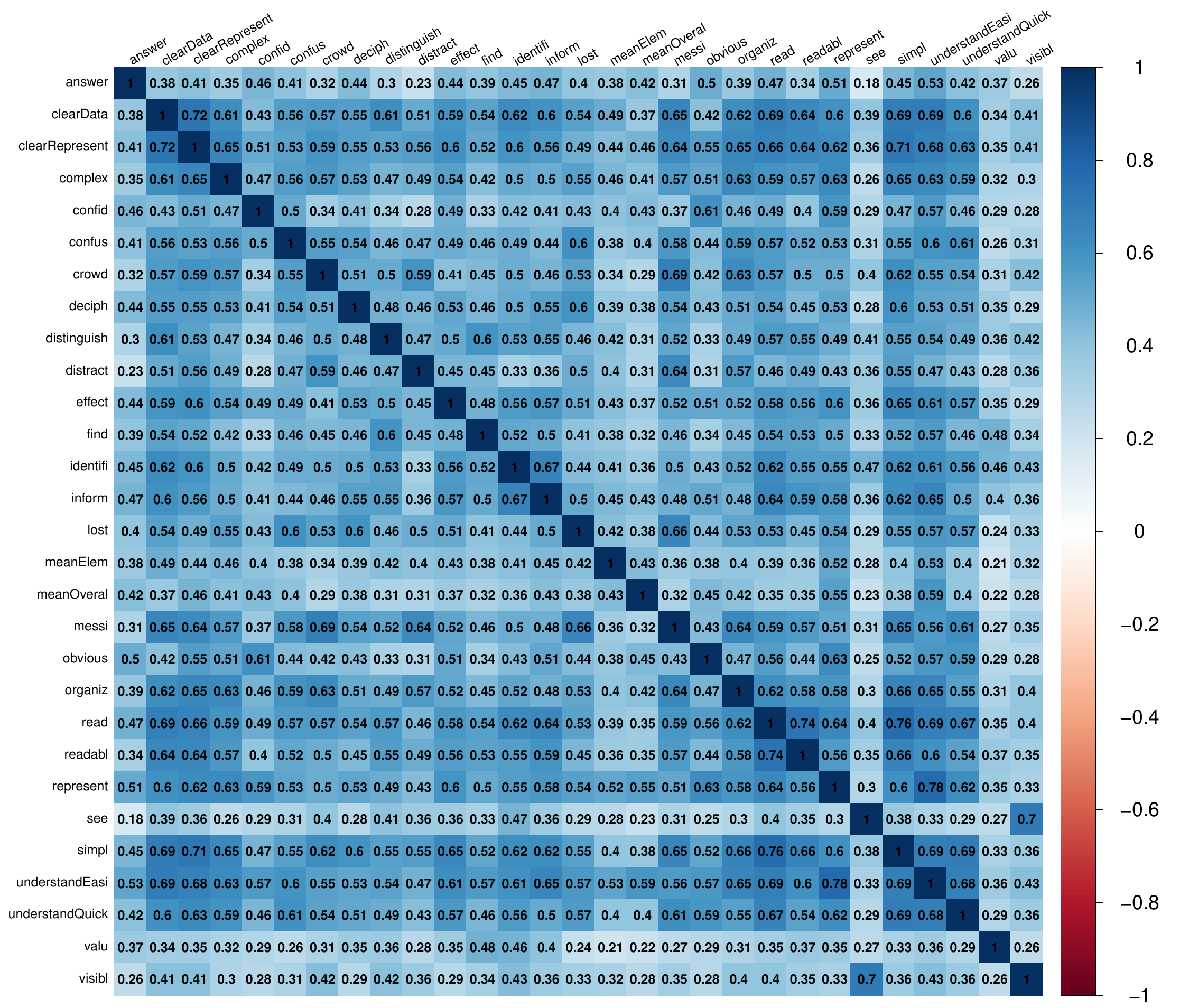}
    \caption{Correlation matrix from our survey data for \textbf{stimulus \stimB} (line chart) after missing data treatement detailed in \autoref{app:sec:missing_data}.}
    \label{app:fig:corrMatrix_B}
\end{figure}

\begin{figure} []
    \centering
    \includegraphics[width=0.7\linewidth]{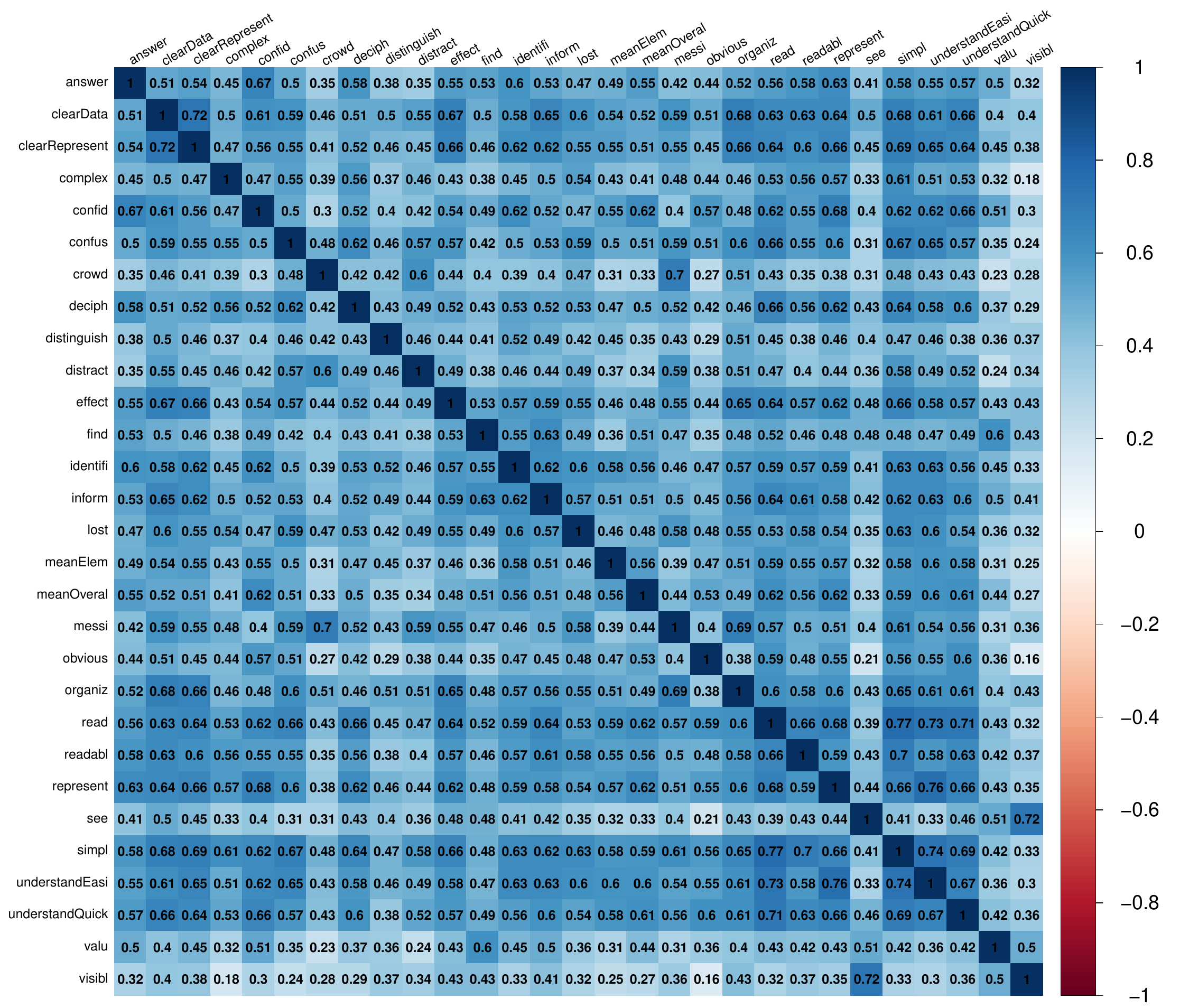}
    \caption{Correlation matrix from our survey data for \textbf{stimulus \stimC} (choropleth map) after missing data treatement detailed in \autoref{app:sec:missing_data}.}
    \label{app:fig:corrMatrix_C}
\end{figure}

\begin{figure} []
    \centering
    \includegraphics[width=0.7\linewidth]{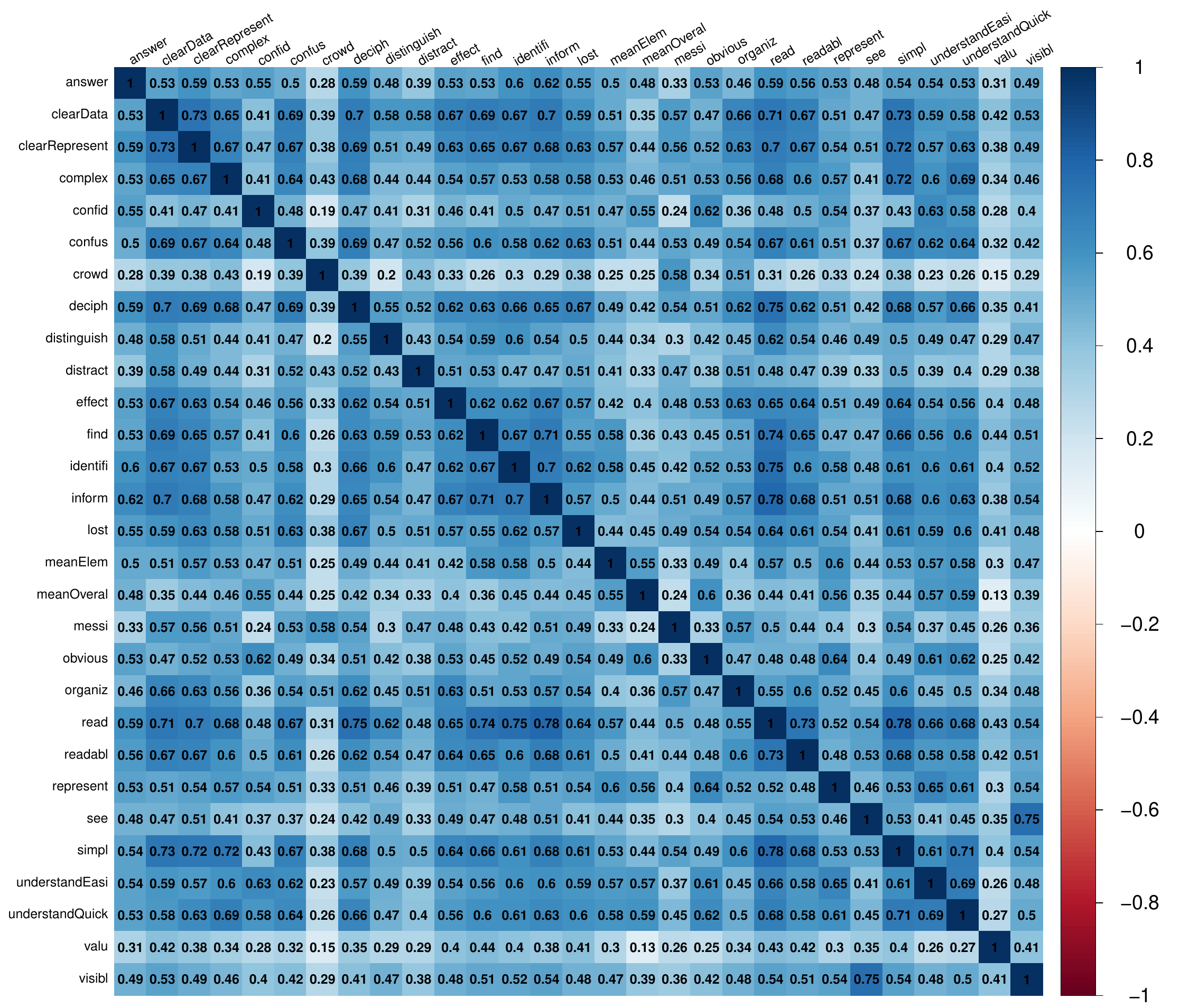}
    \caption{Correlation matrix from our survey data for \textbf{stimulus \stimD} (pie chart) after missing data treatement detailed in \autoref{app:sec:missing_data}.}
    \label{app:fig:corrMatrix_D}
\end{figure}

\begin{figure} []
    \centering
    \includegraphics[width=0.7\linewidth]{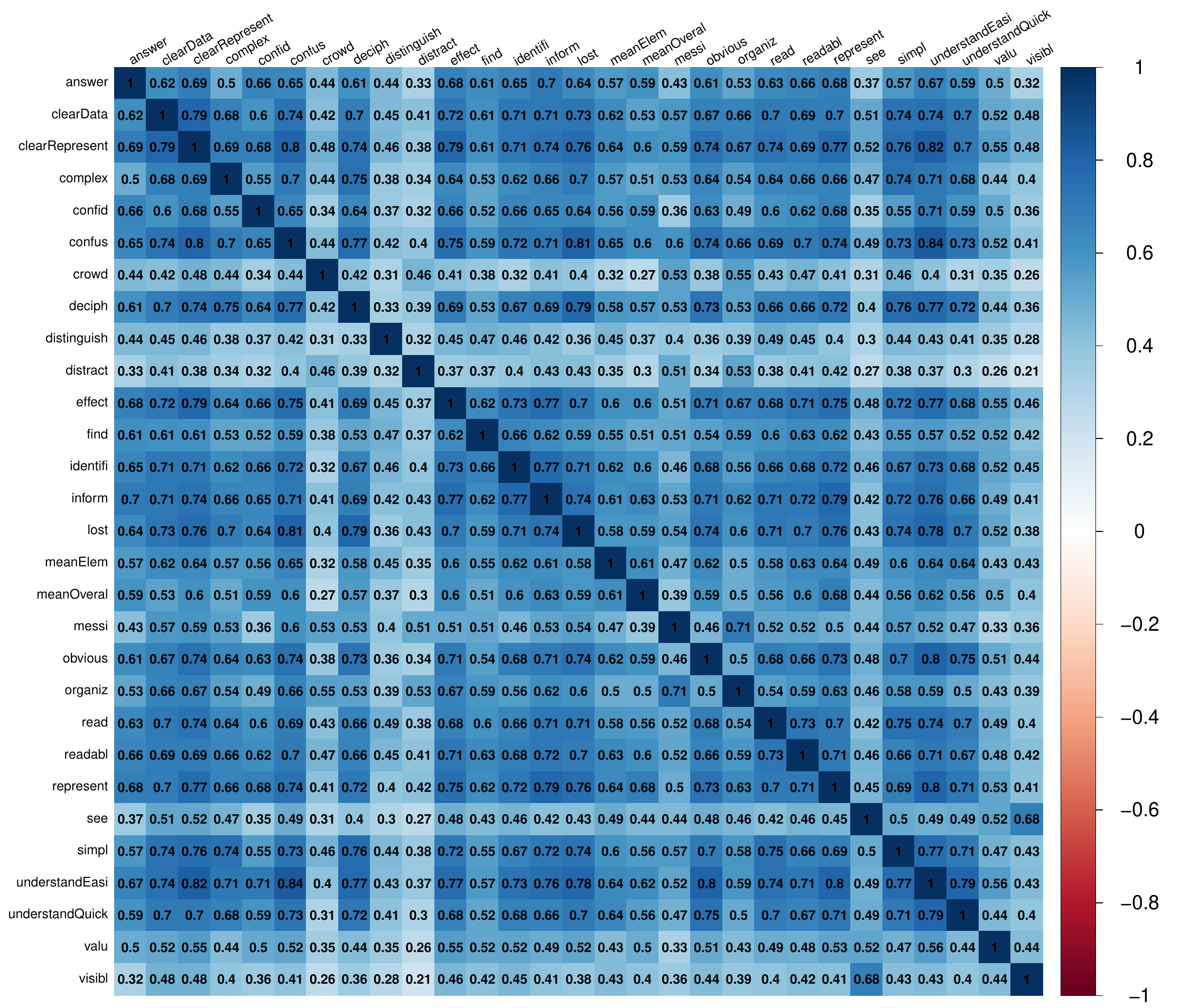}
    \caption{Correlation matrix from our survey data for \textbf{stimulus \stimE} (geneaQuilts) after missing data treatement detailed in \autoref{app:sec:missing_data}.}
    \label{app:fig:corrMatrix_E}
\end{figure}

\begin{figure} []
    \centering
    \includegraphics[width=0.7\linewidth]{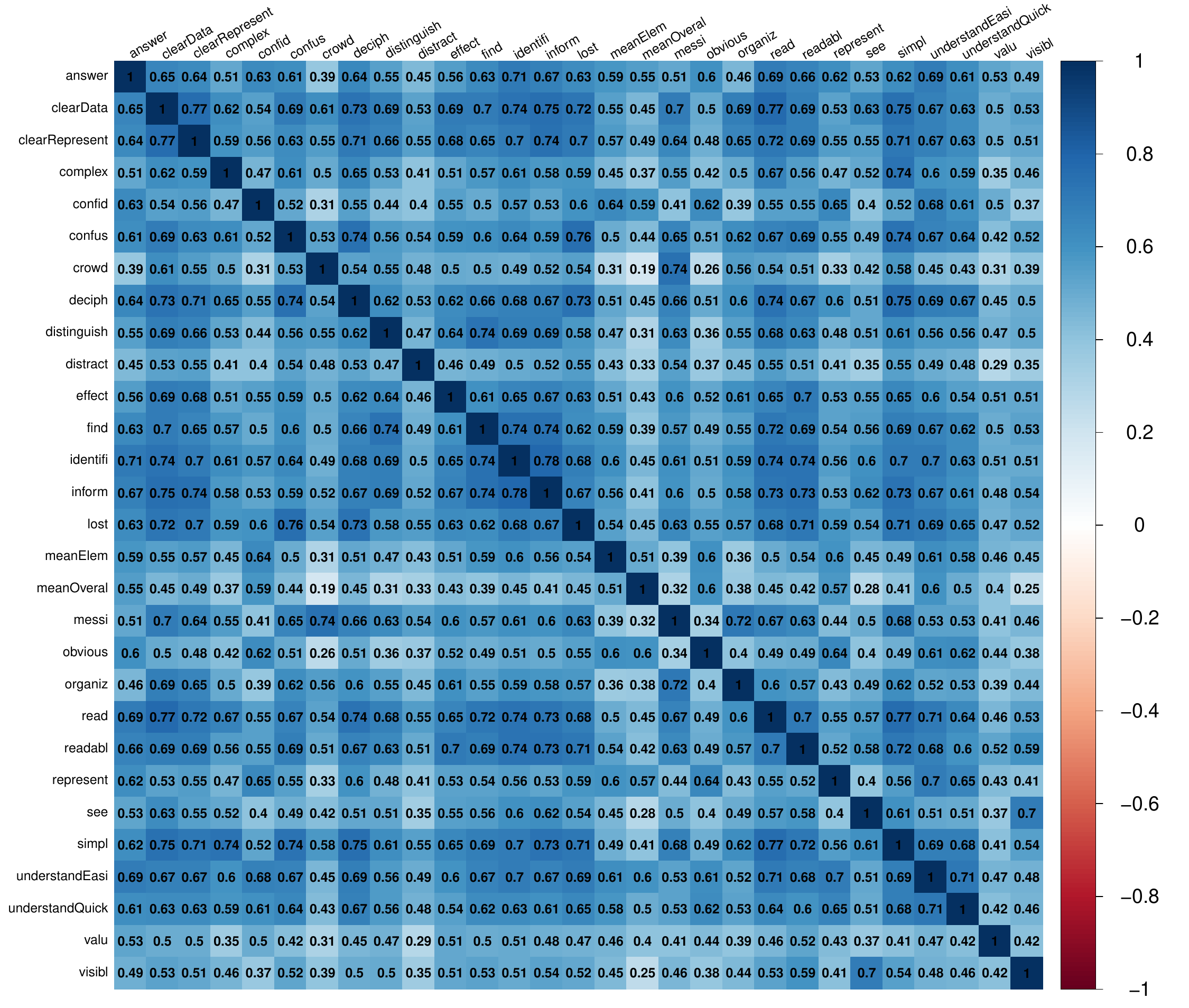}
    \caption{Correlation matrix from our survey data for \textbf{stimulus \stimF} (many lines) after missing data treatement detailed in \autoref{app:sec:missing_data}.}
    \label{app:fig:corrMatrix_F}
\end{figure}

Here, we present the correlation matrix for the full dataset (\autoref{app:fig:corrMatrix_Agg}), and then individual correlation matrices for each of our 6 stimuli in the survey (\autoref{app:fig:corrMatrix_A} to \autoref{app:fig:corrMatrix_F}). We can make a few preliminary observations on these matrices:
\begin{itemize}[nosep,left=0pt .. \parindent]
    \item correlations are all positive and $>$ .3, which is considered moderate. Most correlations are $>$ .5, which is good, ans some $>$ .7, which is considered strong.
    \item correlations seem overall lower for situations that we expected to be of ``higher readability'' (stimuli A, B and C). It might be related to a ceiling effect from the Likert scales, where people chose the maximum ratings more often. As a result, relational structures are less visible in these matrices. We see structure more clearly in ``bad'' readability conditions than in good ones. Studying readability might be better achieved with ``bas readability'' than with good ones?
    \item although these matrices are organized in alphabetical order (on purpose so that the order remains stable to compare multiple analyses), making it more difficult to observe correlation groupings, we can see similar pattern
    \item a few groups of items always highly correlate across stimuli, which means we might expect to find them grouped in factor analyses with more than one factor:
    \begin{itemize}
        \item \texttt{clearData + clearRepresent}
        \item \texttt{read + readabl}
        \item \texttt{understandQuick + understandEasi (+ simpl)}
        \item \texttt{see + visibl}, which also generally do not correlate with other items. Therefore we expect these two form a single factor, unrelated to other items.
    \end{itemize}
\end{itemize}

\section{Appropriateness of data for EFA}
\label{app:sec:data_checks_EFA}
Before conducting the analysis, 
we needed to confirm whether our data was suitable for EFA. Following recommendations  We assessed the univariate normality with a Shapiro Wilk's test, and multivariate normality with a Mardia test.
Both tests showed significant results, indicating that our data violated the normality assumption. In such cases it is recommended to use a Principal Axis (PA) factoring method, which do not entail distributional assumptions \cite{fabrigar_1999_EvaluatingUsea}. We then tested the factorability of our correlation matrix using Bartlett's test of sphericity, as well as the Kaiser Meyer-Olkin (KMO) test, following reference work \cite{taherdoost_2014_ExploratoryFactor}. Bartlett's test yielded a $p$-value of 0, and all individual items' KMO values were above 0.7. Based on these results, we confirmed that our data's correlation matrix was factorable using a Principal Axis factoring method.

\section{Exploratory Factor Analysis: scree plots}
\label{app:sec:EFA_screeplots}
\label{}
Here, we provide all scree plots generated as part of our EFA described in \autoref{subsubsec:EFA}. Scree plots show how much of the data can be explained with 1 to N factors, N being the number of measured variables (in our dataset: 29 items), and using \textit{eigenvalues} of factors as values on the y axis. In the context of EFA, an eigenvalue represents the amount of signal (\ie, information) captured by a factor \cite{Devellis:2021:ScaleDevelopment}. We plot on the same graph the results of parallel analyses, which show how much variance the same number of factors would capture for a randomly generated dataset of the same size as ours.

\begin{figure*} []
    \centering
    \includegraphics[width=\textwidth]{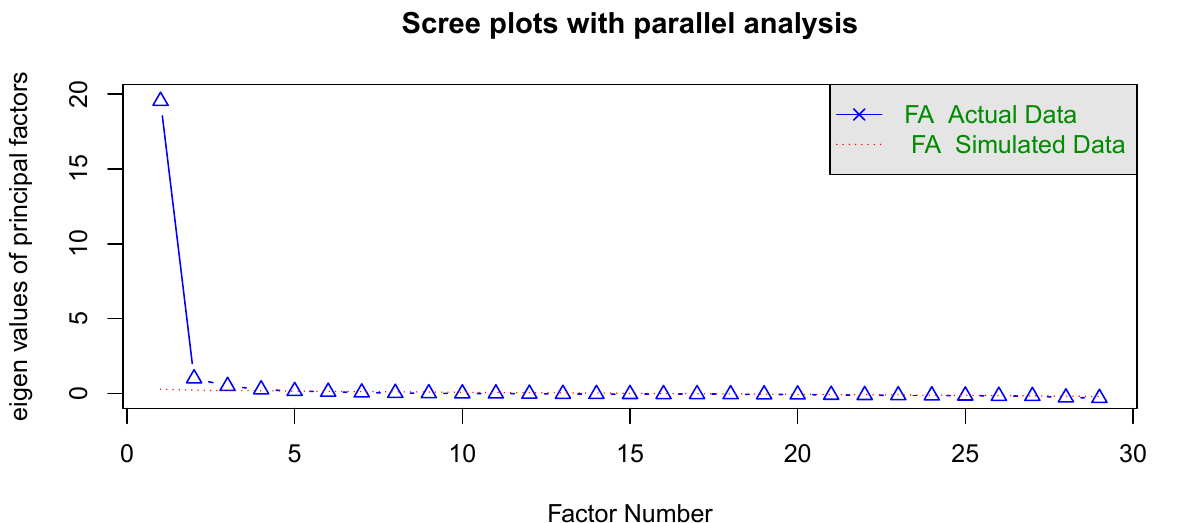}
    \caption{Scree plot from parallel analysis for our \textbf{complete survey data} with actual (blue) and simulated random data (dotted red) lines. Although there is a clear elbow after 1 factor, the blue line and the red dotted line start to overlap at factor 5. This tells us that, although it would be possible to create a unidimensional scale, statistical tests show that 5 factors better explain the data structure. Therefore, before deciding on how many factor we would retain, we explored whether 2-factors, 3-factors, 4-factors and 5-factors solutions could make conceptual sense.}
    \label{app:fig:screeplot_Agg}
\end{figure*}

\begin{figure} []
    \centering
    \includegraphics[width=\linewidth]{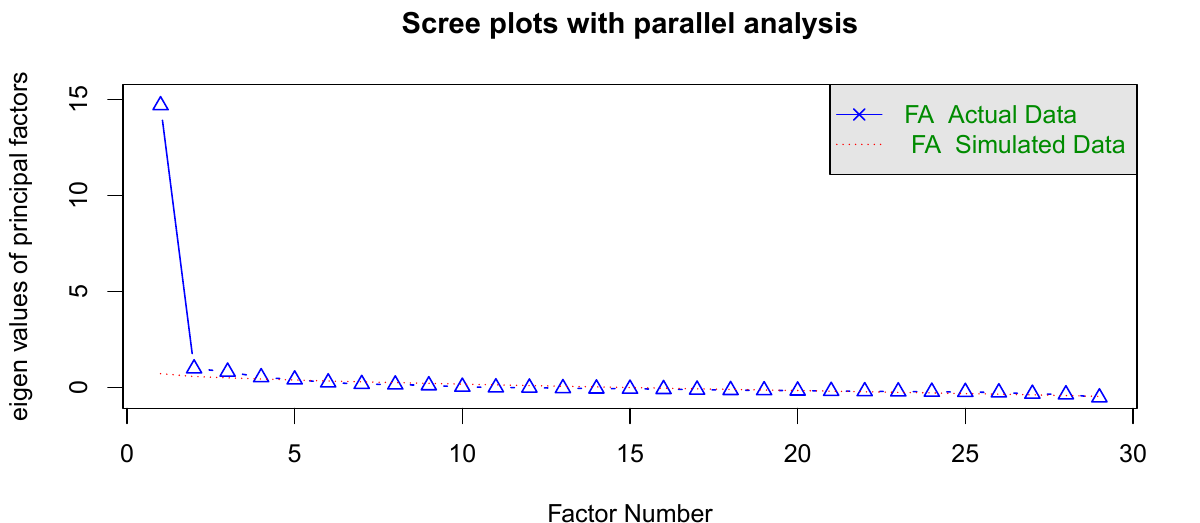}
    \caption{Scree plot from parallel analysis for \textbf{stimulus A}
    .}
    \label{app:fig:screeplot_A}
\end{figure}

\begin{figure} []
    \centering
    \includegraphics[width=\linewidth]{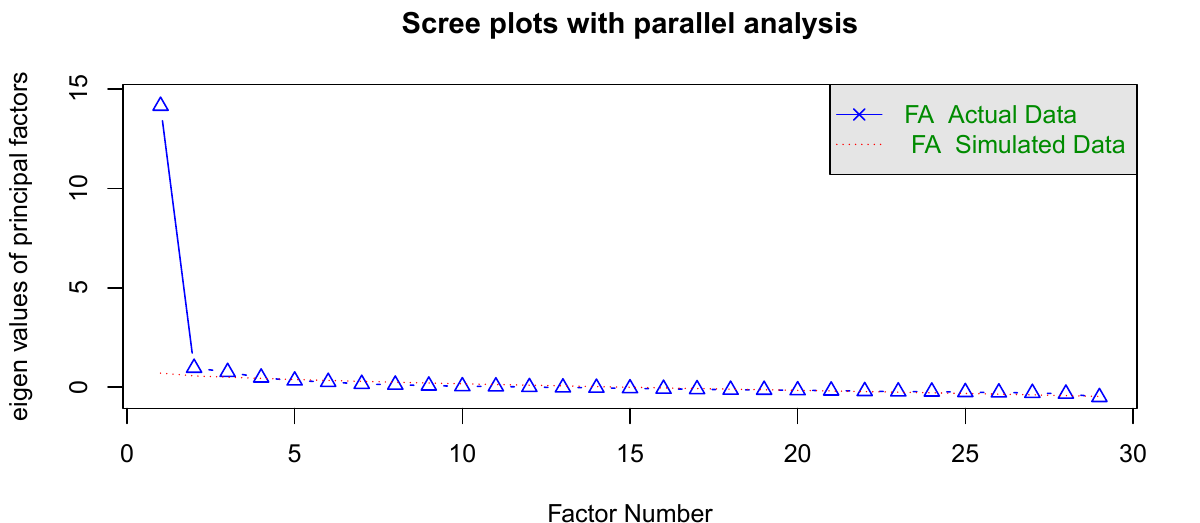}
    \caption{Scree plot from parallel analysis for \textbf{stimulus \stimB}
    .}
    \label{app:fig:screeplot_B}
\end{figure}

\begin{figure} []
    \centering
    \includegraphics[width=\linewidth]{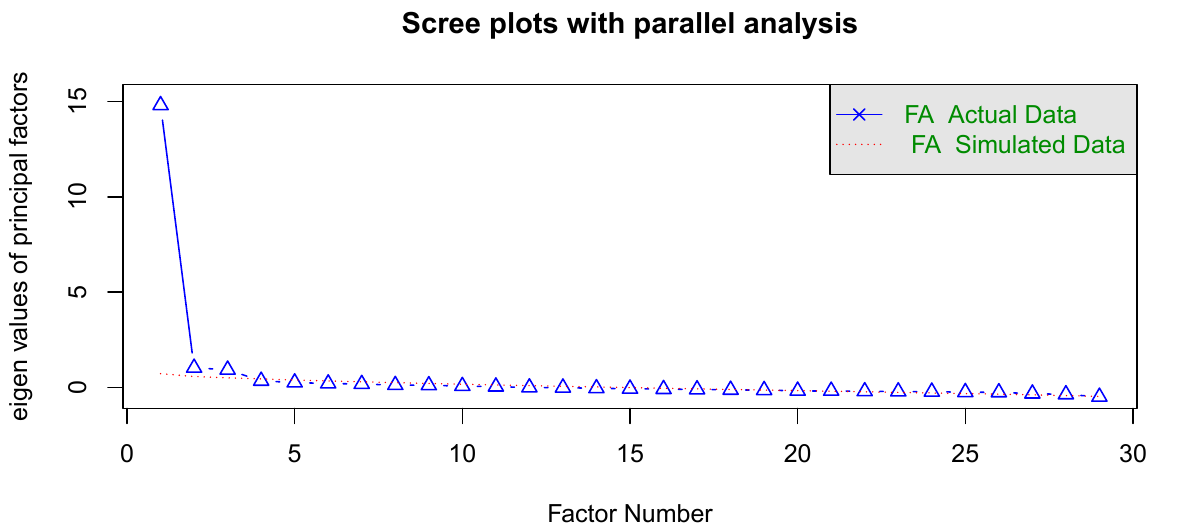}
    \caption{Scree plot from parallel analysis for \textbf{stimulus \stimC}
    .}
    \label{app:fig:screeplot_C}
\end{figure}

\begin{figure} []
    \centering
    \includegraphics[width=\linewidth]{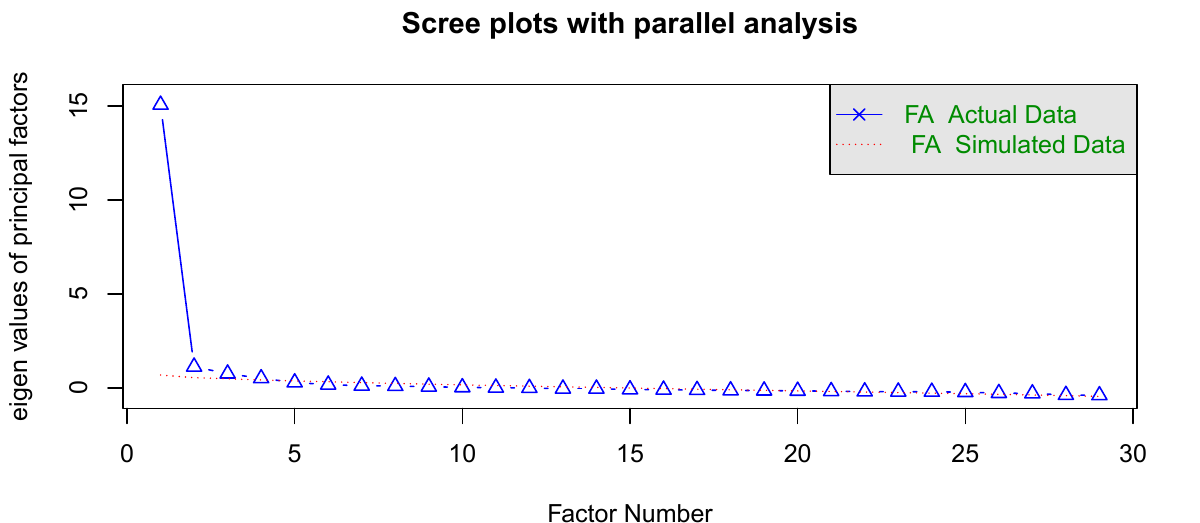}
    \caption{Scree plot from parallel analysis for \textbf{stimulus \stimD}
    .}
    \label{app:fig:screeplot_D}
\end{figure}

\begin{figure} []
    \centering
    \includegraphics[width=\linewidth]{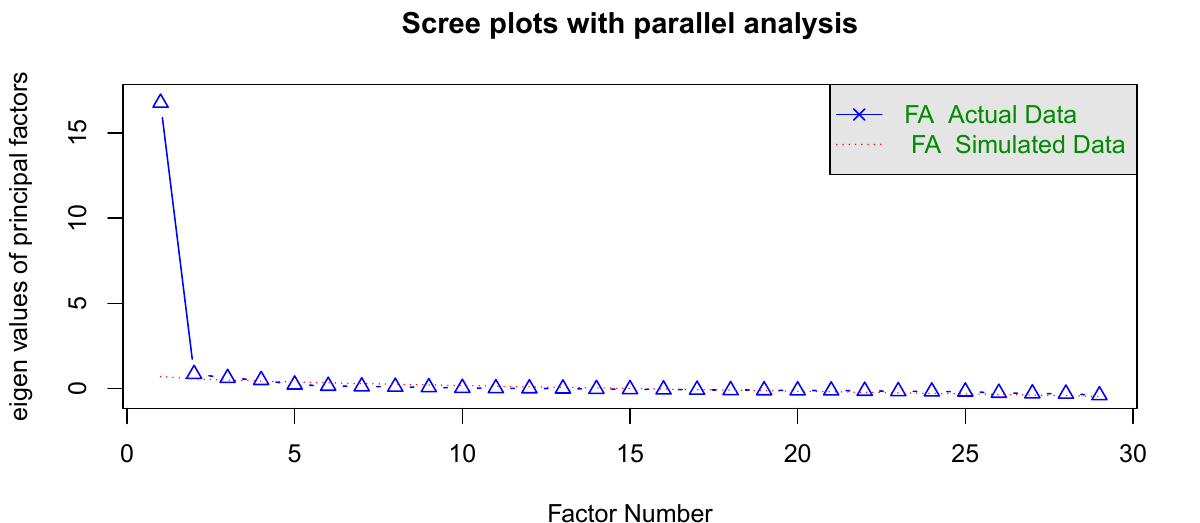}
    \caption{Scree plot from parallel analysis for \textbf{stimulus \stimE}
    .}
    \label{app:fig:screeplot_E}
\end{figure}

\begin{figure} []
    \centering
    \includegraphics[width=\linewidth]{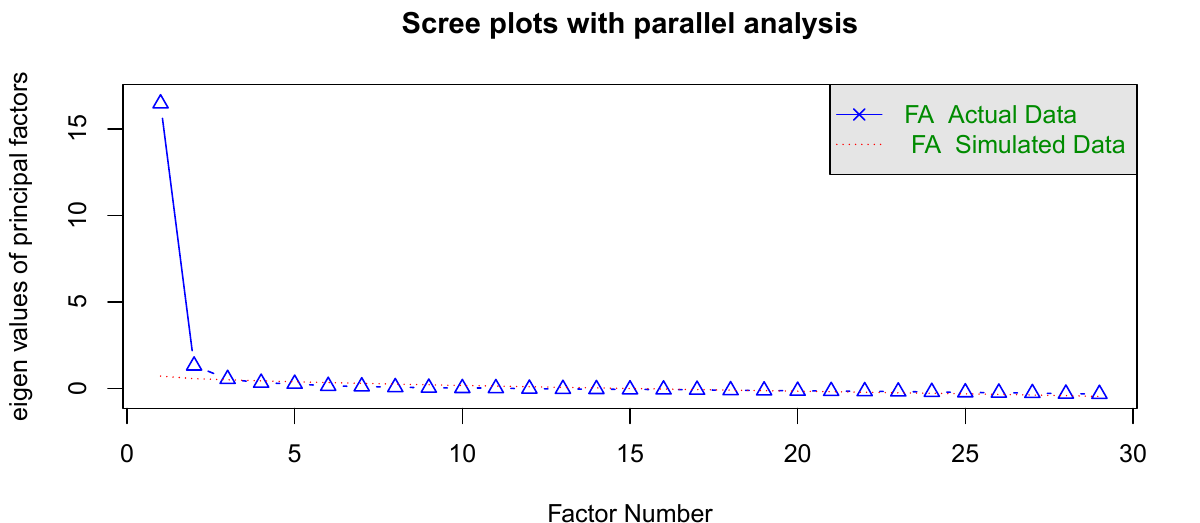}
    \caption{Scree plot from parallel analysis for \textbf{stimulus \stimF}
    .}
    \label{app:fig:screeplot_F}
\end{figure}

Visual analysis of such scree plots is two-fold:
\begin{itemize}
    \item \textbf{Identifying elbows:} the elbow is the point in the slope after which a line begins to level off, meaning that adding new factors does not explain considerably more variance in the data. It is often used as a heuristic for determining the number of factors to retain.
    \item \textbf{Identifying a crossing point between lines:} this is the visual representation of a parallel analysis. We can observe when the line plotted from EFA on our dataset crosses the line from similiar EFA on a randomly generated dataset of same size. The logic to this approach is that the eigenvalue of the last retained factor should exceed that of an eigenvalue from random data.
\end{itemize}
In this appendix, we first present the scree plot for the full dataset (\autoref{app:fig:screeplot_Agg}) on which we based on main analysis; and then individual scree plots for each of our 6 stimuli in the survey (\autoref{app:fig:screeplot_A} to \autoref{app:fig:screeplot_F}), which we examined to confirm that the slopes exhibited similar characteristics across stimuli.

\clearpage

\clearpage

\section{EFA: factors loadings and number of factors to retain}
\label{app:sec:EFA_loadings}

In this section, we present and discuss factor loading tables generated as part of our Exploratory Factor Analyses in \autoref{subsubsec:EFA}.

A 1-factor solution (\autoref{tab:factor_loadings_1factor}) would have provided us with the opportunity to develop a simple instrument; furthermore, it seems like a viable candidate according to visual analysis of scree plots (see \autoref{app:sec:EFA_screeplots}), because all plots show a clear break in the slop after 1 factor. However, there were many limitations in retaining a 1-factor solution for our final instrument:
\begin{itemize}
    \item A single scale would collapse the amount of information that our instrument can measure on a single average score, which might not help other researchers shed light on their empirical results. As parallel analyses and model fit metrics show, the construct appears to be better explained with 3 to 5 factors.
    \item As we describe in our \researchLog, loadings in a 1-factor solution appeared to be very different from one stimulus to another, suggesting that different reasons lead people to find visualization easy or difficult to read depending on the representation they saw. Therefore, deciding on which item should be included in a final, parsimonious scale would require many trade-offs, ultimately lowering the precision and usefulness of the instrument for other researchers.
    \item As the readability domain lacks formal definition, a detailed instrument would provide more opportunities for further research work on understanding and explaining readability.
\end{itemize}

Regarding multiple factors solutions, we discarded the 2-factor solution (\autoref{tab:factor_loadings_2factors}) as it did not make much conceptual sense. Similarly, the last factor in the 5-factors structure (\autoref{tab:factor_loadings_5factors}) was vague and weak in explained variance, so we discarded this solution as well.

In the 3-factors structures, we found the grouping of items to be meaningful. The first factor related to the ability of the reader to understand the visualization, with the following 3 first items in terms of loading importance: ``It is obvious for me how to read this visualization'', ``I am confident in my understanding of this visualization'', and ``I can easily understand the overall meaning of this data visualization''. The second factor related to the visual clarity of the layout, with the first 3 following items: ``I don't find this visualization crowded'', ``I don't find this visualization messy'', and ``I don't find distracting parts in this visualization''. Finally, the two items relating to reading data features were set together: ``I find data features (for example, a minimum, or an outlier, or a trend) visible in this visualization'' and ``I can clearly see data features (for example, a minimum, or an outlier, or a trend) in this visualization''.

However, some items were distributed across factors in this solution, for which me thought that they might better belong together, in a separate factor. For example: “This visualization effectively shows the data to me” in the first factor and “This visualization shows the data in a clear way for me” in the second factor, or “I can easily read this visualization” in the first factor and “I find this visualization readable” in the second factor. These problems did not appear anymore in the 4-factor structure---in which all items cited in the previous sentence belong to the 4th factor of ``DataReading''.

Together with model fit indices described in our \researchLog, we concluded our analysis by choosing the 4-factors solution and conducted a Multi-Group Confirmatory Factor Analysis (MG-CFA) to validate this choice, as described in \autoref{app:sec:EFA_CFA_4factors}.

\begin{figure*} [h]
    \centering
    \includegraphics[width=\textwidth]{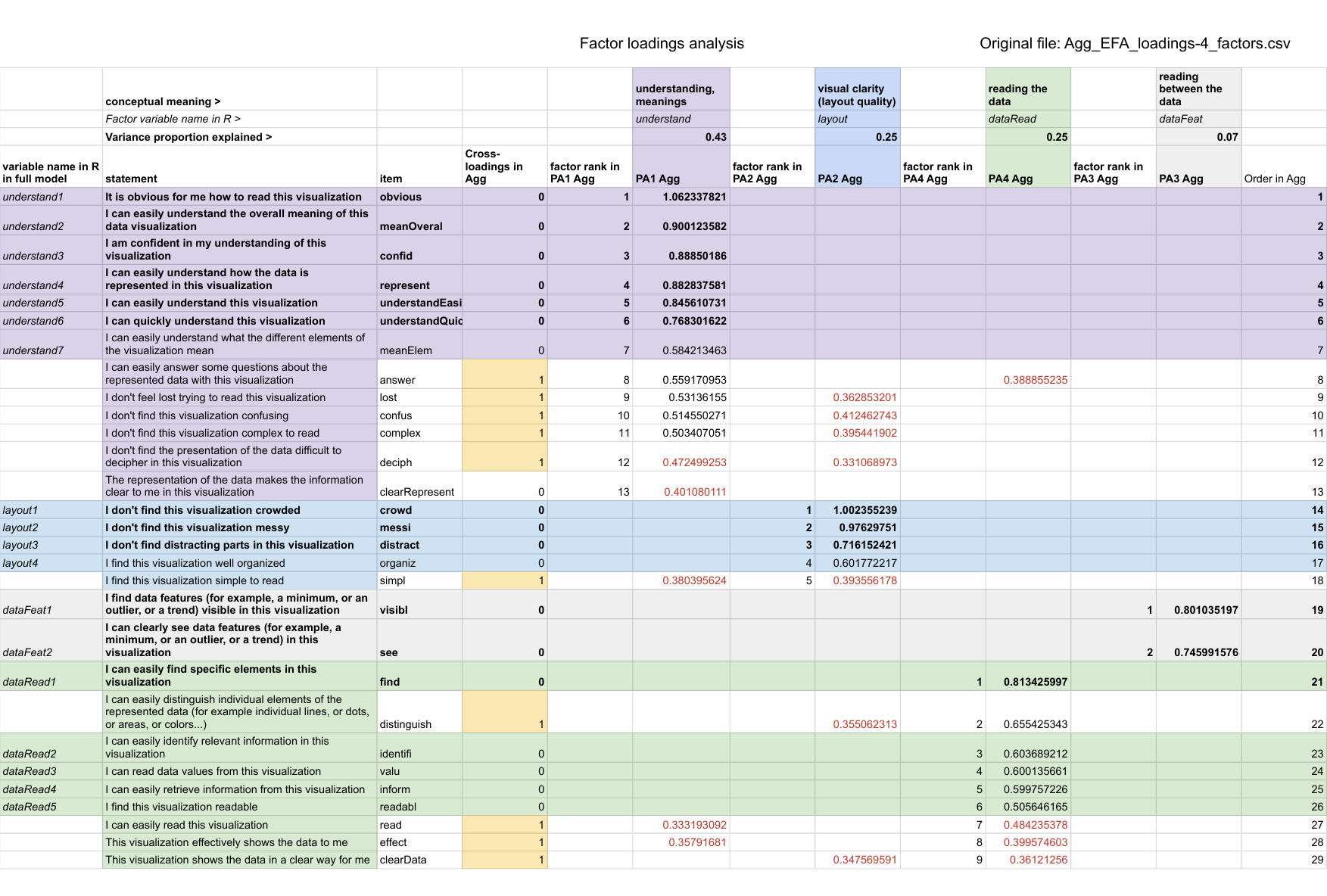}
    \caption{Table of factors loadings generated for the 4-factors solution found with a Principal Axis method. Fully colored lines indicate items that we screened as candidates for further tests of model fit and reliability, because they met the following selection criteria: items should not have loadings $>$ 0.3 in more than 1 factor (no cross-loadings), and items should have a loading value $>$ 0.5 for their associated factor (strong loading value).
    This table was generated following 3 steps. We provide all data and code in the \suppmat{supplemental material}{9cg8j} for steps 1 and 2, and a detailed explanation for all 3 steps in this description.
    \textbf{Step 1: Generate raw factor loadings.} We used our \texttt{EFA-+-all.Rmd} notebook in \texttt{R} to conduct EFA on the collected data. We generate a raw factor loading table for a 4-factors structure with the Principal Axis (PA) method \texttt{PA1 Stimuli 4 factors - Agg.csv}.
    \textbf{Step 2: Factor loadings pre-processing.} We used this table as an input for our \texttt{EFA factor loading analysis.ipynb} notebook in \texttt{Python} for pre-analysis processing. The pre-processing consisted in 3 interventions: a) filtering poor loadings values out, b) ranking the items for easier ordering of the table, and c) managing cross-loadings. In a) we replaced all loadings $<$ 0.3 with \texttt{NA} values. In b) we created the 4 ``factor ranks'' columns, so that, for each factor, all items item belonging to the factor were ranked by loading importance (1 being the highest). When an item exhibited loadings $>$ 0.3 for multiple factors, we assigned it to the factor where its loading was the highest (\eg, in the table above, the item ``lost'' had loadings $>$0.3 in both factors PA1 and PA2, but its factor loading value was higher for PA1; therefore it was assigned to PA1 for ranking). Additionally, we created the "Order in Agg" to facilitate the sorting of the complete spreadsheet, by incrementing all ranks factor after factor. In c) we counted, for each item, how many times it cross-loaded (number of factors after the first where the item had a loading $>$ 0.3) and saved this value in the ``Cross-loadings in Agg'' column.
    \textbf{Step 3: Table visual formatting and annotations (manual).} To facilitate analysis and discussion, we created a spreadsheet with the following layout rules: a) yellow background where cross-loadings counts were not null, b) bold face on items with factor loading values $>$ 0.7, and c) red color on factor loading values $<$ 0.5. We imported the statements corresponding to item codes to facilitate our conceptual analysis and reported the variance proportion explained from the EFA \texttt{R} output. We then annotated the table with our understanding of chich concepts the 4 factors represented, based on the associated items. We attributed one color to each factor to visually bring the structure up. Finally, we fully colored lines for items that met our selection criteria for the next step of item reduction.
    \emph{Note: steps 1 and 2 were conducted for 1-factor, 2-factors, 3-factors and 4-factors solutions. For 3-factors and 4-factors solutions, we also conducted EFA with the Maximum-Likelihood and observed that the factor loadings were not significantly different.
    We performed step 3) only for 3-factors adn and 4-factors solutions.}}
    \label{app:fig:4factors_loadings_commented}
    \vspace{5cm}
\end{figure*}

\begin{table}[t]
\centering
\fontsize{7pt}{7pt}\selectfont
\caption{EFA with \textbf{1 factor}: factor loadings for 29 items, and proportion of the total variance in our survey data explained by the factor.}
\label{tab:factor_loadings_1factor}
\tabulinesep=0.7mm
\begin{tabu} to 0.7\linewidth {X[1,l]X[1,l]}
\toprule
\textbf{terms} &\textbf{PA1} \\
\midrule
answer &0.811 \\
clearData &0.915 \\
clearRepresent &0.909 \\
complex &0.852 \\
confid &0.771 \\
confus &0.891 \\
crowd &0.650 \\
deciph &0.877 \\
distinguish &0.717 \\
distract &0.724 \\
effect &0.872 \\
find &0.822 \\
identifi &0.886 \\
inform &0.890 \\
lost &0.879 \\
meanElem &0.795 \\
meanOveral &0.718 \\
messi &0.788 \\
obvious &0.789 \\
organiz &0.815 \\
read &0.907 \\
readabl &0.874 \\
represent &0.842 \\
see &0.707 \\
simpl &0.913 \\
understandEasi &0.888 \\
understandQuick &0.875 \\
valu &0.633 \\
visibl &0.673 \\
\midrule
\textbf{Proportion of total variance explained}  & \textbf{0.67} \\
\bottomrule
\end{tabu}
\vspace{0.3cm}
\end{table}

\begin{table}[]
\centering
\fontsize{7pt}{7pt}\selectfont
\caption{EFA with \textbf{2 factors}: factor loadings for 29 items, and proportion of the total variance explained by each factor in our survey data.}
\label{tab:factor_loadings_2factors}
\tabulinesep=0.7mm
\begin{tabu} to 0.67\linewidth {X[1.5,l]X[1,l]X[1,l]}
\toprule
\textbf{items} &\textbf{PA1} &\textbf{PA2} \\
\midrule
answer &0.756 &0.088 \\
clearData &0.402 &0.567 \\
clearRepresent &0.536 &0.421 \\
complex &0.479 &0.419 \\
confid &0.943 &-0.145 \\
confus &0.494 &0.444 \\
crowd &-0.278 &0.994 \\
deciph &0.495 &0.429 \\
distinguish &0.038 &0.734 \\
distract &-0.065 &0.850 \\
effect &0.566 &0.350 \\
find &0.312 &0.560 \\
identifi &0.585 &0.345 \\
inform &0.525 &0.411 \\
lost &0.527 &0.397 \\
meanElem &0.737 &0.090 \\
meanOveral &0.915 &-0.174 \\
messi &-0.121 &0.980 \\
obvious &1.003 &-0.188 \\
organiz &0.162 &0.710 \\
read &0.502 &0.453 \\
readabl &0.453 &0.469 \\
represent &0.926 &-0.053 \\
see &0.484 &0.258 \\
simpl &0.443 &0.521 \\
understandEasi &0.884 &0.039 \\
understandQuick &0.793 &0.119 \\
valu &0.479 &0.183 \\
visibl &0.473 &0.233 \\
\midrule
\textbf{Proportion of total variance explained}  & \textbf{0.42} & \textbf{0.30} \\
\bottomrule
\end{tabu}
\end{table}

\begin{table}[]
\centering
\fontsize{7pt}{7pt}\selectfont
\caption{EFA with \textbf{3 factors}: factor loadings for 29 items, and proportion of the total variance explained by each factor in our survey data.}
\label{tab:factor_loadings_3factors}
\tabulinesep=0.7mm
\begin{tabu} to 0.8\linewidth {X[1.5,l]X[1,l]X[1,l]X[1,l]}
\toprule
\textbf{terms} &\textbf{PA1} &\textbf{PA2} &\textbf{PA3} \\\midrule
answer &0.669 &0.046 &0.145 \\
clearData &0.305 &0.520 &0.157 \\
clearRepresent &0.458 &0.385 &0.126 \\
complex &0.507 &0.448 &-0.059 \\
confid &0.956 &-0.135 &-0.020 \\
confus &0.533 &0.480 &-0.078 \\
crowd &-0.211 &1.065 &-0.151 \\
deciph &0.526 &0.460 &-0.065 \\
distinguish &-0.120 &0.656 &0.257 \\
distract &-0.050 &0.877 &-0.046 \\
effect &0.431 &0.279 &0.228 \\
find &0.109 &0.454 &0.340 \\
identifi &0.456 &0.279 &0.216 \\
inform &0.394 &0.344 &0.219 \\
lost &0.548 &0.421 &-0.045 \\
meanElem &0.655 &0.050 &0.138 \\
meanOveral &0.929 &-0.164 &-0.020 \\
messi &-0.086 &1.026 &-0.090 \\
obvious &1.044 &-0.169 &-0.059 \\
organiz &0.099 &0.682 &0.099 \\
read &0.441 &0.428 &0.096 \\
readabl &0.340 &0.412 &0.186 \\
represent &0.911 &-0.056 &0.025 \\
see &-0.020 &-0.057 &0.903 \\
simpl &0.418 &0.518 &0.032 \\
understandEasi &0.901 &0.053 &-0.028 \\
understandQuick &0.796 &0.127 &-0.009 \\
valu &0.195 &0.010 &0.503 \\
visibl &-0.042 &-0.094 &0.928 \\
\midrule
\textbf{Proportion of total variance}  & \textbf{0.36} & \textbf{0.27} & \textbf{0.11}\\
\bottomrule
\end{tabu}
\end{table}

\begin{table}[h!]
\vspace{0.2cm}
\centering
\fontsize{7pt}{7pt}\selectfont
\caption{EFA with \textbf{4 factors}: factor loadings for 29 items, and proportion of the total variance explained by each factor in our survey data.}
\label{tab:factor_loadings_4factors}
\tabulinesep=0.7mm
\begin{tabu} to \linewidth {X[1.5,l]X[1,l]X[1,l]X[1,l]X[1,l]}
\toprule
\textbf{terms} &\textbf{PA1} &\textbf{PA2} &\textbf{PA4} &\textbf{PA3} \\\midrule
answer &0.565 &-0.083 &0.381 &-0.012 \\
clearData &0.255 &0.346 &0.361 &0.039 \\
clearRepresent &0.402 &0.246 &0.310 &0.025 \\
complex &0.503 &0.397 &0.024 &-0.001 \\
confid &0.890 &-0.122 &0.057 &-0.024 \\
confus &0.517 &0.407 &0.060 &-0.026 \\
crowd &-0.135 &1.001 &-0.113 &0.006 \\
deciph &0.476 &0.333 &0.190 &-0.068 \\
distinguish &-0.227 &0.353 &0.658 &0.010 \\
distract &-0.038 &0.711 &0.158 &-0.020 \\
effect &0.360 &0.129 &0.397 &0.061 \\
find &-0.057 &0.120 &0.819 &0.010 \\
identifi &0.322 &0.035 &0.604 &-0.024 \\
inform &0.269 &0.097 &0.598 &-0.015 \\
lost &0.533 &0.361 &0.057 &-0.003 \\
meanElem &0.586 &-0.017 &0.240 &0.040 \\
meanOveral &0.897 &-0.097 &-0.063 &0.019 \\
messi &-0.011 &0.974 &-0.098 &0.049 \\
obvious &1.066 &-0.036 &-0.242 &0.055 \\
organiz &0.133 &0.599 &0.091 &0.105 \\
read &0.335 &0.204 &0.480 &-0.059 \\
readabl &0.242 &0.191 &0.509 &-0.003 \\
represent &0.885 &-0.009 &-0.026 &0.050 \\
see &0.064 &0.063 &0.071 &0.748 \\
simpl &0.385 &0.392 &0.209 &0.004 \\
understandEasi &0.845 &0.037 &0.071 &-0.018 \\
understandQuick &0.770 &0.128 &0.014 &0.023 \\
valu &0.098 &-0.157 &0.587 &0.184 \\
visibl &0.038 &0.040 &0.039 &0.808 \\
\midrule
\textbf{Proportion of total variance}  & \textbf{0.32} & \textbf{0.19} & \textbf{0.19} & \textbf{0.06}\\
\bottomrule
\end{tabu}
\end{table}

\begin{table}[h!]
\centering
\fontsize{7pt}{7pt}\selectfont
\caption{EFA with \textbf{5 factors}: factor loadings for 29 items, and proportion of the total variance explained by each factor in our survey data.}
\label{tab:factor_loadings_5factors}
\tabulinesep=0.7mm
\begin{tabu} to \linewidth {X[1.5,l]X[1,l]X[1,l]X[1,l]X[1,l]X[1,l]}
\toprule
\textbf{terms} &\textbf{PA1} &\textbf{PA4} &\textbf{PA5} &\textbf{PA2} &\textbf{PA3} \\
\midrule
answer &0.544 &0.401 &-0.041 &-0.010 &-0.021 \\
clearData &0.101 &0.330 &0.356 &0.182 &0.051 \\
clearRepresent &0.250 &0.286 &0.311 &0.124 &0.034 \\
complex &0.256 &-0.024 &0.519 &0.184 &0.014 \\
confid &0.858 &0.096 &-0.073 &-0.008 &-0.038 \\
confus &0.317 &0.038 &0.422 &0.231 &-0.018 \\
crowd &-0.004 &-0.075 &-0.032 &0.931 &-0.003 \\
deciph &0.248 &0.150 &0.477 &0.143 &-0.062 \\
distinguish &-0.100 &0.693 &-0.135 &0.359 &-0.002 \\
distract &0.061 &0.203 &-0.035 &0.658 &-0.034 \\
effect &0.248 &0.376 &0.212 &0.056 &0.070 \\
find &-0.009 &0.812 &-0.028 &0.122 &0.007 \\
identifi &0.280 &0.595 &0.074 &0.034 &-0.026 \\
inform &0.186 &0.571 &0.177 &0.041 &-0.012 \\
lost &0.350 &0.041 &0.373 &0.211 &0.004 \\
meanElem &0.593 &0.277 &-0.092 &0.068 &0.031 \\
meanOveral &0.921 &-0.016 &-0.165 &0.049 &0.004 \\
messi &-0.002 &-0.060 &0.190 &0.782 &0.048 \\
obvious &0.880 &-0.218 &0.183 &-0.028 &0.057 \\
organiz &0.114 &0.108 &0.146 &0.481 &0.109 \\
read &0.155 &0.440 &0.380 &0.056 &-0.054 \\
readabl &0.152 &0.484 &0.211 &0.107 &0.002 \\
represent &0.802 &0.006 &0.034 &0.045 &0.046 \\
see &0.021 &0.054 &0.026 &0.024 &0.800 \\
simpl &0.128 &0.152 &0.557 &0.154 &0.020 \\
understandEasi &0.642 &0.063 &0.292 &-0.020 &-0.014 \\
understandQuick &0.528 &-0.010 &0.401 &0.007 &0.034 \\
valu &0.080 &0.562 &-0.001 &-0.133 &0.195 \\
visibl &0.039 &0.038 &-0.066 &0.037 &0.852 \\
\midrule
\textbf{Proportion of total variance}  & \textbf{0.25} & \textbf{0.18} & \textbf{0.14} & \textbf{0.13} & \textbf{0.06} \\
\bottomrule
\end{tabu}
\vspace{12.47cm}
\end{table}

\clearpage

\section{Multi-Group Confirmatory Factor Analysis with a 4-factors model}
\label{app:sec:EFA_CFA_4factors}
Confirmatory Factor Analysis (CFA) is used to examine goodness of fit of a model to a dataset by comparing it to a baseline (null) model. Multi-Group CFA (MG-CFA) is a variation of CFA with an added grouping variable (in our case, the 6 independent groups of participant who rated a specific visualization stimulus). MG-CFA allows to assess to what extent the factor structure (patterns of factor loadings and factor covariances) is consistent across different groups of respondents. The output of CFA is examined through the model fit metrics it produces.

We performed MG-CFA using the \texttt{lavaan} package in \texttt{R} with the following full 4-factor structure extracted from \autoref{app:fig:4factors_loadings_commented}:
\begin{verbatim}
full_Readability_factors <- list(
  understand = c('obvious', 'meanOveral', 'confid',
                 'represent', 'understandEasi', 
                 'understandQuick', 'meanElem'),
  layout = c('crowd', 'messi', 'distract', 'organiz'),
  dataRead = c('find', 'identifi', 'valu', 'inform',
               'readabl'),
  dataFeat = c('visibl', 'see')
)
\end{verbatim}

We report a partial view of resulting fit metrics in \autoref{tab:exp_fit_CFA_fullModel}, and we share the \texttt{R} notebook as well as the complete output in our \suppmat{supplemental material}{9cg8j}. Notably, we obtained the following metrics:
\begin{itemize}
    \item \textbf{Measures of fit indices} (values closer to 1 are better): the Tucker–Lewis Index (TLI) was .94 and the Comparative Fit Index (CFI) was .95;
    \item \textbf{Measures of covariance discrepancies} between observed and the model-implied data (values closer to 0 are better): The Standardized Root Mean square Residual (SRMR) was .046 and the Root Mean Square Error of Approximation (RMSEA) was .067.
\end{itemize} 

In their reference work on cutoff criteria for fit indices, Hu and Bentler \cite{hu_1999_CutoffCriteria} recommend a combination of two criteria to retain a model: fit indices such as TLI or CFI should be higher than .95, and SRMR should be lower than 0.9. They add that a combinational cutoff criterion of RMSEA at .06 and SRMR at .09 is possible but less desirable because it tends to reject models that are, in fact, good fits. 

With those results, we were satisfied that our 4-factors model was a good fit to explain our survey's data. However, CFA conducted on the same dataset from which the factors were extracted is expected to produce good results, and further validation is required by conducting CFA on a set of observations from an independent group \cite{Boateng:2018:BestPractices}. We conducted such analysis in the validation phase of our work in \autoref{subsec:valid_results}.

\begin{table}[t]
\centering
\fontsize{7pt}{7pt}\selectfont
\caption{Fit metrics from Multi-Group Confirmatory Factor Analysis at the scale development stage. In this analysis we fit the full 4-factors model extracted from our EFA to our survey data, using the stimulus as grouping variable. The resulting fit metrics allow us to estimate how appropriate the 4-factor structure is for explaining not only the entire aggregate set of answers, but also subsets of data in our independent groups, as each participant had been randomly assigned to rate 1 of our 6 stimuli visualizations.}
\label{tab:exp_fit_CFA_fullModel}
\tabulinesep=0.8mm
\begin{tabu} to 0.6\linewidth {X[1.2,l]X[1,l]}
\toprule
\textbf{Fit metric} & \textbf{Value from CFA}\\
\midrule
chisq &1802 \\
df &774 \\
pvalue &0 \\
baseline.chisq &21465 \\
baseline.df &918 \\
baseline.pvalue &0 \\
\midrule
cfi &0.950 \\
tli &0.941 \\
cfi.robust &0.950 \\
tli.robust &0.940 \\
\midrule
rmsea &0.067 \\
rmsea.ci.lower &0.063 \\
rmsea.ci.upper &0.071 \\
rmsea.ci.level &0.9 \\
rmsea.pvalue &4.2886e-12 \\
\midrule
srmr &0.0456 \\
srmr\_bentler &0.0456 \\
srmr\_bentler\_nomean &0.0479 \\
crmr &0.0479 \\
crmr\_nomean &0.0506 \\
srmr\_mplus &0.0456 \\
srmr\_mplus\_nomean &0.0479 \\
\bottomrule
\end{tabu}
\end{table} 

\section{Item-subscale reliability}
\label{app:sec:exp_reliability}

We provide in \autoref{tab:expl_items_reliability} reliability and related statistics for each individual item from the 4-factors structure we identified in \autoref{app:sec:EFA_loadings} and tested in \autoref{app:sec:EFA_CFA_4factors}. These values inform us about the importance of each item regarding in ensuring reliability in measures of the factor it relates to. We use these values to build model\_2 in the following \autoref{app:sec:items_reduction}.
\begin{table*}[t]
\centering
\caption{Reliability and related statistics for each individual item. It provides reliability metrics for each factor, if an item were removed. In such a table, the more an alpha, G6, r (items correlation), or S$/$N value drops, the more important an item is to a given factor's reliability.
}
\label{tab:expl_items_reliability}
\fontsize{7pt}{7pt}\selectfont
\tabulinesep=0.8mm
\begin{tabu} to \textwidth{ X[1.5, l] X[1, l] X[1, l] X[1, l] X[1, l] X[1, l] X[1, l] X[1, l] X[1, l] X[1, l]}
\toprule
\textbf{Items} &\textbf{Factor} &\textbf{raw\_alpha} &\textbf{std.alpha} &\textbf{alpha se} &\textbf{G6(smc)} &\textbf{var.r} &\textbf{med.r} &\textbf{average\_r} &\textbf{S/N} \\\midrule
\textbf{obvious} &understand &0.941 &0.943 &0.002 &0.935 &0.003 &0.728 &0.733 &16.459 \\
\textbf{meanOveral} &understand &0.947 &0.948 &0.002 &0.941 &0.002 &0.740 &0.753 &18.314 \\
\textbf{confid} &understand &0.944 &0.945 &0.002 &0.938 &0.003 &0.731 &0.743 &17.332 \\
\textbf{represent} &understand &0.939 &0.941 &0.002 &0.933 &0.003 &0.719 &0.725 &15.817 \\
\textbf{understandEasi} &understand &0.937 &0.939 &0.002 &0.930 &0.002 &0.719 &0.719 &15.384 \\
\textbf{understandQuick} &understand &0.941 &0.942 &0.002 &0.933 &0.002 &0.731 &0.729 &16.101 \\
\textbf{meanElem} &understand &0.946 &0.947 &0.002 &0.940 &0.002 &0.740 &0.750 &18.019 \\
\textbf{crowd} &layout &0.887 &0.887 &0.005 &0.846 &0.004 &0.723 &0.724 &7.869 \\
\textbf{messi} &layout &0.855 &0.856 &0.006 &0.798 &0.000 &0.665 &0.664 &5.933 \\
\textbf{distract} &layout &0.891 &0.892 &0.005 &0.856 &0.005 &0.762 &0.733 &8.223 \\
\textbf{organiz} &layout &0.885 &0.885 &0.005 &0.840 &0.002 &0.723 &0.720 &7.708 \\
\textbf{find} &dataRead &0.900 &0.899 &0.004 &0.883 &0.016 &0.683 &0.690 &8.899 \\
\textbf{identifi} &dataRead &0.895 &0.894 &0.004 &0.874 &0.011 &0.675 &0.678 &8.429 \\
\textbf{valu} &dataRead &0.935 &0.936 &0.003 &0.917 &0.001 &0.778 &0.784 &14.505 \\
\textbf{inform} &dataRead &0.893 &0.893 &0.004 &0.871 &0.010 &0.675 &0.675 &8.310 \\
\textbf{readabl} &dataRead &0.900 &0.899 &0.004 &0.883 &0.013 &0.685 &0.691 &8.945 \\
\textbf{visibl} &dataFeat &0.765 &0.784 &NA &0.614 &0.000 &0.784 &0.784 &3.627 \\
\textbf{see} &dataFeat &0.804 &0.784 &NA &0.614 &0.000 &0.784 &0.784 &3.627 \\
\bottomrule
\end{tabu}
\vspace{2.2cm}
\end{table*}


\section{Reducing items in subscales}
\label{app:sec:items_reduction}
In this section we provide details regarding the final step of scale development: the selection of items that will compose our final subscales.

Reference authors emphasize that item reduction is the heart of scale development \cite{Devellis:2021:ScaleDevelopment}. It entails the use of tools from Classical Test Theory (CTT) and Item Response Theory (IRT). While IRT has a more refined approach in assessing items characteristics that might affect their measuring performance, IRT analysis for less than 10 items might be unreliable \cite{valdivia_2024_NumberResponse}. We did run an IRT analysis, with the goal to provide additional information in case other decision criteria appeared inconclusive, but we only used it to confirm our final selection of subscales' items.

In CTT, researchers use reliability indicators such as Cronbach's alpha to select items that will minimize the final scale's measurement error \cite{Devellis:2021:ScaleDevelopment}.  Additionally, researchers can consider factor loadings \cite{Boateng:2018:BestPractices} for selecting items that capture a good amount of information about the factor's underlying construct.

\subsection{Creating 3 combinations of items}
\label{app:subsec:test_models}

To explore possible combinations of items, we calculated items ranks based on factor loadings and on reliability indicators for each subscale. We produced 3 candidate combinations of items: a reliability optimizer, a factor loadings optimizer, and a mixed approach based on average ranks. 
For reliability optimization (model 1), we calculated the effect of dropping items on reliability indicators for each scale using the \texttt{psych} package in \texttt{R}. The lower the alpha drops if an item were removed, the more important the item is to ensure reliability of the final scale.
We thus ranked the items based on how much the scale' alpha would lowered if they were dropped.
For loadings optimization (model 2), we retrieved the factors loadings from our EFA. We then ranked each item accordingly to their loading values, higher being better.
For each item, we also calculate an average of the two ranks and use it as a basis for a mixed approach (model 3), which might offer a useful trade-off between reliability and factor representation.

For each approach, we selected the 3 higher ranking items in \textsc{Understand}, \textsc{Layout}, and \textsc{DataRead} to build a model. \textsc{DataFeat} has only two items which remain the same across all models.
We then put the 3 models to test. We provide the ranks and the selected items in \autoref{tab:exp_models_reliability_loadings}.
\begin{table*}[t]
\centering
\fontsize{7pt}{7pt}\selectfont
\caption{Ranking of candidate items for each subscale by factor loading and reliability indices to build 3 possible models based on: rank in factor loadings, rank in effect on reliability if dropped, and an average of the two ranks.}
\label{tab:exp_models_reliability_loadings}
\tabulinesep=0.8mm 
\begin{tabu} to 1\textwidth {X[0.6,c]|X[0.7,c]|X[0.6,c]|X[1.5,l]|X[0.8,l]|X[0.6,l]|X[0.6,l]X[0.5,l]X[0.5,l]}
\toprule
\textbf{Model 3} &\textbf{Model 2} &\textbf{Model 1} &\textbf{Items} &\textbf{Factor loadings} &\textbf{alpha with all items} &\multicolumn{2}{m{3cm}}{\textbf{Reliability of subscale if item was dropped}} \\
Mean ranking &Effect on reliability if dropped rank &Factor loadings rank &factor\_name.item\_name &From EFA &raw\_alpha of full factor &raw\_alpha &std.alpha \\
\midrule
\textbf{2.5} \checkmark &4 &\textbf{1} \checkmark &understand.obvious &1.062 &0.950 &0.941 &0.94 \\
\textbf{3} \checkmark &\textbf{2} \checkmark &4 &understand.represent &0.883 &0.950 &0.939 &0.94 \\
\textbf{3} \checkmark &\textbf{1} \checkmark &5 &understand.understandEasi &0.846 &0.950 &0.937 &0.94 \\
4 &6 &\textbf{2} \checkmark &understand.meanOveral &0.900 &0.950 &0.947 &0.95 \\
4 &5 &\textbf{3} \checkmark &understand.confid &0.889 &0.950 &0.944 &0.95 \\
4.5 &\textbf{3} \checkmark &6 &understand.understandQuick &0.768 &0.950 &0.941 &0.94 \\
7 &7 &7 &understand.meanElem &0.584 &0.950 &0.946 &0.95 \\
\midrule
\textbf{1.5} \checkmark &\textbf{1} \checkmark &\textbf{2} \checkmark &layout.messi &0.976 &0.907 &0.855 &0.86 \\
\textbf{2} \checkmark &\textbf{3} \checkmark &\textbf{1} \checkmark &layout.crowd &1.002 &0.907 &0.887 &0.89 \\
\textbf{3} \checkmark &\textbf{2} \checkmark &4 &layout.organiz &0.602 &0.907 &0.885 &0.89 \\
3.5 &4 &\textbf{3} \checkmark &layout.distract &0.716 &0.907 &0.891 &0.89 \\
\midrule
\textbf{2} \checkmark &\textbf{3} \checkmark &\textbf{1} \checkmark &dataRead.find &0.813 &0.923 &0.900 &0.90 \\
\textbf{2} \checkmark &\textbf{2} \checkmark &\textbf{2} \checkmark &dataRead.identifi &0.604 &0.923 &0.895 &0.89 \\
\textbf{2.5} \checkmark &\textbf{1} \checkmark &4 &dataRead.inform &0.600 &0.923 &0.893 &0.89 \\
3.5 &4 &\textbf{3} \checkmark &dataRead.valu &0.600 &0.923 &0.935 &0.94 \\
4 &3 &5 &dataRead.readabl &0.506 &0.923 &0.900 &0.90 \\
\midrule
\textbf{1} \checkmark &\textbf{1} \checkmark &\textbf{1} \checkmark &dataFeat.visibl &0.801 &0.879 &0.765 &0.78 \\
\textbf{1.5} \checkmark &\textbf{1} \checkmark &\textbf{2} \checkmark &dataFeat.see &0.746 &0.879 &0.804 &0.78 \\
\bottomrule
\end{tabu}
\vspace{2cm}
\end{table*}

\subsection{Comparing results: reliability and model fit}

\begin{table}[]
\centering
\caption{Cronbach's alpha and McDonald's omega reliability coefficients for Model 1 (reliability-based) in \autoref{app:sec:items_reduction}}
\label{tab:exp_reliability_model1}
\fontsize{7pt}{7pt}\selectfont
\tabulinesep=0.8mm
\begin{tabu} to \linewidth{ X[1.8, l] X[2.4, l]| X[1, c] X[1, c] X[1, c] X[1, c] X[1, c] X[1, c] X[2, c] }
\toprule
  & \textbf{Subscale}  & A & B & C & D & E & F & \textbf{Full survey} \\
\midrule
\multirow[t]{4}{*}{\textbf{omega tot}} & dataFeat & 0.870 & 0.826 & 0.837 & 0.858 & 0.807 & 0.822 & 0.879 \\
 & dataRead & 0.758 & 0.741 & 0.783 & 0.767 & 0.804 & 0.820 & 0.855 \\
 & layout & 0.768 & 0.845 & 0.838 & 0.752 & 0.753 & 0.822 & 0.886 \\
 & understand & 0.811 & 0.755 & 0.803 & 0.814 & 0.820 & 0.820 & 0.878 \\
\midrule
\multirow[t]{4}{*}{\textbf{raw alpha}} & dataFeat & 0.868 & 0.824 & 0.835 & 0.859 & 0.815 & 0.822 & 0.879 \\
 & dataRead & 0.736 & 0.739 & 0.773 & 0.741 & 0.802 & 0.809 & 0.850 \\
 & layout & 0.760 & 0.844 & 0.834 & 0.748 & 0.751 & 0.801 & 0.885 \\
 & understand & 0.809 & 0.745 & 0.796 & 0.812 & 0.821 & 0.818 & 0.874 \\
\midrule
\multirow[t]{4}{*}{\textbf{std alpha}} & dataFeat & 0.868 & 0.825 & 0.835 & 0.859 & 0.816 & 0.823 & 0.879 \\
 & dataRead & 0.757 & 0.740 & 0.776 & 0.749 & 0.802 & 0.808 & 0.849 \\
 & layout & 0.765 & 0.845 & 0.837 & 0.747 & 0.751 & 0.810 & 0.885 \\
 & understand & 0.811 & 0.747 & 0.801 & 0.813 & 0.823 & 0.820 & 0.878 \\
\bottomrule
\end{tabu}
\end{table}

\begin{table}[]
\centering
\caption{Cronbach's alpha and McDonald's omega reliability coefficients for Model 2 (loadings-based) in \autoref{app:sec:items_reduction}}
\label{tab:exp_reliability_model2}
\fontsize{7pt}{7pt}\selectfont
\tabulinesep=0.8mm
\begin{tabu} to \linewidth{ X[1.8, l] X[2.4, l]| X[1, c] X[1, c] X[1, c] X[1, c] X[1, c] X[1, c] X[2, c] }
\toprule
  & \textbf{Subscale}  & A & B & C & D & E & F & \textbf{Full survey} \\
\midrule
\multirow[t]{4}{*}{\textbf{omega tot}}  & dataFeat & 0.872 & 0.823 & 0.839 & 0.857 & 0.811 & 0.822 & 0.877 \\
 & dataRead & 0.828 & 0.799 & 0.819 & 0.870 & 0.870 & 0.903 & 0.920 \\
 & layout & 0.792 & 0.850 & 0.847 & 0.791 & 0.821 & 0.867 & 0.894 \\
 & understand & 0.858 & 0.875 & 0.874 & 0.852 & 0.910 & 0.868 & 0.931 \\
\midrule
\multirow[t]{4}{*}{\textbf{raw alpha}} & dataFeat & 0.868 & 0.824 & 0.835 & 0.859 & 0.815 & 0.822 & 0.879 \\
 & dataRead & 0.820 & 0.791 & 0.811 & 0.870 & 0.868 & 0.904 & 0.920 \\
 & layout & 0.773 & 0.841 & 0.827 & 0.789 & 0.815 & 0.854 & 0.891 \\
 & understand & 0.855 & 0.858 & 0.869 & 0.848 & 0.909 & 0.865 & 0.927 \\
\midrule
\multirow[t]{4}{*}{\textbf{std alpha}} & dataFeat & 0.868 & 0.825 & 0.835 & 0.859 & 0.816 & 0.823 & 0.879 \\
 & dataRead & 0.826 & 0.794 & 0.817 & 0.870 & 0.868 & 0.904 & 0.920 \\
 & layout & 0.774 & 0.850 & 0.837 & 0.789 & 0.814 & 0.860 & 0.892 \\
 & understand & 0.857 & 0.871 & 0.874 & 0.849 & 0.910 & 0.868 & 0.930 \\
\bottomrule
\end{tabu}
\end{table}

\begin{table}[]
\centering
\caption{Cronbach's alpha and McDonald's omega reliability coefficients for Model 3 (average ranks from model 1 and 2) in \autoref{app:sec:items_reduction}}
\label{tab:exp_reliability_model3}
\fontsize{7pt}{7pt}\selectfont
\tabulinesep=0.8mm
\begin{tabu} to \linewidth{ X[1.8, l] X[2.4, l]| X[1, c] X[1, c] X[1, c] X[1, c] X[1, c] X[1, c] X[2, c] }
\toprule
  & \textbf{Subscale}  & A & B & C & D & E & F & \textbf{Full survey} \\
\midrule
\multirow[t]{4}{*}{\textbf{omega tot}} & dataFeat & 0.872 & 0.822 & 0.838 & 0.858 & 0.815 & 0.822 & 0.878 \\
 & dataRead & 0.828 & 0.799 & 0.819 & 0.869 & 0.871 & 0.903 & 0.920 \\
 & layout & 0.792 & 0.850 & 0.847 & 0.791 & 0.821 & 0.868 & 0.895 \\
 & understand & 0.869 & 0.859 & 0.838 & 0.839 & 0.914 & 0.848 & 0.923 \\
\midrule
\multirow[t]{4}{*}{\textbf{raw alpha}} & dataFeat & 0.868 & 0.824 & 0.835 & 0.859 & 0.815 & 0.822 & 0.879 \\
 & dataRead & 0.820 & 0.791 & 0.811 & 0.870 & 0.868 & 0.904 & 0.920 \\
 & layout & 0.773 & 0.841 & 0.827 & 0.789 & 0.815 & 0.854 & 0.891 \\
 & understand & 0.867 & 0.852 & 0.826 & 0.837 & 0.914 & 0.847 & 0.921 \\
\midrule
\multirow[t]{4}{*}{\textbf{std alpha}} & dataFeat & 0.868 & 0.825 & 0.835 & 0.859 & 0.816 & 0.823 & 0.879 \\
 & dataRead & 0.826 & 0.794 & 0.817 & 0.870 & 0.868 & 0.904 & 0.920 \\
 & layout & 0.774 & 0.850 & 0.837 & 0.789 & 0.814 & 0.860 & 0.892 \\
 & understand & 0.868 & 0.853 & 0.830 & 0.838 & 0.914 & 0.848 & 0.922 \\
\bottomrule
\end{tabu}
\end{table}

For each of the produced models, we used the \texttt{psych} package to calculate Cronbach's alpha and McDonald's omega. We provide the code and data sources for this analysis in our \suppmat{supplemental material}{9cg8j}.

\autoref{tab:exp_reliability_model1} to \autoref{tab:exp_reliability_model3} show Cronbach's alpha coefficients calculated with \texttt{R's psych} package for each option. We used the \texttt{lavaan} package for conducting Multi-Group Confirmatory Factor Analysis (MG-CFA) for each model, and compared model fit values. We provide code and complete data from this analysis in our \suppmat{supplemental material}{9cg8j}.

\begin{table}[H]
\centering
\fontsize{7pt}{7pt}\selectfont
\caption{Fit metrics from Multi-Group Confirmatory Factor Analysis on our 3 candidate models, as described in \autoref{app:subsec:test_models}.}
\label{tab:item_reduction_models_fit}
\tabulinesep=0.8mm
\begin{tabu} to \linewidth {X[1,l]X[1.3,l]X[1,l]X[1,l]X[1,l]X[1.1,l]}
\toprule
\textbf{Fit indices} &\textbf{Full model (all items)} &\textbf{Model\_1} &\textbf{Model\_2} &\textbf{Model\_3} &\textbf{Model\_final} \\
\midrule
\textbf{chisq} &1802 &502 &540 &537 &478 \\
\textbf{cfi} &0.950 &0.970 &0.974 &0.974 &0.978 \\
\textbf{tli} &0.941 &0.956 &0.963 &0.962 &0.968 \\
\textbf{srmr} &0.046 &0.040 &0.036 &0.036 &0.035 \\
\textbf{rmsea} &0.067 &0.064 &0.068 &0.068 &0.061 \\
\bottomrule
\end{tabu}
\end{table}

Results showed that all models performed similarly, although the 3d model from average rankings performed slightly better(see Model 1, Model 2 and Model 3 in \autoref{tab:item_reduction_models_fit}). We provide the comparison of all 3 aforementioned models as well as the ``full'' version (containing all items selected from), the final, adjusted model in \autoref{tab:item_reduction_models_fit}.

\subsection{Final selection of items}
We used model 3 as our final selection, with one modification: in the \textsc{Layout} subscale, we exchanged the item ``I find this visualization well organized'' with ``I don’t find distracting parts in this visualization'' in order to ensure better phrasing consistency with other items of the subscale.
We obtained the finat set of items described in \autoref{tab:PREVis_items}.

\begin{table}[H]
\centering
\footnotesize
\tabulinesep=0.8mm
\caption{Our final selection of items forming the PREVis instrument across 4 subscales: \SUn, \SLa, \SDF, and \SDR.}
\label{tab:PREVis_items}
\begin{tabu} to \linewidth{ X[0.7, l] X[2, l] }
\toprule
\textbf{Item code} &\textbf{Item statement} \\\midrule
\SUnDot obvious &It is obvious for me how to read this visualization \\
\SUnDot represent &I can easily understand how the data is represented in this visualization \\
\SUnDot understandEasi &I can easily understand this visualization \\
\SLaDot messi &I don't find this visualization messy \\
\SLaDot crowd &I don't find this visualization crowded \\
\SLaDot distract &I don't find parts of the visualization distracting \\
\SDFDot visibl &I find data features (for example, a minimum, or an outlier, or a trend) visible in this visualization \\
\SDFDot see &I can clearly see data features (for example, a minimum, or an outlier, or a trend) in this visualization \\
\SDRDot inform &I can easily retrieve information from this visualization \\
\SDRDot identifi &I can easily identify relevant information in this visualization \\
\SDRDot find &I can easily find specific elements in this visualization \\
\bottomrule
\end{tabu}
\end{table}

We submitted our final selection to the same MG-CFA and obtained improved model fit as shown in \autoref{tab:item_reduction_models_fit}.
The reliability measures for this final subscales composition are displayed in \autoref{tab:exp_reliability_model_final}.


\section{Ratings plots in exploratory survey}
\label{app:sec:exp_ratings}
\begin{table*}[t]
\centering
\caption{Cronbach's alpha and McDonald's omega reliability coefficients for our final subscales \PREVisColors.}
\label{tab:exp_reliability_model_final}
\fontsize{8pt}{8pt}\selectfont
\tabulinesep=0.8mm
\begin{tabu} to 0.7\textwidth{ X[1.8, l] X[2.4, l]| X[1, c] X[1, c] X[1, c] X[1, c] X[1, c] X[1, c] X[2, c] }
\toprule
  & \textbf{Subscale}  & A & B & C & D & E & F & \textbf{Full survey} \\
\midrule
\multirow[t]{4}{*}{\textbf{omega tot}} & dataFeat & 0.872 & 0.826 & 0.837 & 0.858 & 0.811 & 0.822 & 0.879 \\
 & dataRead & 0.829 & 0.799 & 0.819 & 0.870 & 0.871 & 0.904 & 0.920 \\
 & layout & 0.768 & 0.844 & 0.838 & 0.751 & 0.753 & 0.823 & 0.886 \\
 & understand & 0.869 & 0.859 & 0.839 & 0.839 & 0.914 & 0.849 & 0.922 \\
\midrule
\multirow[t]{4}{*}{\textbf{raw alpha}} & dataFeat & 0.868 & 0.824 & 0.835 & 0.859 & 0.815 & 0.822 & 0.879 \\
 & dataRead & 0.820 & 0.791 & 0.811 & 0.870 & 0.868 & 0.904 & 0.920 \\
 & layout & 0.760 & 0.844 & 0.834 & 0.748 & 0.751 & 0.801 & 0.885 \\
 & understand & 0.867 & 0.852 & 0.826 & 0.837 & 0.914 & 0.847 & 0.921 \\
\midrule
\multirow[t]{4}{*}{\textbf{std alpha}} & dataFeat & 0.868 & 0.825 & 0.835 & 0.859 & 0.816 & 0.823 & 0.879 \\
 & dataRead & 0.826 & 0.794 & 0.817 & 0.870 & 0.868 & 0.904 & 0.920 \\
 & layout & 0.765 & 0.845 & 0.837 & 0.747 & 0.751 & 0.810 & 0.885 \\
 & understand & 0.868 & 0.853 & 0.830 & 0.838 & 0.914 & 0.848 & 0.922 \\
\bottomrule
\end{tabu}
\end{table*}

\begin{figure*} [h]
    \centering
    \includegraphics[width=0.8\textwidth]{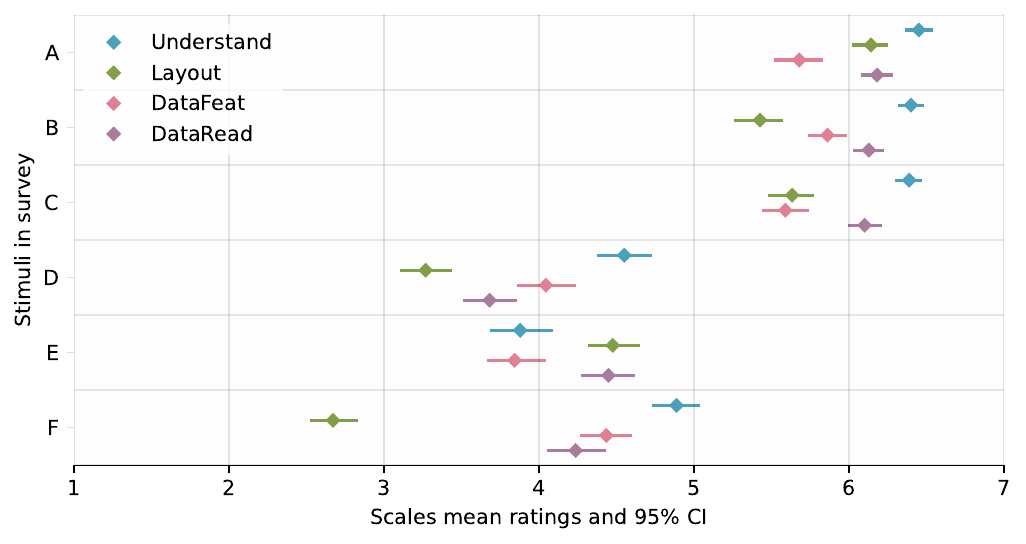}
    \caption{Average ratings (from 1 = ``Strongly disagree'' to 7 = ``Strongly agree'') using the four PREVis subscales \mbox{\raisebox{0.05\baselineskip}{\includegraphics[height=0.15cm,keepaspectratio]{figures/4_subscales_d3.pdf}}} on 6 visualizations of different readability: (\stimA $/$ \stimB $/$ \stimC) $>$ (\stimD $/$ \stimE $/$ \stimF). For a given subscale, ratings and 95\% CI do not overlap and fit the readability ranking.}
    \label{app:fig:Exp_ratings_PREVis}
\end{figure*}

In this section we present average ratings of stimuli with 95\% CI from our exploratory survey:
\begin{itemize}
    \item aggregate ratings from each of our final \PREVisColors subscales in \autoref{app:fig:Exp_ratings_PREVis};
    \item for each subscale and each stimulus, the subscale's average rating with the subscale's aggregate score; and
    \item individual item averages across all stimuli for all items in the original 4-factors structure identified in \autoref{app:sec:EFA_CFA_4factors}---including items that we did not retain for the final PREVis instrument.
\end{itemize}

\clearpage

\begin{figure} []
    \centering
    \includegraphics[width=\linewidth]{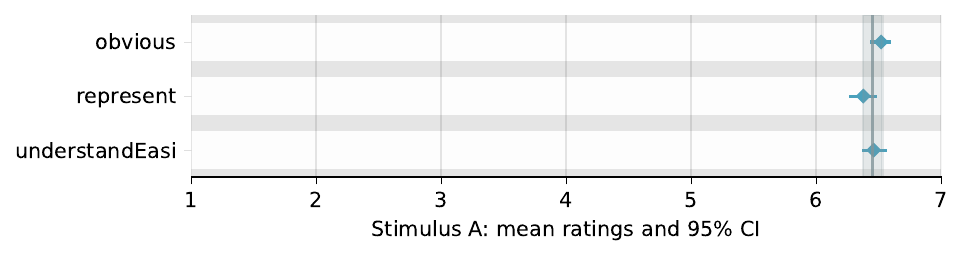}
    \caption{Comparison of average ratings from \SUn items, compared to the subscale's average values and 95\% CI (vertical line and rectangle) for stimulus \stimA.}
    \label{app:fig:Exp_ratings_SUn_A}
\end{figure}

\begin{figure} []
    \centering
    \includegraphics[width=\linewidth]{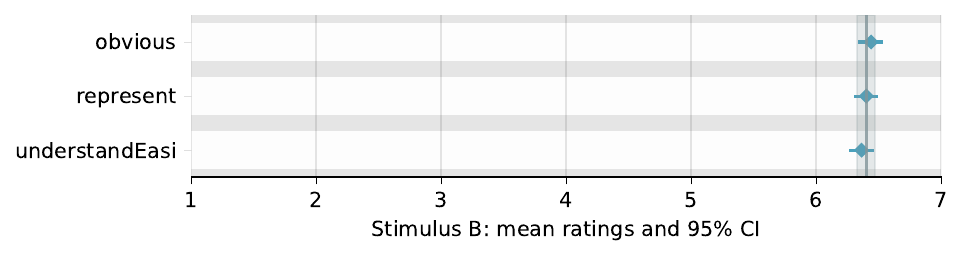}
    \caption{Comparison of average ratings from \SUn items, compared to the subscale's average values and 95\% CI (vertical line and rectangle) for stimulus \stimB.}
    \label{app:fig:Exp_ratings_SUn_B}
\end{figure}

\begin{figure} []
    \centering
    \includegraphics[width=\linewidth]{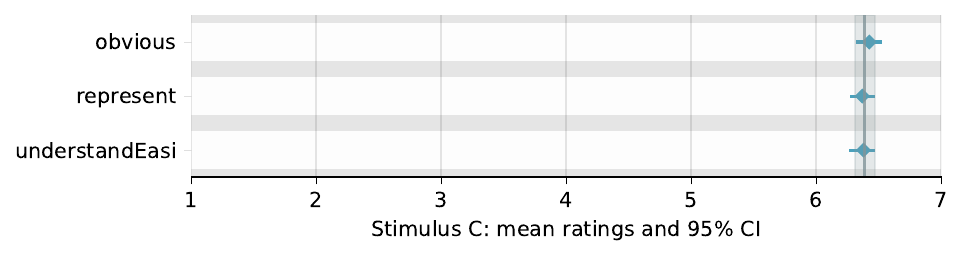}
    \caption{Comparison of average ratings from \SUn items, compared to the subscale's average values and 95\% CI (vertical line and rectangle) for stimulus \stimC.}
    \label{app:fig:Exp_ratings_SUn_C}
\end{figure}

\begin{figure} []
    \centering
    \includegraphics[width=\linewidth]{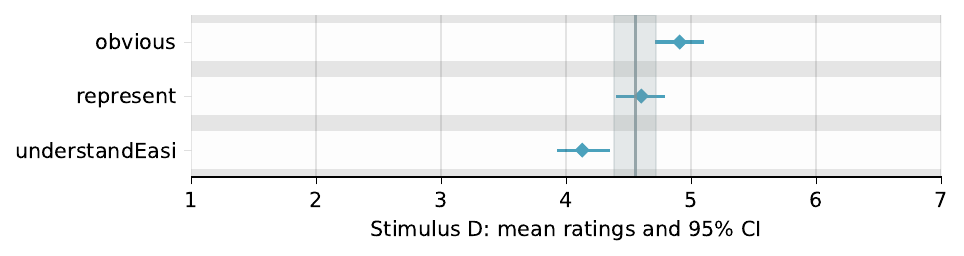}
    \caption{Comparison of average ratings from \SUn items, compared to the subscale's average values and 95\% CI (vertical line and rectangle) for stimulus \stimD.}
    \label{app:fig:Exp_ratings_SUn_D}
\end{figure}

\begin{figure} []
    \centering
    \includegraphics[width=\linewidth]{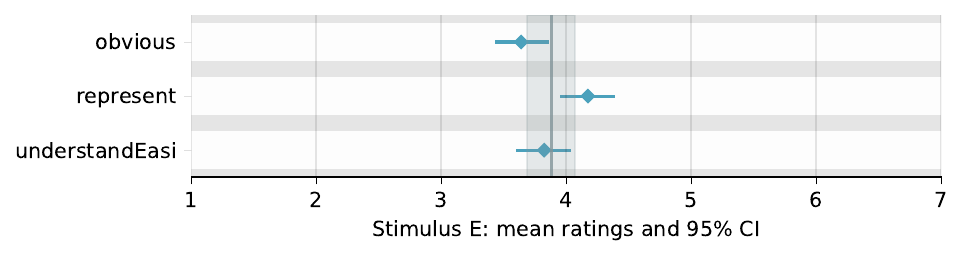}
    \caption{Comparison of average ratings from \SUn items, compared to the subscale's average values and 95\% CI (vertical line and rectangle) for stimulus \stimE.}
    \label{app:fig:Exp_ratings_SUn_E}
\end{figure}

\begin{figure} []
    \centering
    \includegraphics[width=\linewidth]{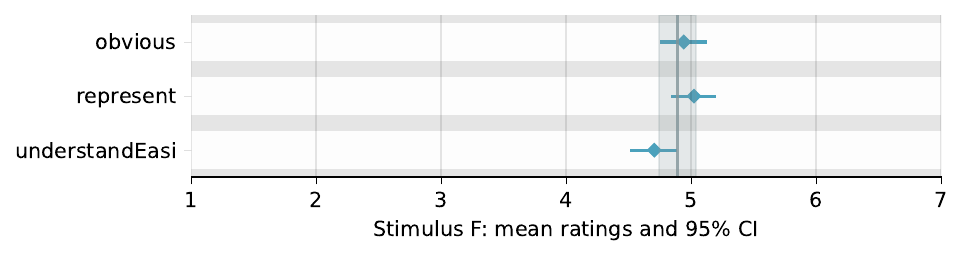}
    \caption{Comparison of average ratings from \SUn items, compared to the subscale's average values and 95\% CI (vertical line and rectangle) for stimulus \stimF.}
    \label{app:fig:Exp_ratings_SUn_F}
\end{figure}

\begin{figure} []
    \centering
    \includegraphics[width=\linewidth]{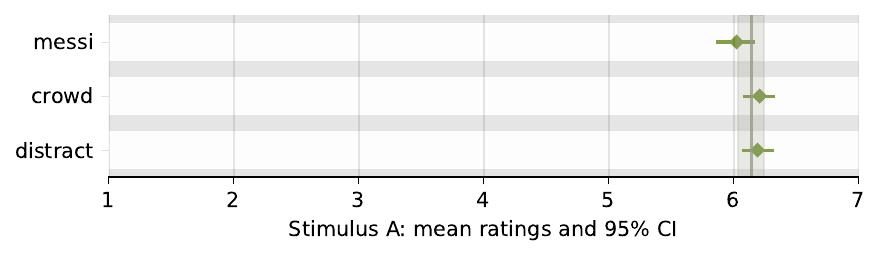}
    \caption{Comparison of average ratings from \SLa items, compared to the subscale's average values and 95\% CI (vertical line and rectangle) for stimulus \stimA.}
    \label{app:fig:Exp_ratings_SLa_A}
\end{figure}

\begin{figure} []
    \centering
    \includegraphics[width=\linewidth]{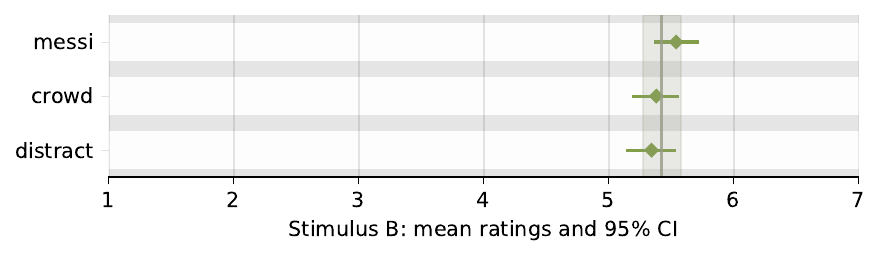}
    \caption{Comparison of average ratings from \SLa items, compared to the subscale's average values and 95\% CI (vertical line and rectangle) for stimulus \stimB.}
    \label{app:fig:Exp_ratings_SLa_B}
\end{figure}

\begin{figure} []
    \centering
    \includegraphics[width=\linewidth]{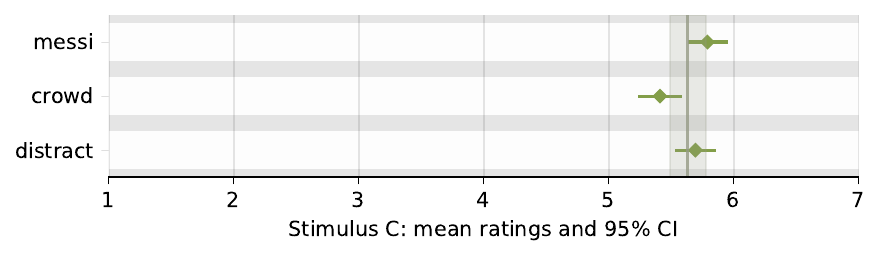}
    \caption{Comparison of average ratings from \SLa items, compared to the subscale's average values and 95\% CI (vertical line and rectangle) for stimulus \stimC.}
    \label{app:fig:Exp_ratings_SLa_C}
\end{figure}

\begin{figure} []
    \centering
    \includegraphics[width=\linewidth]{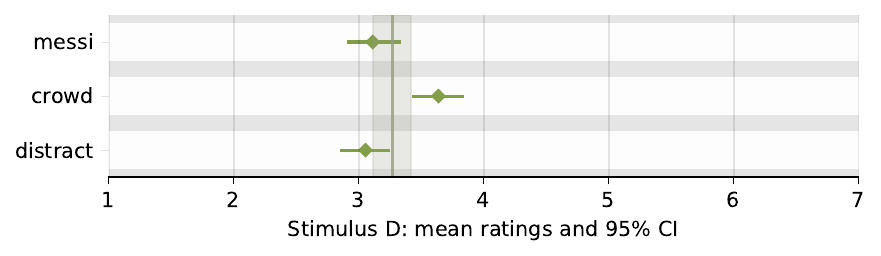}
    \caption{Comparison of average ratings from \SLa items, compared to the subscale's average values and 95\% CI (vertical line and rectangle) for stimulus \stimD.}
    \label{app:fig:Exp_ratings_SLa_D}
\end{figure}

\begin{figure} []
    \centering
    \includegraphics[width=\linewidth]{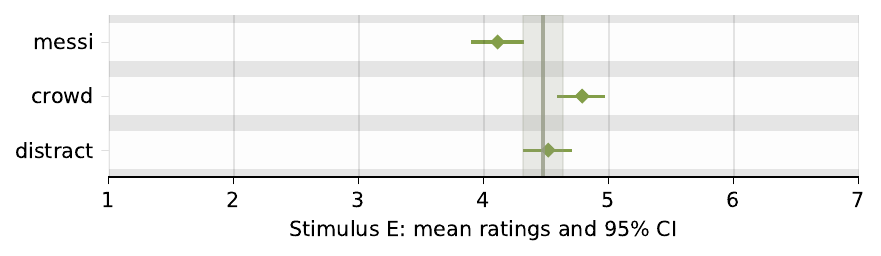}
    \caption{Comparison of average ratings from \SLa items, compared to the subscale's average values and 95\% CI (vertical line and rectangle) for stimulus \stimE.}
    \label{app:fig:Exp_ratings_SLa_E}
\end{figure}

\begin{figure} []
    \centering
    \includegraphics[width=\linewidth]{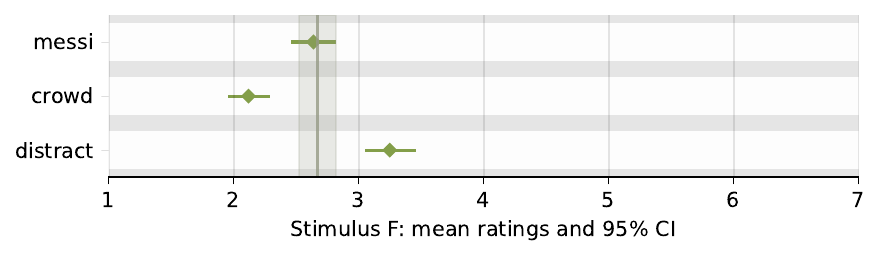}
    \caption{Comparison of average ratings from \SLa items, compared to the subscale's average values and 95\% CI (vertical line and rectangle) for stimulus \stimF.}
    \label{app:fig:Exp_ratings_SLa_F}
\end{figure}

\begin{figure} []
    \centering
    \includegraphics[width=\linewidth]{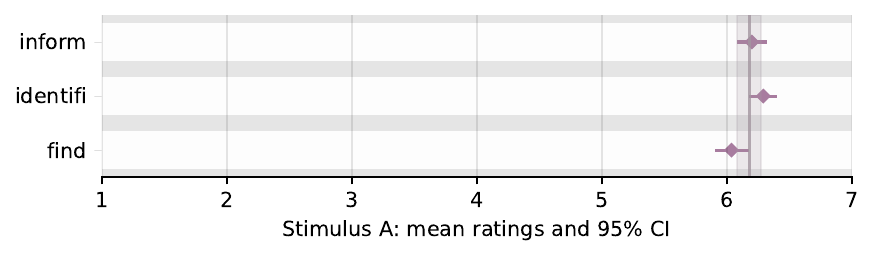}
    \caption{Comparison of average ratings from \SDR items, compared to the subscale's average values and 95\% CI (vertical line and rectangle) for stimulus \stimA.}
    \label{app:fig:Exp_ratings_SDR_A}
    \vspace{-0.5cm}
\end{figure}

\begin{figure} []
    \centering
    \includegraphics[width=\linewidth]{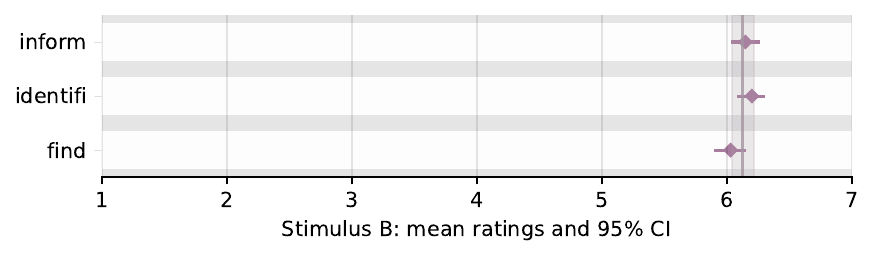}
    \caption{Comparison of average ratings from \SDR items, compared to the subscale's average values and 95\% CI (vertical line and rectangle) for stimulus \stimB.}
    \label{app:fig:Exp_ratings_SDR_B}
    \vspace{-0.5cm}
\end{figure}

\begin{figure} []
    \centering
    \includegraphics[width=\linewidth]{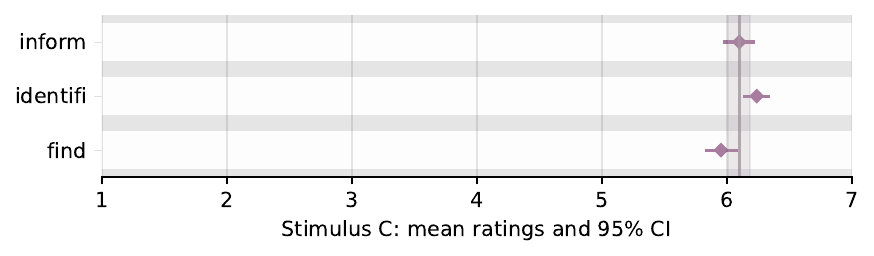}
    \caption{Comparison of average ratings from \SDR items, compared to the subscale's average values and 95\% CI (vertical line and rectangle) for stimulus \stimC.}
    \label{app:fig:Exp_ratings_SDR_C}
    \vspace{-0.5cm}
\end{figure}

\begin{figure} []
    \centering
    \includegraphics[width=\linewidth]{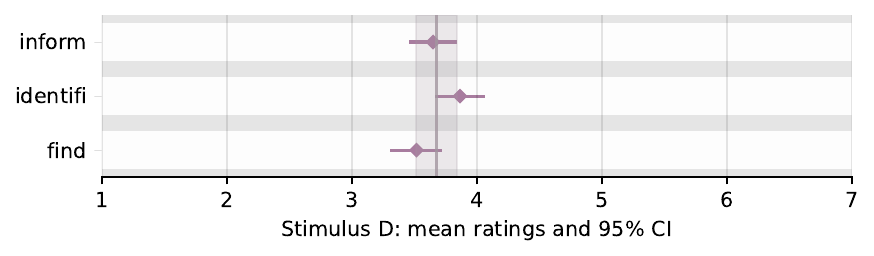}
    \caption{Comparison of average ratings from \SDR items, compared to the subscale's average values and 95\% CI (vertical line and rectangle) for stimulus \stimD.}
    \label{app:fig:Exp_ratings_SDR_D}
    \vspace{-0.5cm}
\end{figure}

\begin{figure} []
    \centering
    \includegraphics[width=\linewidth]{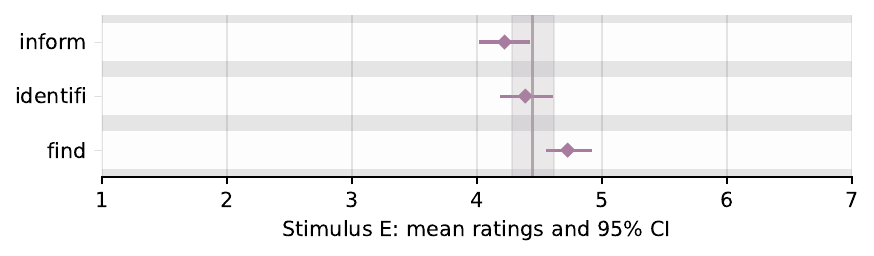}
    \caption{Comparison of average ratings from \SDR items, compared to the subscale's average values and 95\% CI (vertical line and rectangle) for stimulus \stimE.}
    \label{app:fig:Exp_ratings_SDR_E}
    \vspace{-0.5cm}
\end{figure}

\begin{figure} []
    \centering
    \includegraphics[width=\linewidth]{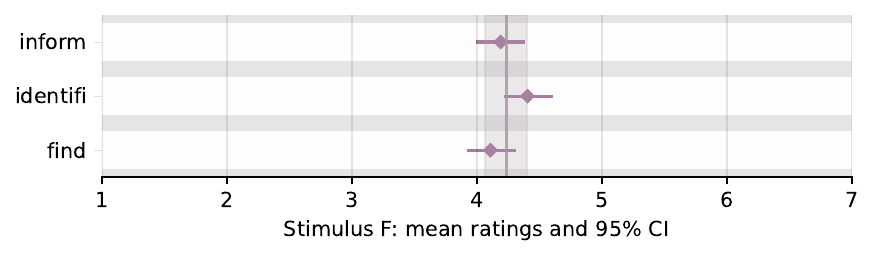}
    \caption{Comparison of average ratings from \SDR items, compared to the subscale's average values and 95\% CI (vertical line and rectangle) for stimulus \stimF.}
    \label{app:fig:Exp_ratings_SDR_F}
    \vspace{-0.5cm}
\end{figure}

\begin{figure} []
    \centering
    \includegraphics[width=\linewidth]{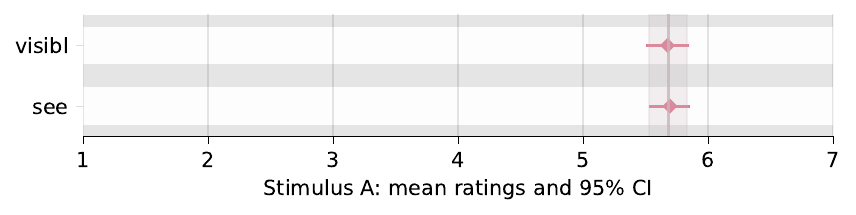}
    \caption{\protect\revis{Comparison of average ratings from \SDF items, compared to the subscale's average values and 95\% CI (vertical line and rectangle) for stimulus \stimA.}}
    \label{app:fig:Exp_ratings_SDF_A}
\end{figure}

\begin{figure} []
    \centering
    \includegraphics[width=\linewidth]{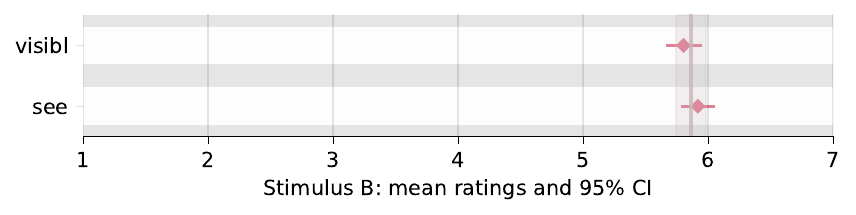}
    \caption{\protect\revis{Comparison of average ratings from \SDF items, compared to the subscale's average values and 95\% CI (vertical line and rectangle) for stimulus \stimB.}}
    \label{app:fig:Exp_ratings_SDF_B}
\end{figure}

\begin{figure} []
    \centering
    \includegraphics[width=\linewidth]{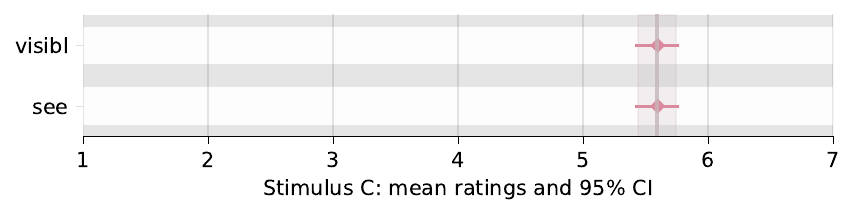}
    \caption{\protect\revis{Comparison of average ratings from \SDF items, compared to the subscale's average values and 95\% CI (vertical line and rectangle) for stimulus \stimC.}}
    \label{app:fig:Exp_ratings_SDF_C}
\end{figure}

\begin{figure} []
    \centering
    \includegraphics[width=\linewidth]{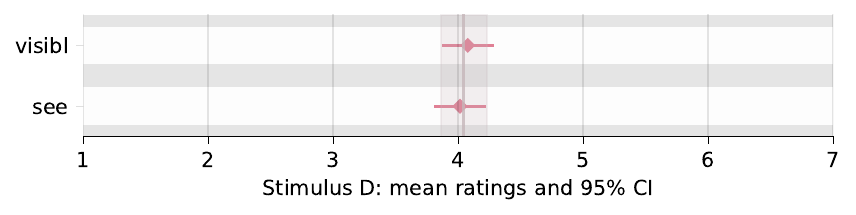}
    \caption{\protect\revis{Comparison of average ratings from \SDF items, compared to the subscale's average values and 95\% CI (vertical line and rectangle) for stimulus \stimD.}}
    \label{app:fig:Exp_ratings_SDF_D}
\end{figure}

\begin{figure} []
    \centering
    \includegraphics[width=\linewidth]{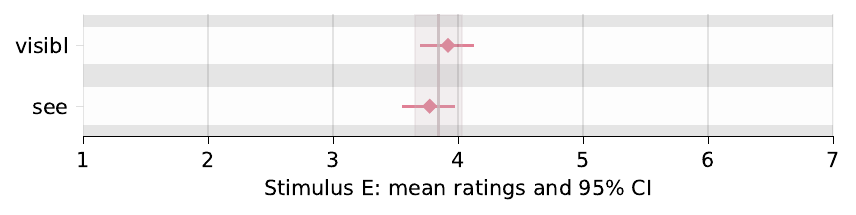}
    \caption{\protect\revis{Comparison of average ratings from \SDF items, compared to the subscale's average values and 95\% CI (vertical line and rectangle) for stimulus \stimE.}}
    \label{app:fig:Exp_ratings_SDF_E}
\end{figure}

\begin{figure} []
    \centering
    \includegraphics[width=\linewidth]{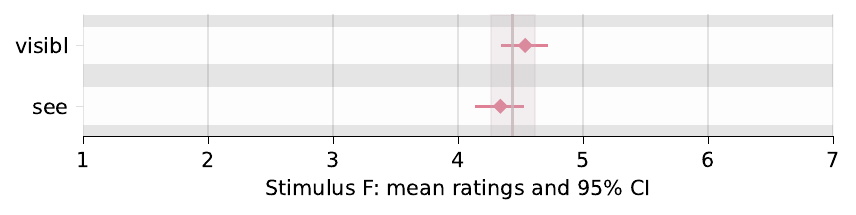}
    \caption{\protect\revis{Comparison of average ratings from \SDF items, compared to the subscale's average values and 95\% CI (vertical line and rectangle) for stimulus \stimF.}}
    \label{app:fig:Exp_ratings_SDF_F}
\end{figure}

\begin{figure} []
    \centering
    \includegraphics[width=\linewidth]{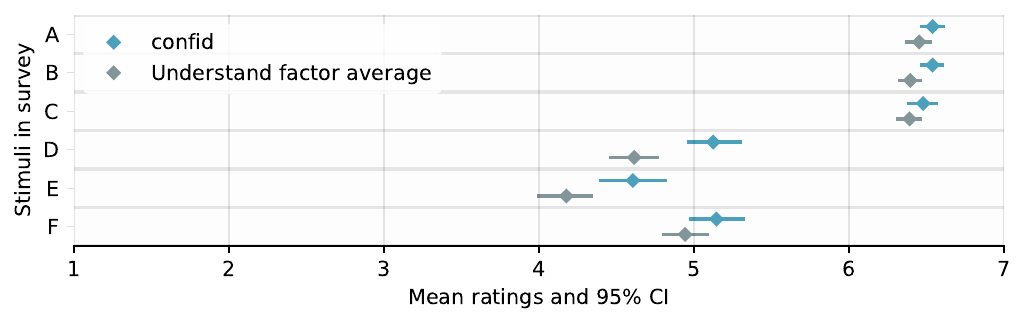}
    \caption{Comparison of ratings from the \emph{\textbf{confid}} item and average ratings from all items in the \emph{Understand} factor, across 6 stimuli in the exploratory survey the.}
    \label{app:fig:Full_factors_ratings-Understand_confid}
\end{figure}

\begin{figure} []
    \centering
    \includegraphics[width=\linewidth]{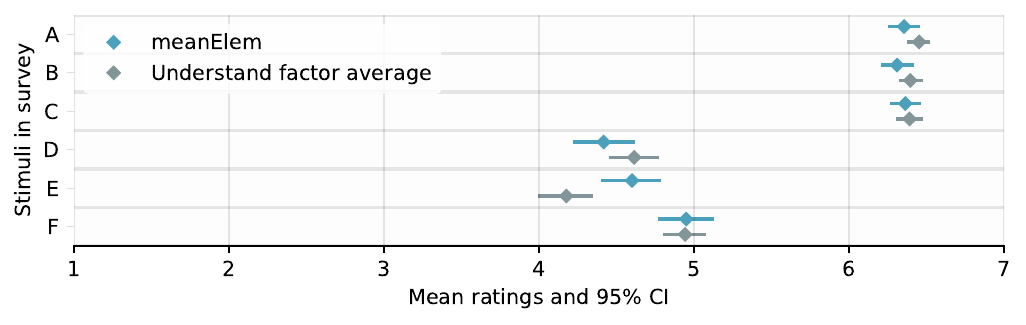}
    \caption{Comparison of ratings from the \emph{\textbf{meanElem}} item and average ratings from all items in the \emph{Understand} factor, across 6 stimuli in the exploratory survey the.}
    \label{app:fig:Full_factors_ratings-Understand_meanElem}
\end{figure}

\begin{figure} []
    \centering
    \includegraphics[width=\linewidth]{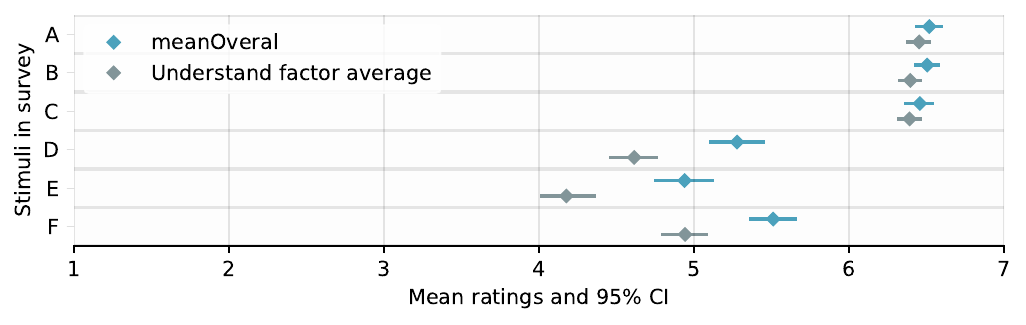}
    \caption{Comparison of ratings from the \emph{\textbf{meanOveral}} item and average ratings from all items in the \emph{Understand} factor, across 6 stimuli in the exploratory survey the.}
    \label{app:fig:Full_factors_ratings-Understand_meanOveral}
\end{figure}

\begin{figure} []
    \centering
    \includegraphics[width=\linewidth]{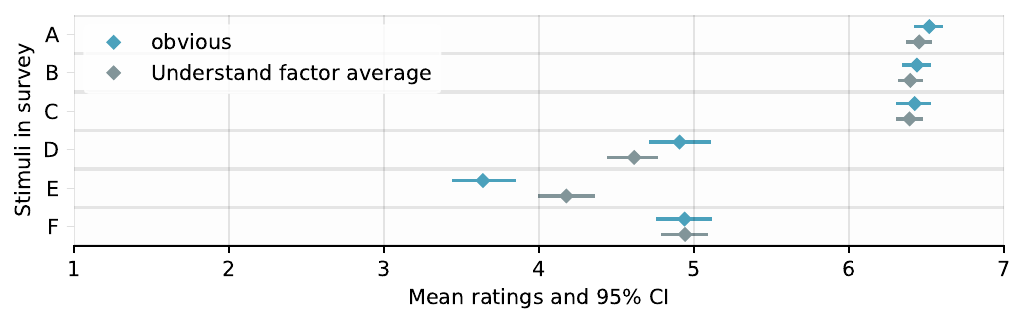}
    \caption{Comparison of ratings from the \emph{\textbf{obvious}} item and average ratings from all items in the \emph{Understand} factor, across 6 stimuli in the exploratory survey the.}
    \label{app:fig:Full_factors_ratings-Understand_obvious}
\end{figure}

\begin{figure} []
    \centering
    \includegraphics[width=\linewidth]{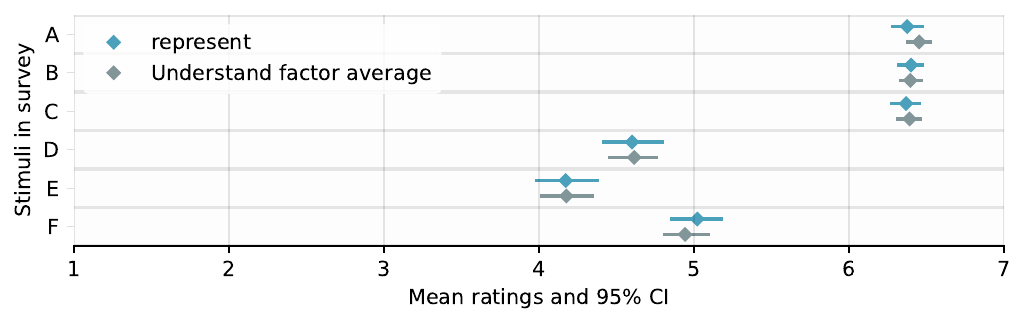}
    \caption{Comparison of ratings from the \emph{\textbf{represent}} item and average ratings from all items in the \emph{Understand} factor, across 6 stimuli in the exploratory survey the.}
    \label{app:fig:Full_factors_ratings-Understand_represent}
\end{figure}

\begin{figure} []
    \centering
    \includegraphics[width=\linewidth]{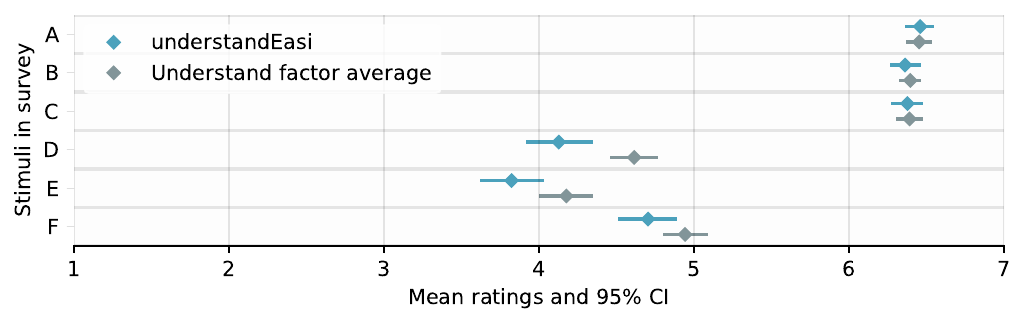}
    \caption{Comparison of ratings from the \emph{\textbf{understandEasi}} item and average ratings from all items in the \emph{Understand} factor, across 6 stimuli in the exploratory survey the.}
    \label{app:fig:Full_factors_ratings-Understand_understandEasi}
\end{figure}

\begin{figure} []
    \centering
    \includegraphics[width=\linewidth]{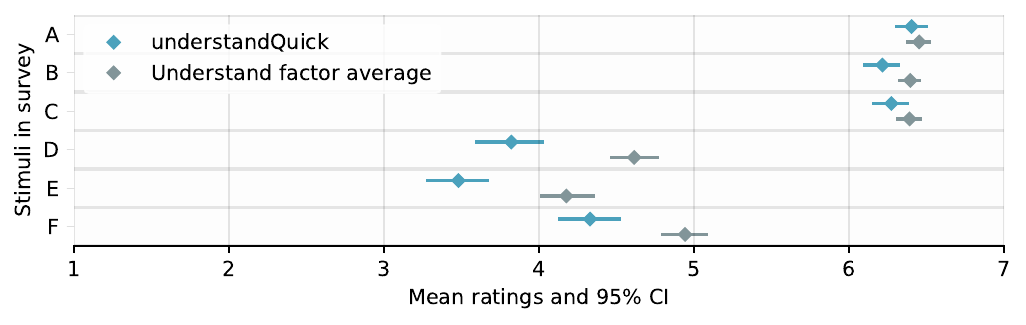}
    \caption{Comparison of ratings from the \emph{\textbf{understandQuick}} item and average ratings from all items in the \emph{Understand} factor, across 6 stimuli in the exploratory survey.}
    \label{app:fig:Full_factors_ratings-Understand_understandQuick}
\end{figure}

\begin{figure} []
    \centering
    \includegraphics[width=\linewidth]{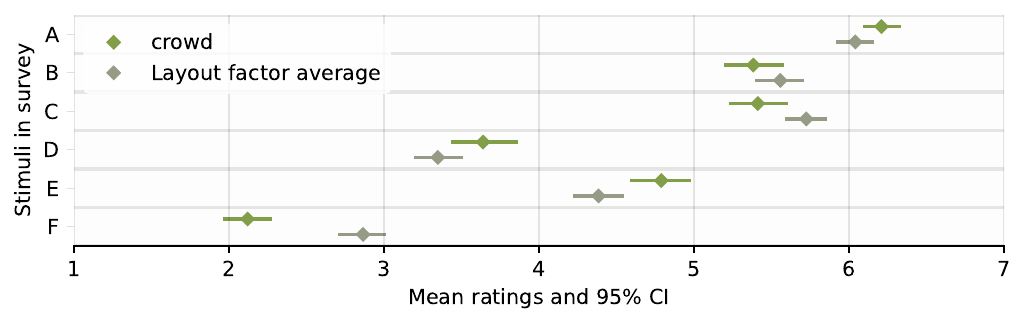}
    \caption{Comparison of ratings from the \emph{\textbf{crowd}} item and average ratings from all items in the \emph{Layout} factor, across 6 stimuli in the exploratory survey the.}
    \label{app:fig:Full_factors_ratings-Layout_crowd}
\end{figure}

\begin{figure} []
    \centering
    \includegraphics[width=\linewidth]{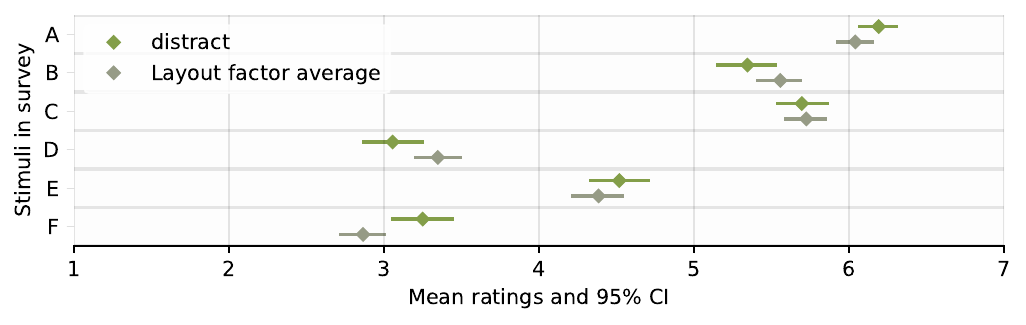}
    \caption{Comparison of ratings from the \emph{\textbf{distract}} item and average ratings from all items in the \emph{Layout} factor, across 6 stimuli in the exploratory survey the.}
    \label{app:fig:Full_factors_ratings-Layout_distract}
\end{figure}

\begin{figure} []
    \centering
    \includegraphics[width=\linewidth]{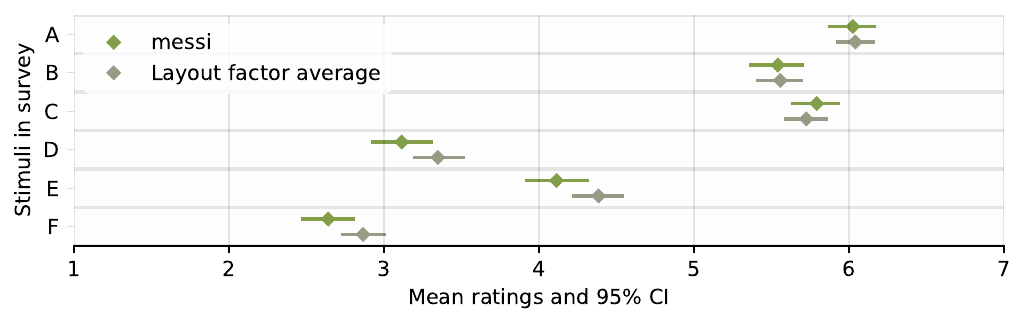}
    \caption{Comparison of ratings from the \emph{messi} item and average ratings from all items in the \emph{Layout} factor, across 6 stimuli in the exploratory survey the.}
    \label{app:fig:Full_factors_ratings-Layout_messi}
\end{figure}

\begin{figure} []
    \centering
    \includegraphics[width=\linewidth]{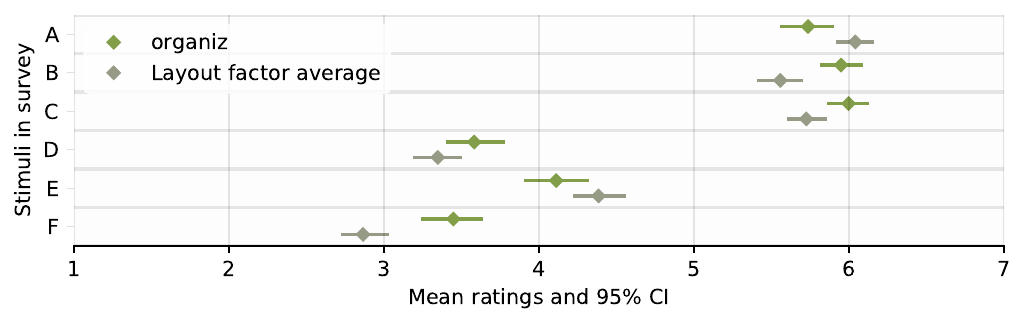}
    \caption{Comparison of ratings from the \emph{\textbf{organiz}} item and average ratings from all items in the \emph{Layout} factor, across 6 stimuli in the exploratory survey the.}
    \label{app:fig:Full_factors_ratings-Layout_organiz}
\end{figure}

\begin{figure} []
    \centering
    \includegraphics[width=\linewidth]{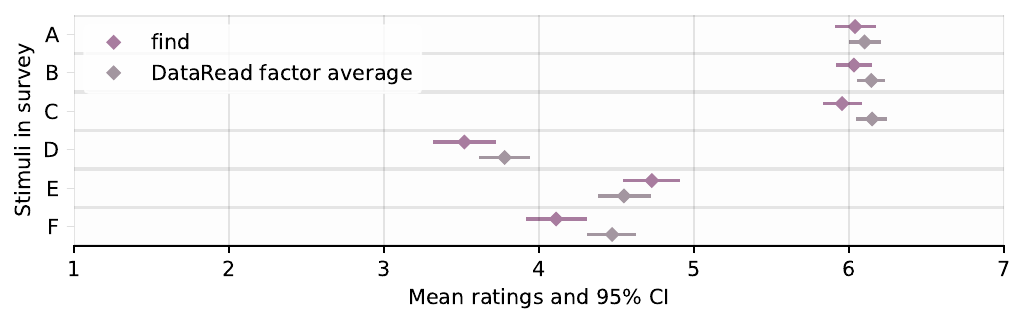}
    \caption{Comparison of ratings from the \emph{\textbf{find}} item and average ratings from all items in the \emph{DataRead} factor, across 6 stimuli in the exploratory survey the.}
    \label{app:fig:Full_factors_ratings-DataRead_find}
\end{figure}

\begin{figure} []
    \centering
    \includegraphics[width=\linewidth]{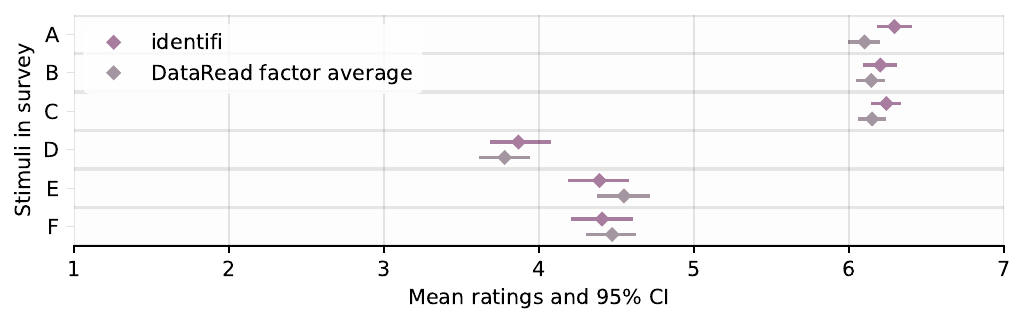}
    \caption{Comparison of ratings from the \emph{\textbf{identifi}} item and average ratings from all items in the \emph{DataRead} factor, across 6 stimuli in the exploratory survey the.}
    \label{app:fig:Full_factors_ratings-DataRead_identifi}
\end{figure}

\begin{figure} []
    \centering
    \includegraphics[width=\linewidth]{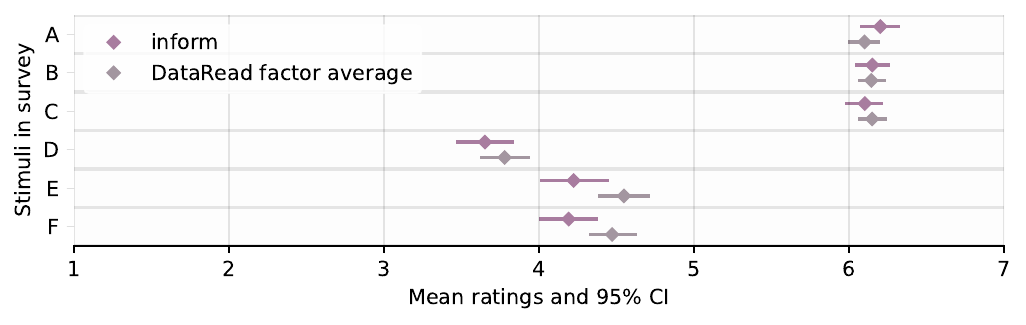}
    \caption{Comparison of ratings from the \emph{\textbf{inform}} item and average ratings from all items in the \emph{DataRead} factor, across 6 stimuli in the exploratory survey the.}
    \label{app:fig:Full_factors_ratings-DataRead_inform}
    \vspace{4cm}
\end{figure}

\begin{figure} []
    \centering
    \includegraphics[width=\linewidth]{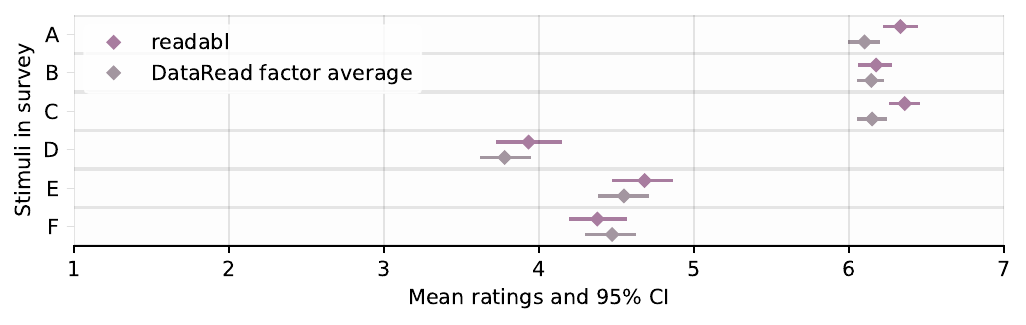}
    \caption{Comparison of ratings from the \emph{\textbf{readabl}} item and average ratings from all items in the \emph{DataRead} factor, across 6 stimuli in the exploratory survey the.}
    \label{app:fig:Full_factors_ratings-DataRead_readabl}
\end{figure}

\begin{figure} []
    \centering
    \includegraphics[width=\linewidth]{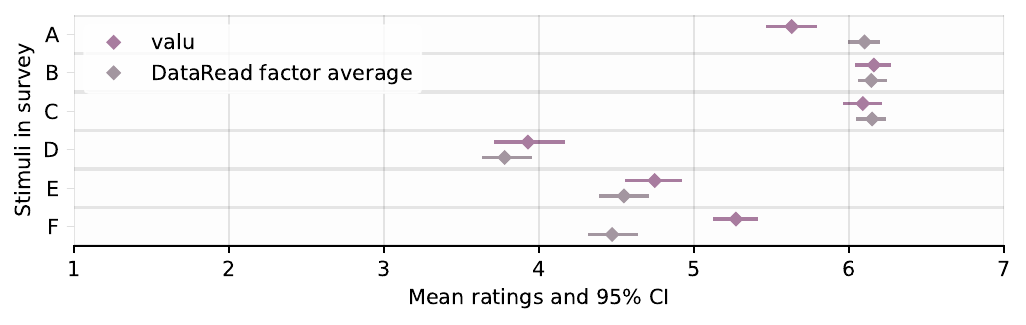}
    \caption{Comparison of ratings from the \emph{\textbf{valu}} item and average ratings from all items in the \emph{DataRead} factor, across 6 stimuli in the exploratory survey the.}
    \label{app:fig:Full_factors_ratings-DataRead_valu}
\end{figure}

\begin{figure} []
    \centering
    \includegraphics[width=\linewidth]{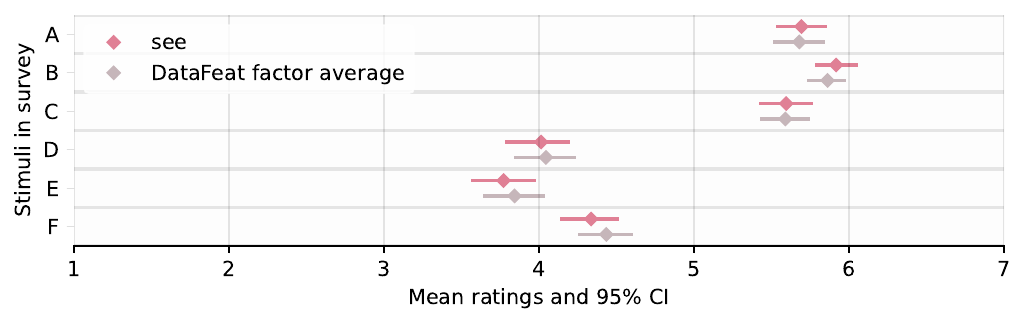}
    \caption{Comparison of ratings from the \emph{\textbf{see}} item and average ratings from all items in the \emph{DataFeat} factor, across 6 stimuli in the exploratory survey the.}
    \label{app:fig:Full_factors_ratings-DataFeat_see}
\end{figure}

\begin{figure} []
    \centering
    \includegraphics[width=\linewidth]{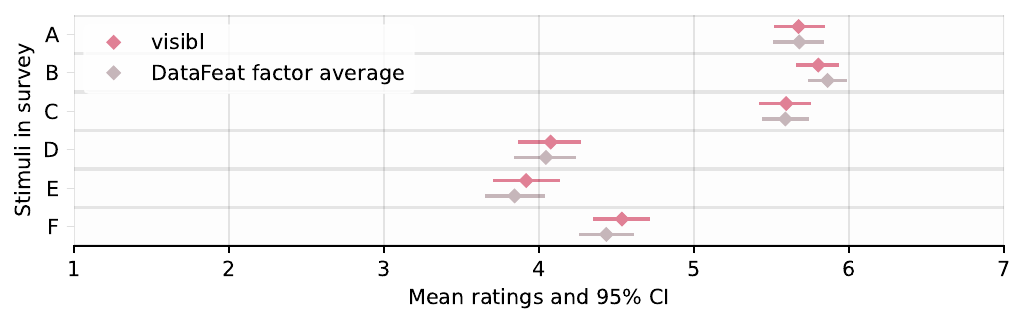}
    \caption{Comparison of ratings from the \emph{\textbf{visibl}} item and average ratings from all items in the \emph{DataFeat} factor, across 6 stimuli in the exploratory survey the.}
    \label{app:fig:Full_factors_ratings-DataFeat_visibl}
    \vspace{4cm}
\end{figure}

\clearpage
\section{Validation survey design}
\label{app:sec:validation_survey_design}

As we stated in \autoref{subsec:validation_study_design}, our validation study design was mainly based on the exploratory study described in \autoref{subsec:exploratory_study_design}, but we made some adjustments. Here, we detail the reason for these changes.

\textbf{All participants rated all stimuli.} This choice was motivated by the fact that future researchers might be interested in running studies where participants rate multiple stimuli. As we had an independent-groups design for our exploratory survey, we decided to run a within-participants study.

\begin{figure}
    \centering
    \includegraphics[width=1\linewidth]{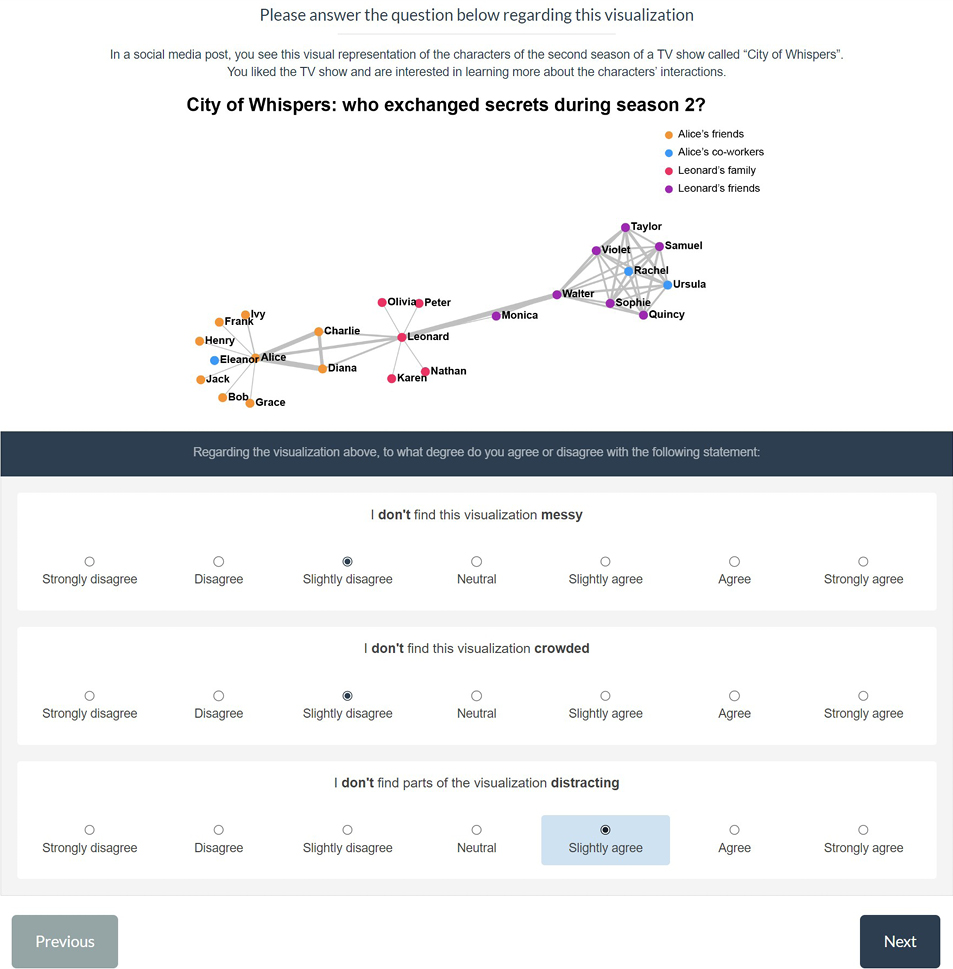}
    \caption{Example screenshot from our validation survey in action: rating stimulus node-link visualization \stimVB with \SUn items.}
    \label{app:fig:validation_survey_screenshot}
\end{figure}

\textbf{Rating items for each subscale were presented together on a single screen.} We expected our study to exceed the 10 minutes target, after which crowd-sourced work can induce fatigue in participants; therefore, we were seeking ways to facilitate participant's task as much as possible. Showing groups of ratings instead of a single item per screen was a solution already successfully implemented in our team's previous work \cite{He:2022:Beauvis} and showed good time performance.  \autoref{app:fig:validation_survey_screenshot} shows an example screenshot from our survey design.
However, with this style of questions in the LimeSurvey platform we used, we lost the possibility to provide an ``I don't know / Not applicable'' answer option with an optional open text field (as we had done in the exploratory survey). Instead, we added a separate, optional comment question for participants who wanted to qualify their answers about each visualization they rated.

\textbf{We used 3 stimuli of a single type (node-link).} As explained in \autoref{subsec:validation_study_design}, in order to produce our Multi-Trait Multi-Method matrix (\autoref{fig:MTMM}), we decided to compare \scalename \PREVisColors scores with graph aesthetics metrics from Gove's work \cite{gove_2018_ItPays} to assess convergent validity of our final instrument. As a result, we only used node-link diagrams. We also wanted to verify that average \PREVisColors scores with 95\% CI allowed us to distinguish between different levels of graph layout metrics (\ie, testing construct validity with the differentiation by the ``known groups'' criterion \cite{Boateng:2018:BestPractices}). To that end, we wanted to have 3 different groups of measures for each participant, so we produced 3 different visualizations with different levels of graph layout metrics.

\subsection{Stimuli visualizations generation and metrics}
\begin{figure}
    \centering
    \includegraphics[width=1\linewidth]{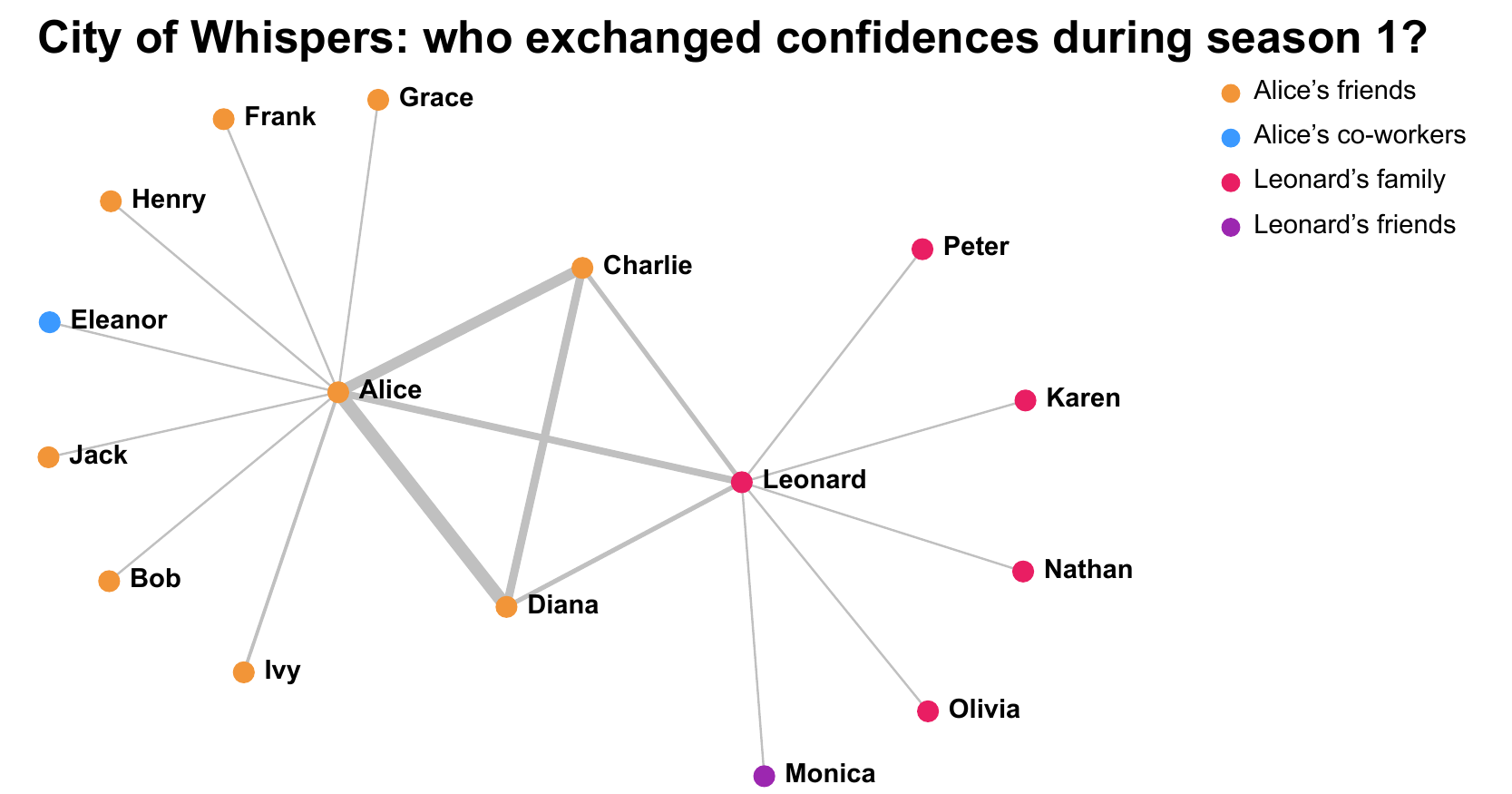}
    \caption{Stimulus \stimVA in our validation survey.}
    \label{fig:valid_stimulus_A}
\end{figure}

\begin{figure}
    \centering
    \includegraphics[width=1\linewidth]{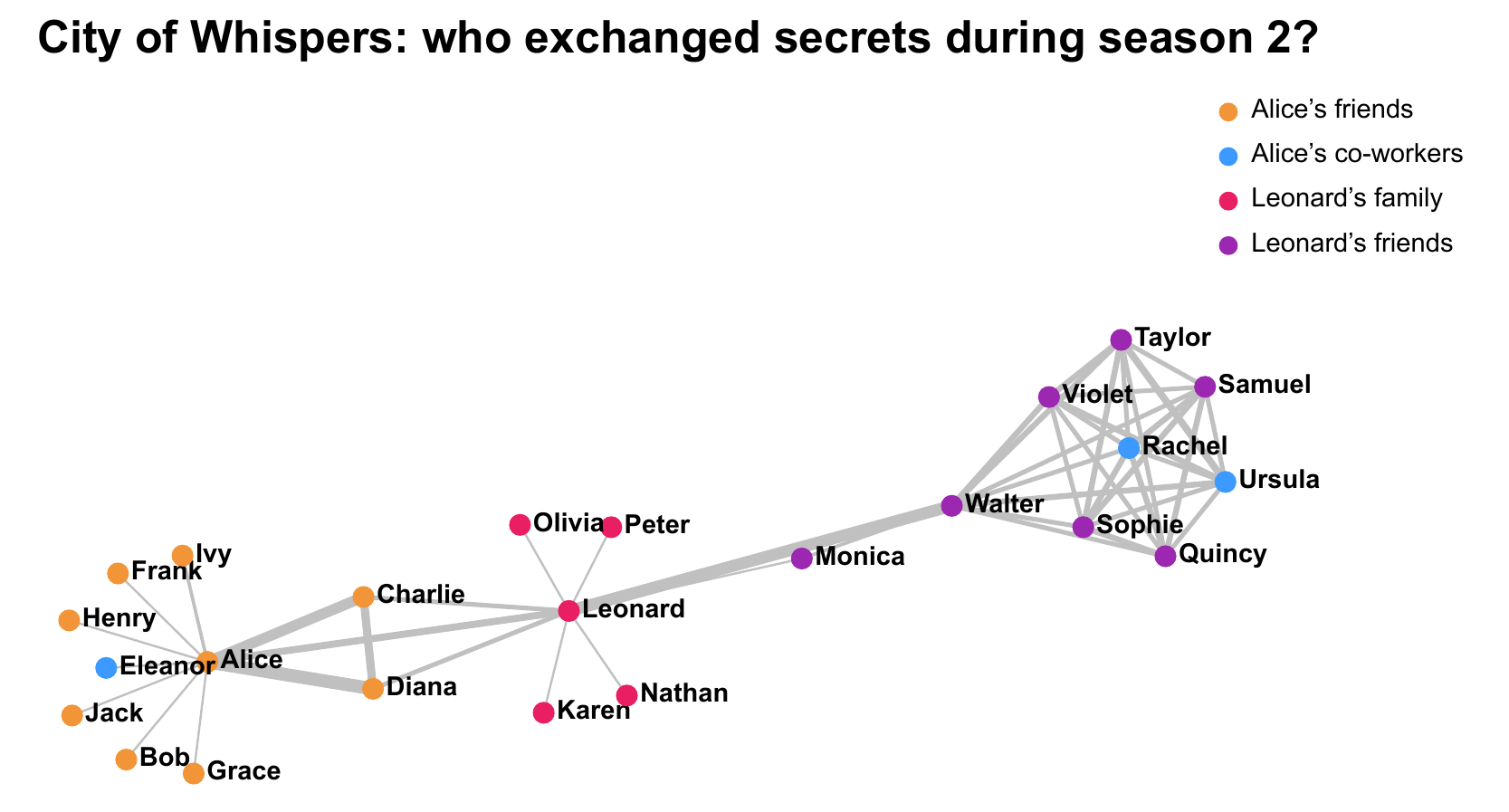}
    \caption{Stimulus \stimVB in our validation survey.}
    \label{fig:valid_stimulus_B}
\end{figure}

\begin{figure}
    \centering
    \includegraphics[width=1\linewidth]{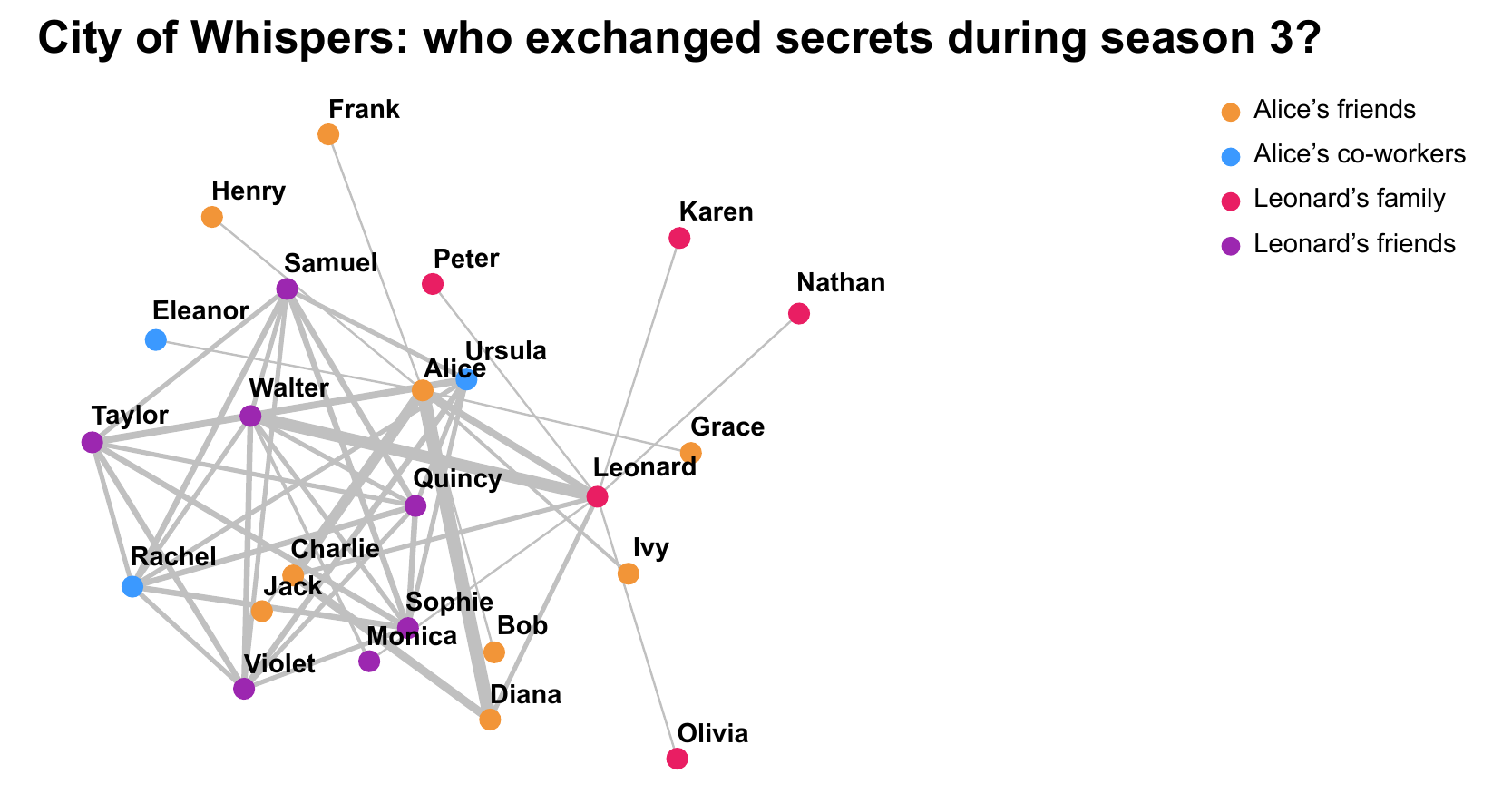}
    \caption{Stimulus \stimVC in our validation survey.}
    \label{fig:valid_stimulus_C}
\end{figure}

We created node-link stimuli based on a transformed version of the ``Les Misérables'' dataset. This network was first created by \href{https://people.sc.fsu.edu/~jburkardt/datasets/sgb/sgb.html}{Donald Knuth} as part of the \href{https://www-cs-faculty.stanford.edu/~knuth/sgb.html}{Stanford Graph Base}. It contains 77 nodes corresponding to characters of the novel, and 254 vertices connecting two characters whenever they appear in the same chapter. We retrieved the dataset from Mike Bostock's \href{https://gist.github.com/mbostock/4062045}{.block} repository. We then replaced all characters' names by new, fictional first names of an imaginary TV show we called ``City of whispers''. Finally, we generated different visualizations using a simple force-directed graph generation script using the \texttt{D3.js} library (v6). We provide our code as \suppmat{supplemental material}{s5w62}.

\subsubsection{Graph layout metrics}
\label{app:subsec:greadability}
As our goal was to compare the scale’s ratings with graph aesthetics metrics, we implemented the \texttt{Greadability.js} library from R. Gove \cite{gove_2018_ItPays} (\texttt{\href{https://github.com/rpgove/greadability}{github\discretionary{}{.}{.}com\discretionary{/}{}{/}rpgove\discretionary{/}{}{/}greadability}}) within our code to calculate 4 layout metrics, which are described on the \href{https://github.com/rpgove/greadability/blob/master/README.md}{Greadability repository} as follow:
\begin{itemize}
    \item \textbf{Edge crossings:} ``measures the fraction of edges that cross (intersect) out of an approximate maximum number that can cross.''
    \item \textbf{Edge crossing angle:} ``measures the mean deviation of edge crossing angles from the ideal edge crossing angle (70 degrees).''
    \item \textbf{Angular resolution (minimum):} ``measures the mean deviation of adjacent incident edge angles from the ideal minimum angles (360 degrees divided by the degree of that node).''
    \item \textbf{Angular resolution (deviation):} ``measures the average deviation of angles between incident edges on each vertex.''
\end{itemize}

To influence the layout, we use the same dataset but we vary three parameters: the number of plotted nodes and two ``force'' attributes in the \texttt{D3 force-directed} graph component. In particular, for the number of nodes we affected the following values: N $=$ 16 nodes in df, and N $=$ 24 nodes in \stimVB and \stimVC. For further details we refer the reader to our \researchLog or to our \suppmat{stimuli generation code folder}{s5w62}.


For \stimVA and \stimVB, we dragged elements to improve the clarity of the layout. We left \stimVC as it was.
We then retrieved the calculated values from the \texttt{Greadability.js} library (which gets calculated again after each manipulation of the layout through user interaction and output in the console) and we save the generated SVG files, which we provide in our \suppmat{supplemental material folder}{9cg8j}.

We obtained the metrics described in \autoref{app:tab:greadability}. We noticed that all metrics corresponded to our presumed readability ranking of \stimVA$>$\stimVB$>$\stimVC except for the crossing angle where \stimVB$>$\stimVA. This metric is described in Gove's work \cite{gove_2018_ItPays} as a score on the average angle value of crossing between edges for non-adjacent nodes. The underlying assumption is that a 70° crossing is the ideal crossing angle for readability. However, stimulus \stimVA (\autoref{fig:valid_stimulus_A}) has only one such edge crossing, with an approximate value of 90°, hence its score is lower than \stimVB.

\begin{table}[h!]
    \centering
    \fontsize{7pt}{7pt}\selectfont
    \caption{Metrics obtained with the Greadability library for each node-link  stimulus in our validation survey.}
    \label{app:tab:greadability}
    \tabulinesep=0.8mm
    \begin{tabu} to \linewidth {X[0.1]  X[0.7]  X[0.7]  X[1]  X[1]  X[1]  }
        \toprule
         & Crossing & Crossing \mbox{angle} & Angular \mbox{resolution} min & Angular \mbox{resolution} dev & Greadability \mbox{average}\\ \midrule
        \stimVA & 0.973 & 0.715 & 0.893 & 0.908 & 0.872\\ 
        \stimVB & 0.879 & 0.846 & 0.641 & 0.776 & 0.786\\ 
        \stimVC & 0.643 & 0.784 & 0.575 & 0.713 & 0.679\\ 
        \bottomrule
    \end{tabu}
\end{table}

Finally, we modified the obtained SVG files in Adobe Illustrator to (1) modify nodes colors and add legends, (2) add titles, and (3) improve image fit in our survey screen---\ie, we rotated the full graph \textit{\textbf{without altering the node-link layout}}, ensuring continued validity of all graph layout metrics. We also provide the AI files in our \suppmat{supplemental material folder}{9cg8j}. \autoref{fig:valid_stimulus_A} to \autoref{fig:valid_stimulus_C} show the resulting images we used in our survey.

\subsubsection{Color impairment simulations}

\begin{figure}
    \centering
    \includegraphics[width=0.7\linewidth]{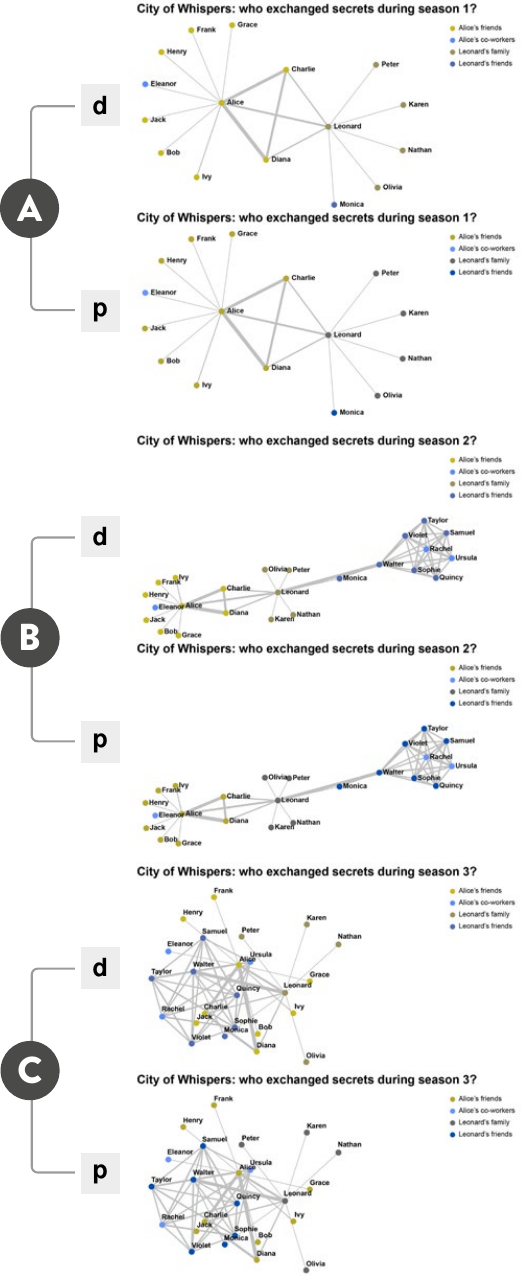}
    \caption{Deuteranipa (d) and protanopia (p) simulations in Adobe Illustrator for our validation survey stimuli.}
    \label{fig:valid_survey_stimuli_color_vision}
\end{figure}

Since we used color to encode groups of nodes in the visualizations, we chose color that would remain distinguishable for people with color vision deficiencies. We assessed the results in Adobe Illustrator using the View $>$ Proof Setup for deuteranopia and protanopia, as shown in \autoref{fig:valid_survey_stimuli_color_vision}.

\subsection{Reading tasks}
\label{app:subsec:validation_survey_tasks}
Topology tasks are specific to node-link networks and were not part of our original analysis for reading tasks in the exploratory survey described in \autoref{subsec:exploratory_study_design}. To decide on reading tasks for the validation survey, we referred to Ghoniem \etal's \cite{Ghoniem:2005:ReadabilityGraphs} work on assessing graph readability. We provide details on our selection of tasks in our \researchLog.

As a result, we selected two types of task, which we repeated for each stimulus visualization in our survey:
\begin{itemize}
    
    \item \textbf{Find the node with the maximum number of adjacent nodes.} \eg, ``\textit{Who is more popular?}''. Using Amar \etal's taxonomy \cite{Amar:2005:LowLevelTasks} as a base to describe low-level visual analytics components, this task could be called a \textit{Find Extremum Node} task. In our survey, we coded it as Find Extremum Node---\texttt{TaskFEN}.
    
    Because we used the same dataset for all stimuli, for each question we specified a subset of nodes in which the extremum had to be identified, which meant that the reader had to visually \textit{Filter} the specified nodes as a preliminary reading step, before performing a \textit{Determine range}, and finally what Amar \etal \cite{Amar:2005:LowLevelTasks} call a \textit{very low-level mathematical comparison task}.

    \item \textbf{Find the set of nodes adjacent to a node.} Instead of finding a set of nodes, we proposed a simplified version of this task, which could be better described as ``Test direct connection'', \eg ``\textit{Is A directly connected to B?}''.
    In Amar \etal's taxonomy \cite{Amar:2005:LowLevelTasks}, this would relates to a series of \textit{Find} tasks, for which a reader could adopt different strategies:
    
    \begin{itemize}
        \item \textit{Find} source Node $+$ \textit{Determine range} on Adjacent Nodes $\rightarrow$ \textit{Find} target Node among Adjacent Nodes, or
        \item \textit{Find} source Node $+$ \textit{Find} target Node $\rightarrow$ \textit{Correlate}.
    \end{itemize}
    
     In our survey, we coded this task as Find Adjacent Node---\texttt{TaskFAN}.
\end{itemize}

\begin{figure}
    \centering
    \includegraphics[width=\linewidth]{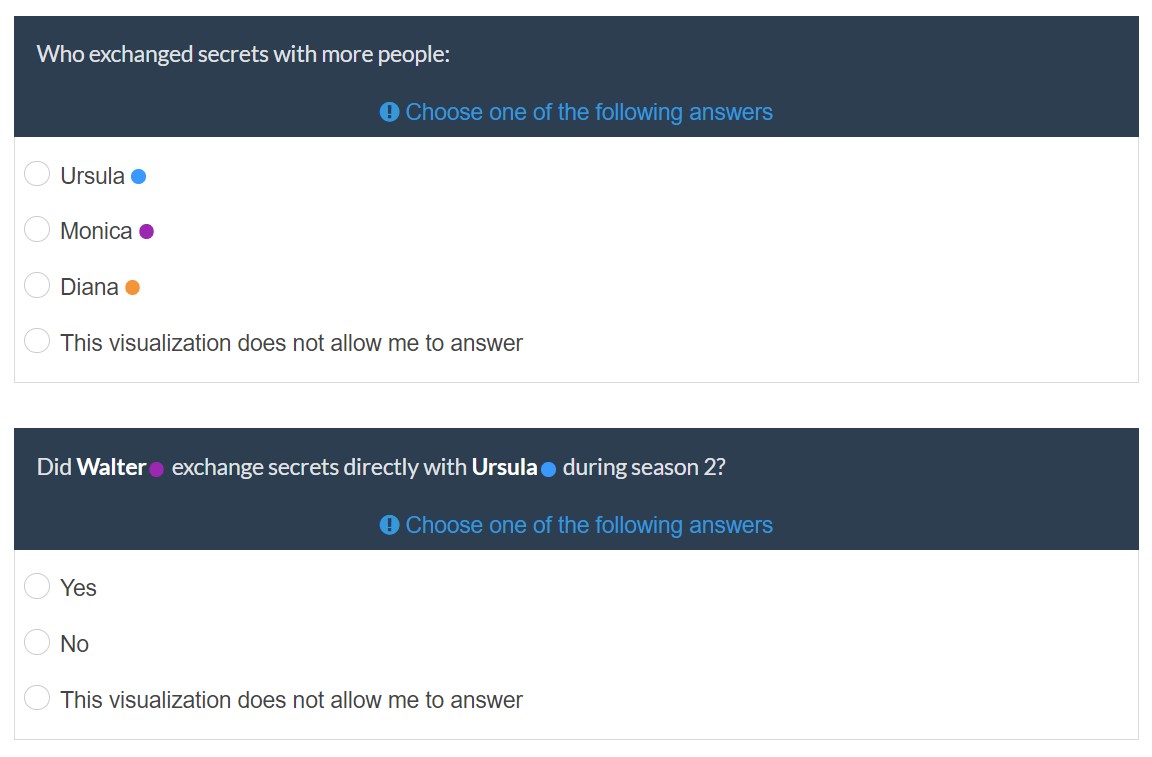}
    \caption{Example screenshots of reading questions in our validation survey, in which we used color dots to facilitate the finding of relevant nodes in the visualization.}
    \label{fig:valid_survey_reading_task_screenshot}
\end{figure}

Although above-mentioned tasks were combinations of low-level tasks, it was possible to facilitate the “Find” task by using pre-attentive encodings to help participants filter the layout, similar to what Ghoniem \etal \cite{Ghoniem:2005:ReadabilityGraphs} did in their experimental design. We would thus ensure as low as possible level for the task. Since we did not want to change the colors for each task, we used cluster colors and references to these colors in the questions (as shown in \autoref{fig:valid_survey_reading_task_screenshot}). As a result, the \textit{Find} tasks were not completely pre-attentive, but we facilitated them as much as possible.

\section{Validation survey: additional procedure details}
\label{app:sec:validation_survey_procedure}
Here we provide additional details about the procedure and conditional design of our validation survey.

Our survey was structured in 4 main steps:
\begin{enumerate}
    \item \textbf{Consent form.} Participants needed to read and approve of our consent form in order to proceed with the questionnaire. All subsequent questions were mandatory, with the exception of comments after ratings scales.

    \item \textbf{Preliminary questions and instructions.} We then asked participants 1 question about color-vision deficiency before showing them an explanatory screen on how to read a node-link diagram such as the ones we used as stimuli visualizations.
    We asked participants to confirm their understanding of our explanations and then provided them with instructions regarding the experimental part of our survey.

    \item \textbf{Survey stimuli reading and rating questions.} We assigned a random order of appearance to each stimulus. We report the resulting distribution of appearance orders further below in \autoref{tab:validation_survey_randomization}, and technical explanations in our \researchLog. For each stimulus, participants aswered:
    \begin{enumerate}
    
        \item \textbf{All reading task questions} (see \autoref{app:subsec:validation_survey_tasks}). In addition, for the \textbf{first} visualization in order of appearance only, we asked a comprehension check question (``\textit{What was the title of the visualization you just saw?}''). In this question we hid the node-link visualization. In line with \href{https://researcher-help.prolific.com/hc/en-gb/articles/360009223553-Prolific-s-Attention-and-Comprehension-Check-Policy}{Prolific attention and comprehension check policy}, we gave participants two opportunities to answer correctly and they had the option of going back to the previous screen to read again the visualization and its accompanying text. Incorrectly answering this question twice was an exclusion criteria (similar to our exploratory survey).

        \item \textbf{\scalename questionnaire.} For each subscale \PREVisColors, we displayed all rating items in a single screen with labeled 7-point Likert scale answer options as shown in \autoref{app:fig:validation_survey_screenshot}. Subscales appeared in a random order, and items within subscales were also randomized.
        
        For each stimulus we also included one Instructional Manipulation Check (IMC, \eg, ``For attention check purposes, please select \textbf{slightly agree} with this item''). We could not randomize the placement of such items, but we scattered them across different subscales and also modified the answer option we asked respondents to select---avoiding extreme values because, as we reported in our \researchLog, participants from our pre-test interviews (exploratory survey) were uncomfortable in using extreme points of the Likert scale. Answering incorrectly more than once to IMCs was an exclusion criterion.
        
    \end{enumerate}
    
    \item \textbf{Extraversion personality trait questions.} As explained in \autoref{subsec:validation_study_design}, in order to produce our Multi-Trait Multi-Method matrix (\autoref{fig:MTMM}), we decided to compare \scalename \PREVisColors scores with a measure of extraversion as personality trait. We used the 2 scale items related to extraversion from a validated 10-items version of the Big Five personality Inventory \cite{rammstedt_2007_MeasuringPersonality}: ``I see myself as \textbf{extraverted, enthusiastic.}'' and ``I see myself as reserved, quiet.'' (for the latter we would reverse the rating before calculating average extraversion scores for participants).
    
    We presented these two items using the same answer options as our \scalename rating items and we specified that these questions were calibration items, unrelated to the visualizations participants had seen before.
    
\end{enumerate}

\section{Validation study results}
\label{app:sec:validation_results}
\subsection{Stimuli randomization order distribution:a}
\label{app:sec:validation_results:a}
We used the LimeSurvey platform to distribute our survey; LimeSurvey provides different levels of randomization functions, on which we do not have control in terms of parameters. We monitored the randomization order and noticed that stimuli orders of appearance were not equally distributed among participants. As such, the most common order was \stimVB, then \stimVC, then \stimVA.



\begin{table}[h!]
\centering
\fontsize{7pt}{7pt}\selectfont
\caption{Order in which participants saw 3 node-link stimuli visualizations in our validation study.}
\label{tab:validation_survey_randomization}
\tabulinesep=0.8mm
\begin{tabu} to 0.8\linewidth {X[1.3,l]X[1,c]X[1,c]X[1,c]}
\toprule
 & Count in \stimVA & Count in \stimVB & Count in \stimVC \\
\midrule
\textbf{Appeared 1s}t & 34 & 65 & 49 \\
\textbf{Appeared 2d} & 53 & 38 & 57 \\
\textbf{Appeared 3d} & 61 & 45 & 42 \\
\bottomrule
\end{tabu}
\end{table}

\subsection{Dimensionality, reliability and construct validity tests}
\label{app:sec:validation_results:b}

We thoroughly validated \scalename by conducting different tests of dimensionality, reliability, and construct validity. We report in the main paper the key findings; in this section we describe the methods and provide some more details on our results. We provide all code and data to conduct these analyses in our \suppmat{supplemental material folder}{9cg8j}.

\subsubsection{Tests of dimensionality}
\begin{figure}[h]
    \centering
    \includegraphics[width=\linewidth]{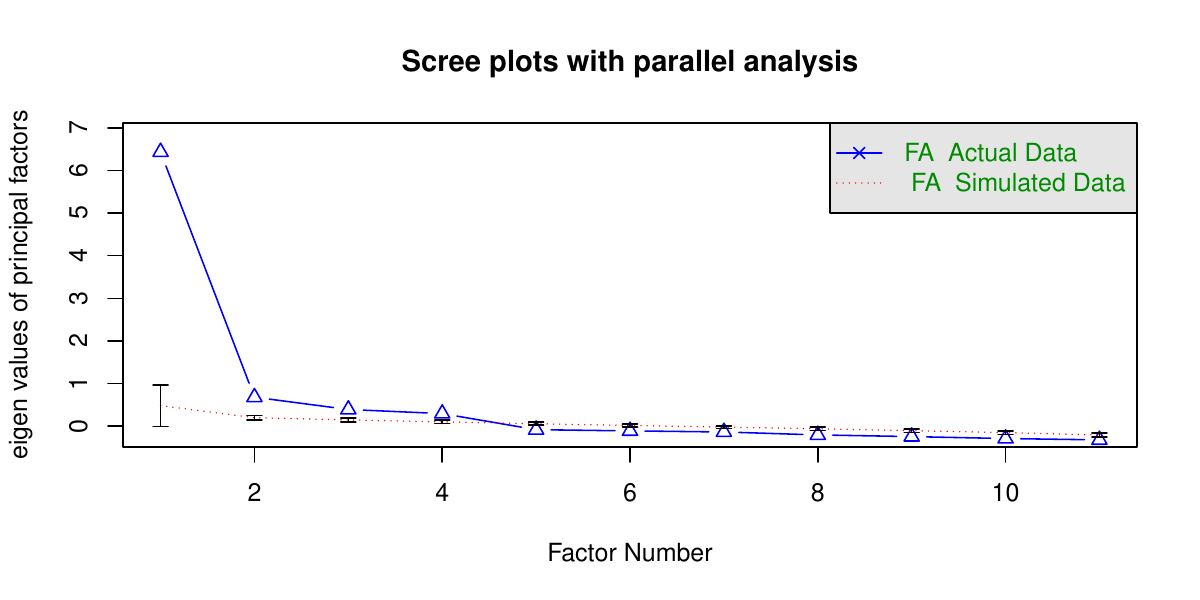}
    \caption{Scree plot from parallel analysis with collected ratings from our validation survey.}
    \label{app:fig:valid_screeplot}
\end{figure}

To validate the 4-factors dimensions, we followed the same steps as in the exploratory approach. We started with a parallel analysis (which we explained in \autoref{app:sec:EFA_screeplots}). From visual analysis on the scree plot in \autoref{app:fig:valid_screeplot}, we confirmed that 4 factors were appropriate to explain the variance in our data.

We then conducted Multi-Group Confirmatory Analysis (MG-CFA) (see \autoref{app:sec:EFA_CFA_4factors}).
\autoref{tab:valid_fit_metrics} shows the model fit metrics, which allowed us to validate our 4-factor structure. We confirmed goodness of fit of our 4-factors structure with the following metrics: the Tucker–Lewis Index (TLI) was 0.97, the Comparative Fit Index (CFI) was 0.98, the Standardized Root Mean square Residual (SRMR) was 0.034. These metrics were in line with cutoff values recommended from Hu and Bentler \cite{hu_1999_CutoffCriteria}: fit indices such as TLI or CFI should be higher than .95, and SRMR should be lower than 0.9.

\begin{table}[h]
\centering
\caption{\scalename fit metrics from our validation study.}
\label{tab:valid_fit_metrics}
\fontsize{7pt}{7pt}\selectfont
\tabulinesep=0.8mm
\begin{tabu} to \linewidth{ X[1, l] X[1, l] X[1, c] X[1, c] X[1, c] X[1, c]}
\toprule
\textbf{Fit indices} & \textbf{Full survey} & \textbf{Stimulus A} & \textbf{Stimulus B} & \textbf{Stimulus C} \\
\midrule
\textbf{chisq} & 203 & 85 & 57 & 59 \\
\textbf{pvalue} & 0.000 & 0.000 & 0.023 & 0.013 \\
\textbf{cfi} & 0.979 & 0.966 & 0.987 & 0.984 \\
\textbf{tli} & 0.970 & 0.951 & 0.981 & 0.977 \\
\textbf{srmr} & 0.034 & 0.038 & 0.033 & 0.031 \\
\textbf{rmsea} & 0.073 & 0.092 & 0.059 & 0.062 \\
\bottomrule
\end{tabu}
\end{table}

\begin{figure}[b]
    \vspace{0.5cm}
    \centering
    \includegraphics[width=\linewidth]{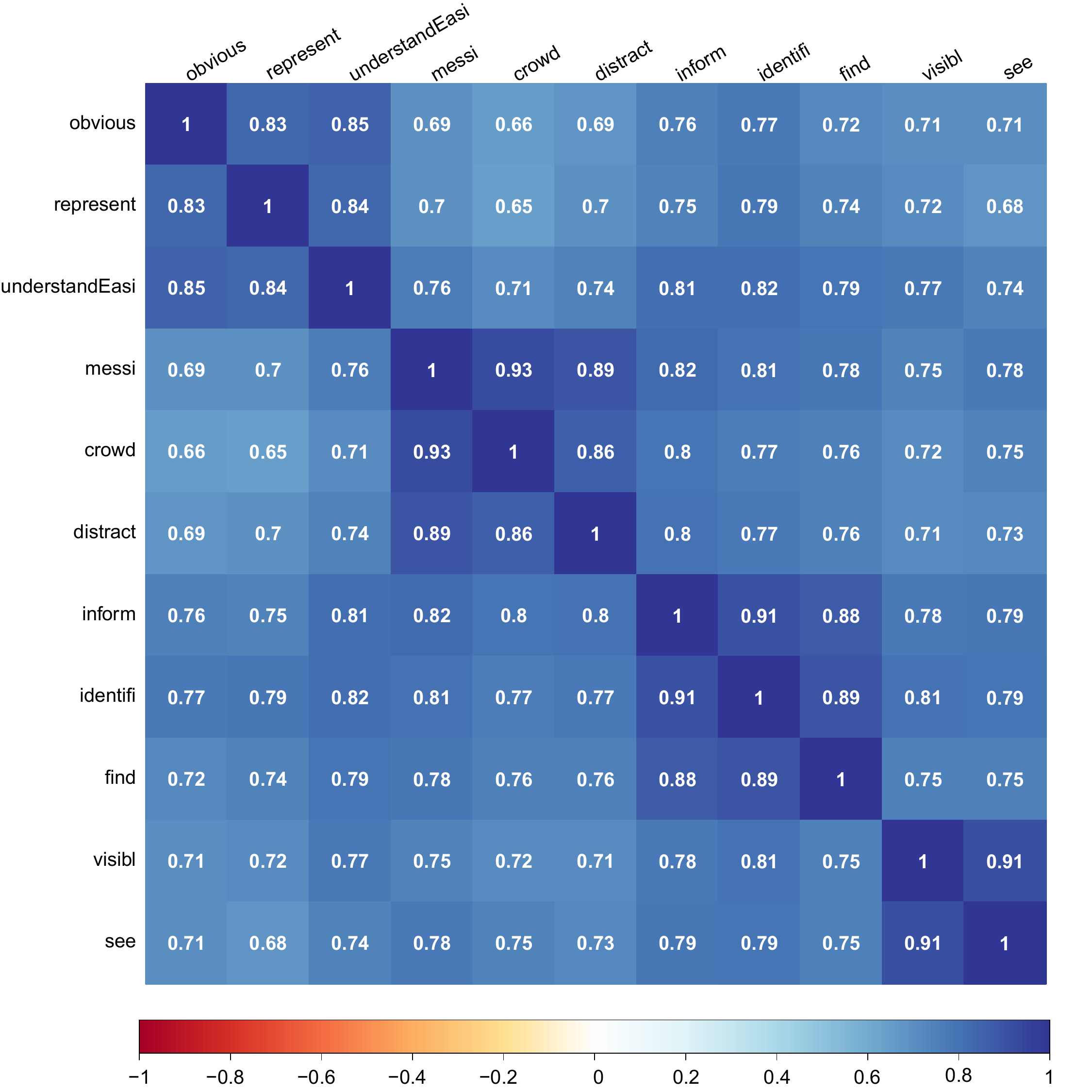}
    \caption{\protect\revis{Repeated measures correlation matrix \scalename \PREVisColors items (see \autoref{app:tab:PREVis_items_variables} for variable name correspondences with items statements). We generated this matrix using the \texttt{rmcorr} package in \texttt{R}.}}
    \label{app:fig:valid_scales_corr}
    \vspace{0.5cm}
\end{figure}

\subsubsection{Tests of reliability}

We conducted a first rough assessment of reliability by plotting the \revis{repeated measures} correlation matrix for our collected data in \autoref{app:fig:valid_scales_corr}, \revis{using the \texttt{rmcorr} package in \texttt{R}}. This matrix also reflected clusters of covariance associated with our 4-factors structure.

\begin{table}[h]
\centering
\caption{\scalename subscales' Cronbach's alpha and McDonald's omega reliability coefficients from our validation study.}
\label{tab:valid_reliabilities}
\fontsize{7pt}{7pt}\selectfont
\tabulinesep=0.8mm
\begin{tabu} to \linewidth{ X[1.1, l] X[1.1, l] X[1, c] X[1, c] X[1, c] X[1, c]}
\toprule
\textbf{Data set} &\textbf{Coefficient} &\textbf{understand} &\textbf{layout} &\textbf{dataRead} &\textbf{dataFeat} \\\midrule
\multirow{3}{*}{\textbf{Full survey}} &\textbf{raw alpha} &0.936 &0.953 &0.955 &0.946 \\
&\textbf{std alpha} &0.936 &0.953 &0.956 &0.946 \\
&\textbf{omega tot} &0.937 &0.953 &0.956 &0.946 \\
\midrule
\multirow{3}{*}{\textbf{Stimulus} \stimVA} &\textbf{raw alpha} &0.926 &0.899 &0.895 &0.911 \\
&\textbf{std alpha} &0.928 &0.900 &0.896 &0.912 \\
&\textbf{omega tot} &0.929 &0.902 &0.901 &0.912 \\
\midrule
\multirow{3}{*}{\textbf{Stimulus} \stimVB} &\textbf{raw alpha} &0.914 &0.884 &0.918 &0.932 \\
&\textbf{std alpha} &0.914 &0.885 &0.918 &0.932 \\
&\textbf{omega tot} &0.914 &0.886 &0.920 &0.932 \\
\midrule
\multirow{3}{*}{\textbf{Stimulus} \stimVC} &\textbf{raw alpha} &0.901 &0.877 &0.933 &0.892 \\
&\textbf{std alpha} &0.901 &0.886 &0.933 &0.892 \\
&\textbf{omega tot} &0.901 &0.891 &0.933 &0.892 \\
\bottomrule
\end{tabu}
\end{table}

Then, we tested the reliability of individual \scalename \PREVisColors subscales by calculating Cronbach's alpha and McDonald's omega coefficient using \texttt{R's psych} package. \autoref{tab:valid_reliabilities} shows the reliability coefficients for each subscale, calculated for the full survey data and individually for each stimulus data subset. We obtained high reliability values for all subscales: we found the lowest value to be raw alpha $=$ 0.877 for \SLa in \stimVC. DeVellis and Thorpe consider such reliability values as ``very good'' \cite{Devellis:2021:ScaleDevelopment}.

\begin{table}[b]
\centering
\footnotesize
\tabulinesep=1.2mm
\caption{Variable names correspondences with PREVis items in our analyses.}
\label{app:tab:PREVis_items_variables}
\begin{tabu} to \linewidth{ X[0.1, l] X[0.6, l] X[1.6, l] }
\toprule
\textbf{} &\textbf{Item code} &\textbf{Item statement} \\\midrule
\multirow{3}{*}{{\rotatebox[origin=c]{90}{ \hspace{1pt}\SUn}}} &obvious &It is obvious for me how to read this visualization \\
&represent &I can easily understand how the data is represented in this visualization \\
&understandEasi &I can easily understand this visualization \\
\midrule
\multirow{3}{*}{{\rotatebox[origin=c]{90}{\SLa}}} &messi &I don't find this visualization messy \\
&crowd &I don't find this visualization crowded \\
&distract &I don't find parts of the visualization distracting \\
\midrule
\multirow{2}{*}{\rotatebox[origin=c]{90}{{\SDF}}} &visibl &I find data features (for example, a minimum, or an outlier, or a trend) visible in this visualization \\
&see &I can clearly see data features (for example, a minimum, or an outlier, or a trend) in this visualization \\
\midrule
\multirow{3}{*}{\rotatebox[origin=c]{90}{{\SDR}}} &inform &I can easily retrieve information from this visualization \\
&identifi &I can easily identify relevant information in this visualization \\
&find &I can easily find specific elements in this visualization \\
\bottomrule
\end{tabu}
\end{table}


\clearpage
\subsubsection{Tests of construct validity}
\label{app:sec:construct_validity}

\hypertarget{appMTMM}{We} conducted multiple tests to confirm the validity of \scalename. Construct validity testing consist in verifying that the instrument correctly targets the construct that it was built to measure: in our case, perceived readability in data visualizations. 
Following recommendations from our reference literature in scale development \cite{Boateng:2018:BestPractices, Devellis:2021:ScaleDevelopment}, and as we state in \autoref{subsec:validation_study_design}, we designed our validation study to test for convergent and discriminant validity \revis{using the multi-traits multi-method (MTMM) approach:
\begin{itemize}
    \item[\MTMMoneText] \textbf{Inter-subscales reliability:} our subscales' scores should highly and positively correlate among themselves, indicating the existence of a shared latent variable in respondents: the \textit{perceived readability} construct.
    \item[\MTMMtwoText] \textbf{Discriminant validity:} our subscales' scores should not correlate with extraversion scores measured with a similar method (\ie, in our, case, a 7-point Likert scale from the Big Five personality Inventory short version \cite{rammstedt_2007_MeasuringPersonality}) because it is a different, unrelated construct.
    \item[\MTMMthreeText] \textbf{Convergent validity:} our subscales' scores should positively correlate with Greadability metrics \cite{gove_2018_ItPays} because they are also indicators related to readability, but measured using a different method.
\end{itemize}
}

\revis{
A MTMM correlation matrix allows us to check all of these criteria at once. We first generated MTMM matrices in \texttt{R} using the \texttt{corr} function in the \texttt{psych} package, which assumes independence of observations. It is a conservative way to analyze our results because, by assuming independence of all 
\scalename measurements, between-participant differences add error to the measures, possibly reducing their covariance across stimuli. Our study's design, however, also allowed for within-participant assessment of \MTMMoneText inter-subscales reliability and \MTMMthreeText convergent validity using repeated measures correlations:
\begin{itemize}
    \item[\MTMMoneText] \textbf{Inter-subscales reliability:} each of our 148 participants rated each stimulus (\stimVA, \stimVB, \stimVC), therefore we collected  $3 \times 148$ measures for each \PREVisColors subscale. This allowed us to conduct a repeated measures correlation analysis (\ie, correlations based on within-participant covariance of \scalename ratings across stimuli).
    \item[\MTMMtwoText] \textbf{Discriminant validity:} because personality traits are stable, we measured extraversion only once per participant, thus collecting 148 observations. 
    Because extraversion scores did not vary across experimental conditions, conducting a repeated measures correlation analysis caused computational errors. These errors stemmed from the underlying model encountering near-zero values during the covariance analysis, leading to numerical precision issues and falsely negative sums of squares.
    \item[\MTMMthreeText] \textbf{Convergent validity:} we calculated Greability measures for each stimulus (see \autoref{app:tab:greadability}) and copied these values in the result table for each participant, therefore we have $3 \times 148$ values. This allowed us to conduct \textbf{a repeated measures correlation analysis} (\ie, correlations based on within-participant covariance to remove between-participant variance).
\end{itemize}
}

\begin{figure}[h]
    \centering
    \includegraphics[width=1\linewidth]{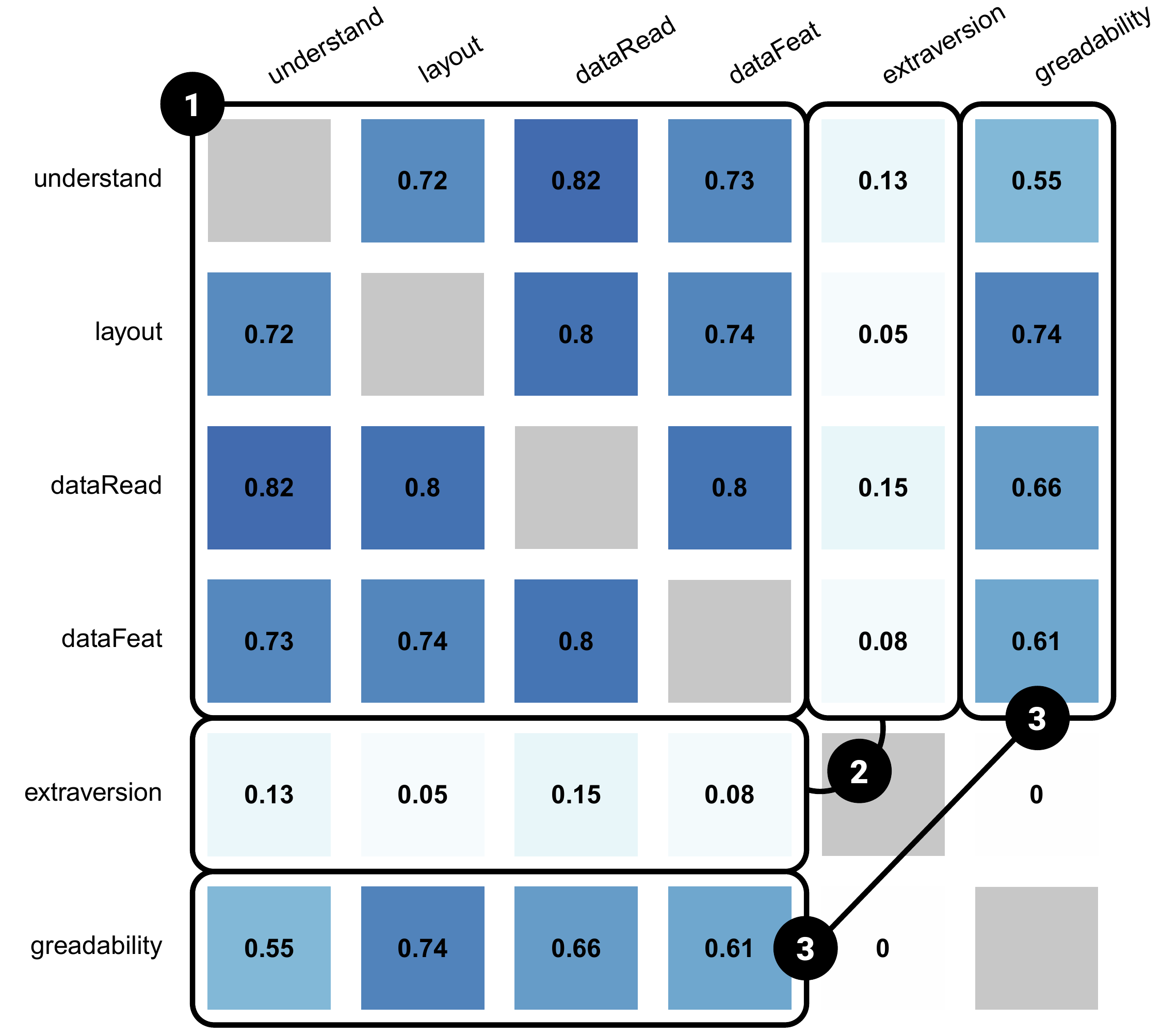}
    \caption{\protect\revis{Multi-trait multi-method (MTMM) \textbf{independent measures} correlation matrix at the scale level: \MTMMoneCaption reliability among \scalename subscales, \MTMMtwoCaption discriminant validity from an unrelated personality trait in respondents, and \MTMMthreeCaption convergent validity with graph layout metrics.}}
    \label{app:fig:MTMM_im}
\end{figure}

\revis{\textbf{Scale-level correlation matrices.} \autoref{app:fig:MTMM_im} shows a scale-level version of the independent measures MTMM matrix generated with the \texttt{cov2cor} function in the \texttt{stat} package in \texttt{R}. This conservative approach confirms all 3 criteria despite the between-participant noise: inter-subscales reliability (positive and high correlations in \MTMMoneText), 
discriminant validity (correlations close to 0 in \MTMMtwoText), 
and convergent validity (positive correlations in \MTMMthreeText).}

\begin{figure}[h]
    \centering
    \includegraphics[width=1\linewidth]{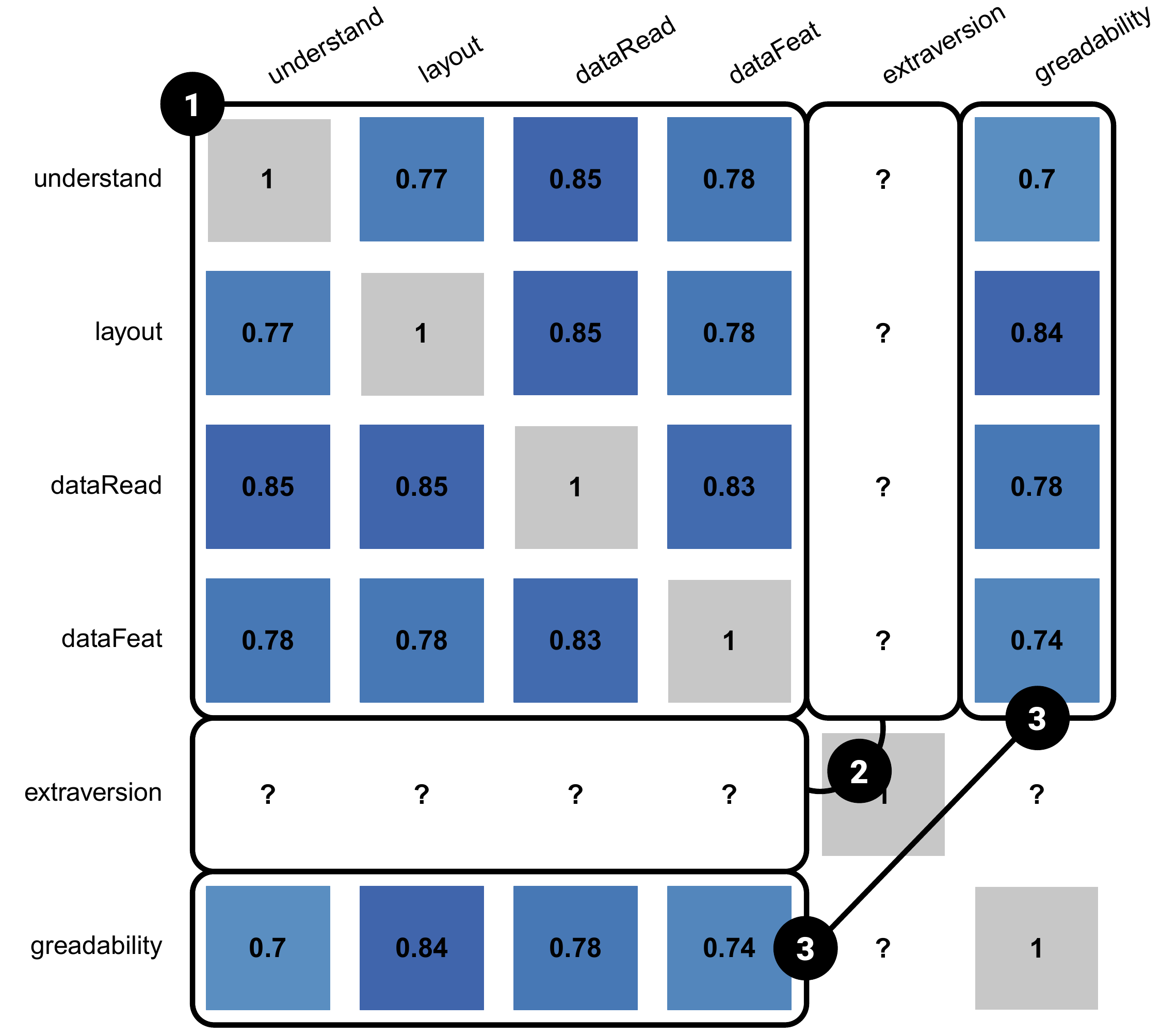}
    \caption{\protect\revis{Multi-trait multi-method (MTMM) \textbf{repeated measures} correlation matrix at the scale level: \MTMMoneCaption reliability among \scalename subscales, \MTMMtwoCaption discriminant validity from an unrelated personality trait in respondents, and \MTMMthreeCaption convergent validity with graph layout metrics.}}
    \label{app:fig:MTMM_rm}
\end{figure}

\revis{\autoref{app:fig:MTMM_rm} shows a scale-level version of the repeated measures MTMM matrix. Eliminating between-subject variance allows to find much higher correlations between \scalename \PREVisColors ratings and Greadability metrics, strengthening convergent validity of our instrument. Inter-subscales reliability also slightly improves. As extraversion was not collected with repeated measures, this tool does not allow us to calculate discriminant validity. While attempting to process the data, the \texttt{rmcorr} package in \texttt{R} returned NA values for correlations between \PREVisColors subscales and Greadability metrics, as well as for correlations between Greadability metrics and extraversion ratings.}

\revis{This is explained technically because Greadability has fixed values across participants within each condition (stimuli \stimVA, \stimVB, and \stimVC), and extraversion ratings have a fixed values within each participant across the three conditions. This can result in very small or zero variance within those groups of variables. When the variance is extremely small, numerical precision issues can lead to negative sum of squares values, which leads to an invalid calculation for the repeated measures correlation coefficient. We document the problem and the code we used to apply a selective adjustment based on a small tolerance value of $1\mathrm{e}{-10}$ it in our \researchLog.}

\begin{figure}[h]
    \centering
    \includegraphics[width=\linewidth]{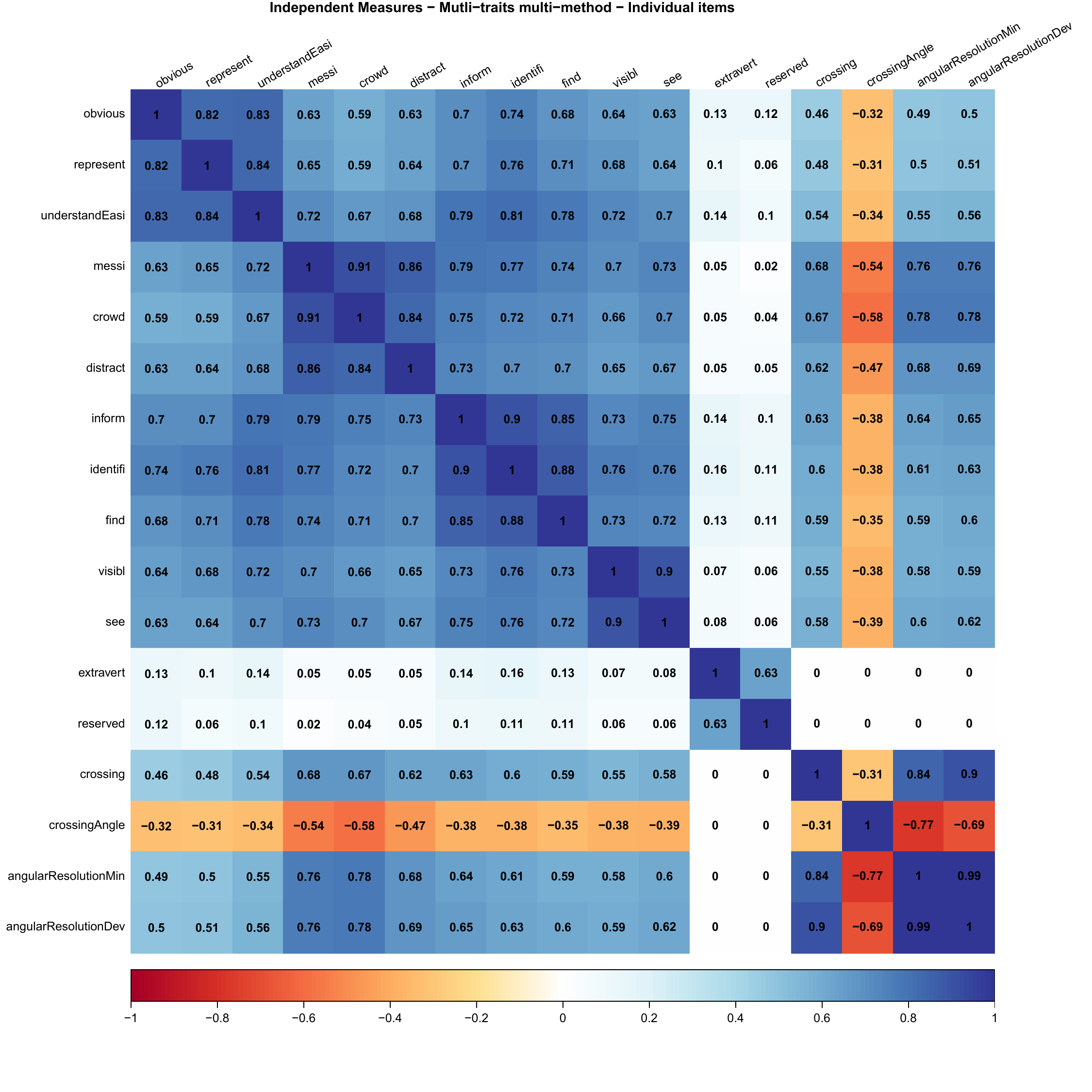}\vspace{-5mm}
    \caption{\protect\revis{Item-level Multi-Trait Multi-Method \textbf{independent measures} correlation matrix from our validation study.}}
    \label{app:fig:full_MTMM_im}
\end{figure}

\begin{figure}[h]
    \centering
    \includegraphics[width=\linewidth]{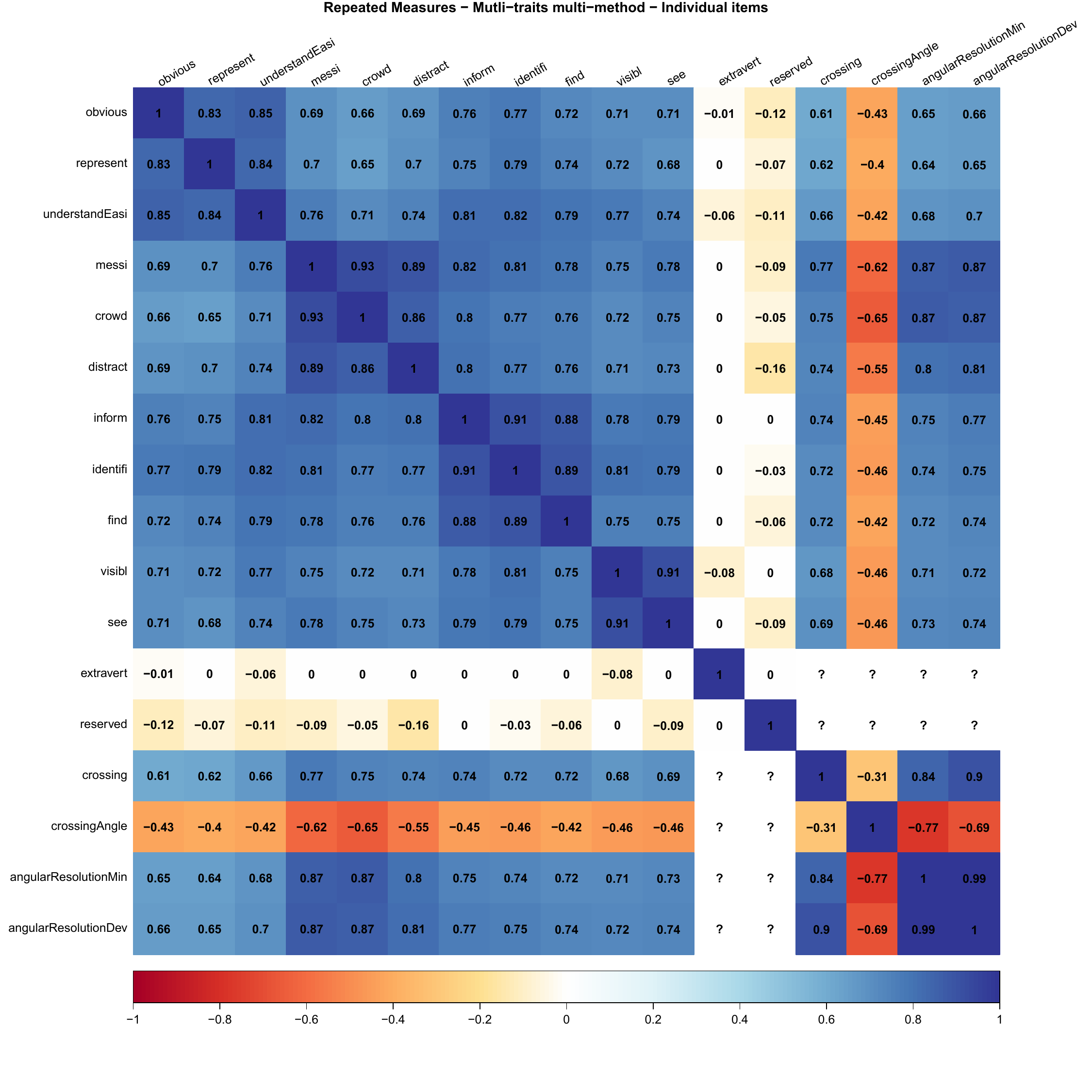}\vspace{-5mm}
    \caption{\protect\revis{Item-level Multi-Trait Multi-Method \textbf{repeated measures }correlation matrix from our validation study.}}
    \label{app:fig:full_MTMM_rm}
\end{figure}

\clearpage
\revis{\textbf{Item-level correlation matrices.} \autoref{app:fig:full_MTMM_im} shows the independent measures MTMM matrix at the item-level.}
The first 11 lines and rows correspond to \scalename items, grouped by subscale, and show high correlations; the next 2 elements correspond to the ``extraversion'' items from the BFI 10-items scale \cite{rammstedt_2007_MeasuringPersonality} and show an absence of correlation with all data collected in this survey; the last 4 rows and columns correspond to the \texttt{Greadability.js} metrics \cite{gove_2018_ItPays} from \autoref{app:tab:greadability}. In this part of the item-level MTMM, one of the items in the set of \texttt{Greadability} metrics shows negative correlation to the others. It corresponds to the \texttt{``crossingAngle''} metric. We discuss this metric above in \autoref{app:subsec:greadability}.
\revis{\autoref{app:fig:full_MTMM_rm} shows the repeated measures MTMM matrix at the item-level.}

\revis{We share in \autoref{fig:MTMM} in the main paper a scale-level \revis{composite} MTMM matrix where correlations across stimuli for PREVis and Greadability are calculated with the repeated measure approach from the \texttt{rmcorr} package in \texttt{R}, and the correlations for the \textit{extraversion} between-participants variable is calculated based on the \texttt{cov2cor} function in \texttt{stat} package from \texttt{R}. }

\begin{figure}[h!]
    \centering
    \includegraphics[width=\linewidth]{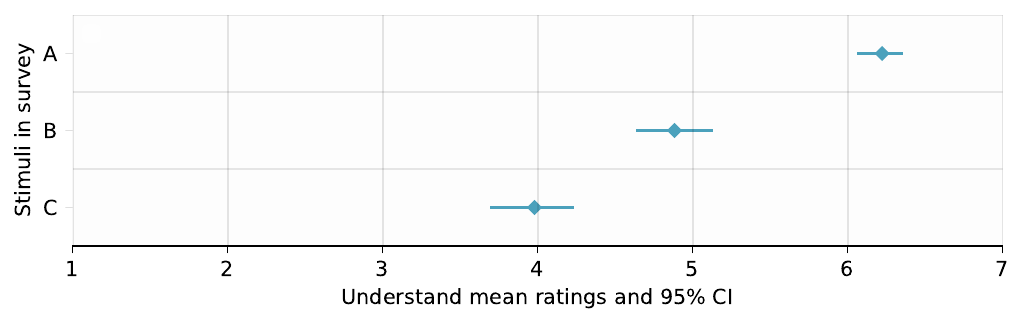}
    \caption{\SUn scores across stimuli in our validation study allow to distinguish and correctly rank \stimVA $>$ \stimVB $>$ \stimVC.}
    \label{app:fig:valid_SUn}
\end{figure}
\begin{figure}[h!]
    \centering
    \includegraphics[width=\linewidth]{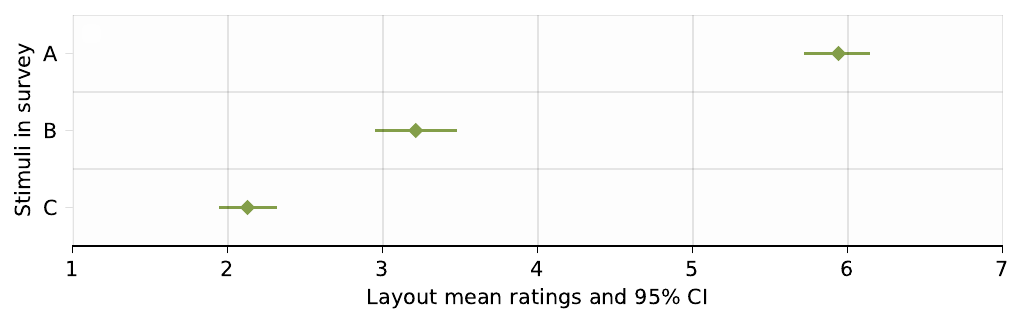}
    \caption{\SLa scores across stimuli in our validation study allow to distinguish and correctly rank \stimVA $>$ \stimVB $>$ \stimVC.}
    \label{app:fig:valid_SLa}
\end{figure}
\begin{figure}[h!]
    \centering
    \includegraphics[width=\linewidth]{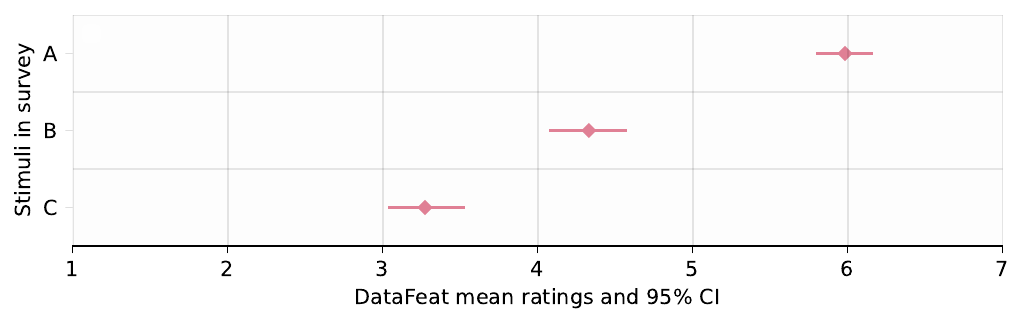}
    \caption{\SDF scores across stimuli in our validation study allow to distinguish and correctly rank \stimVA $>$ \stimVB $>$ \stimVC.}
    \label{app:fig:valid_SDF}
\end{figure}
\begin{figure} [h!]
    \centering
    \includegraphics[width=\linewidth]{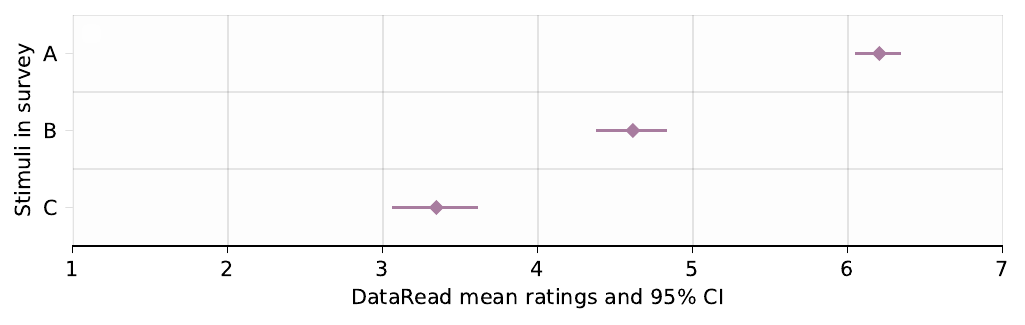}
    \caption{\SDR scores across stimuli in our validation study allow to distinguish and correctly rank \stimVA $>$ \stimVB $>$ \stimVC.}
    \label{app:fig:valid_SDR}
\end{figure}

Finally, we plotted \scalename \PREVisColors subscales' scores with 95\% CI across stimuli to test their validity with what Boateng \etal call ``\textit{differentiation by known groups}'' \cite{Boateng:2018:BestPractices}: 
(\autoref{app:fig:valid_SUn} to \autoref{app:fig:valid_SDR}). We verified that we could distinguish between the 3 node-link, in the expected order: \mbox{\stimVA $>$ \stimVB $>$ \stimVC}.

\begin{figure} [h!]
    \centering
    \includegraphics[width=\linewidth]{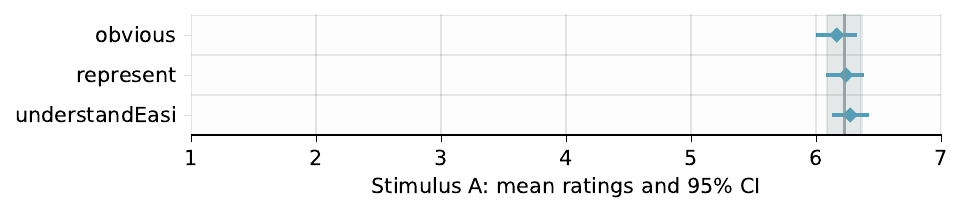}
    \caption{\SUn individual items scores and average score in \stimVA.}
    \label{app:fig:valid_A_SUn}
\end{figure}

\begin{figure} [h!]
    \centering
    \includegraphics[width=\linewidth]{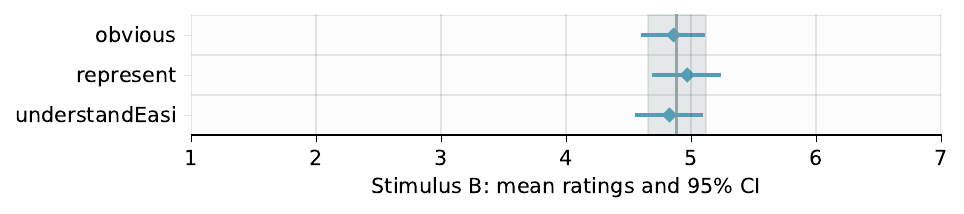}
    \caption{\SUn individual items scores and average score in \stimVB.}
    \label{app:fig:valid_B_SUn}
\end{figure}

\begin{figure} [h!]
    \centering
    \includegraphics[width=\linewidth]{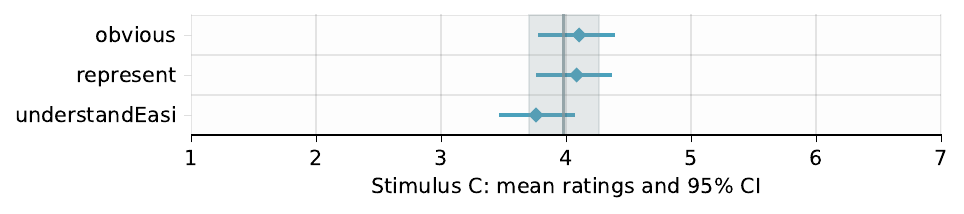}
    \caption{\SUn individual items scores and average score in \stimVC.}
    \label{app:fig:valid_C_SUn}
\end{figure}

\begin{figure} [h!]
    \centering
    \includegraphics[width=\linewidth]{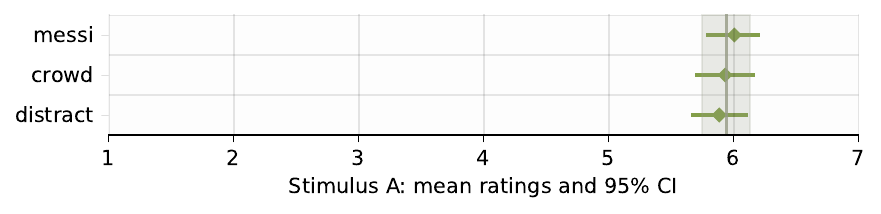}
    \caption{\SLa individual items scores and average score in \stimVA.}
    \label{app:fig:valid_A_SLa}
\end{figure}

\begin{figure} [h!]
    \centering
    \includegraphics[width=\linewidth]{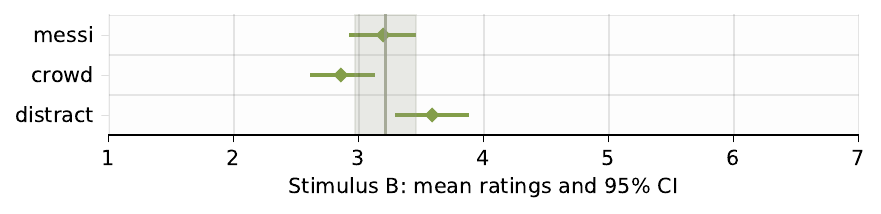}
    \caption{\SLa individual items scores and average score in \stimVB.}
    \label{app:fig:valid_B_SLa}
\end{figure}

\begin{figure} [h!]
    \centering
    \includegraphics[width=\linewidth]{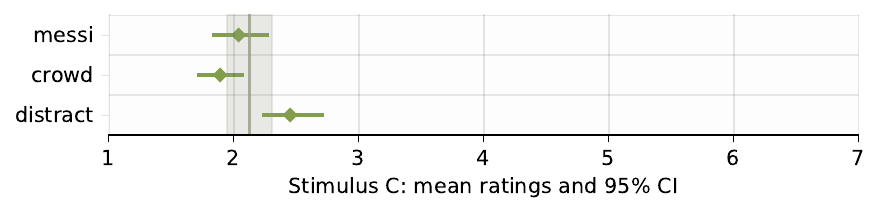}
    \caption{\SLa individual items scores and average score in \stimVC.}
    \label{app:fig:valid_C_SLa}
\end{figure}

\begin{figure} [h!]
    \centering
    \includegraphics[width=\linewidth]{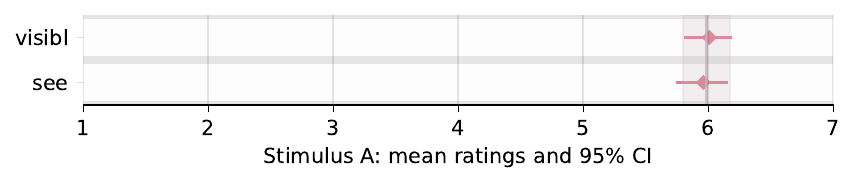}
    \caption{\SDF individual items scores and average score in \stimVA.}
    \label{app:fig:valid_A_SDF}
\end{figure}

\begin{figure} [h!]
    \centering
    \includegraphics[width=\linewidth]{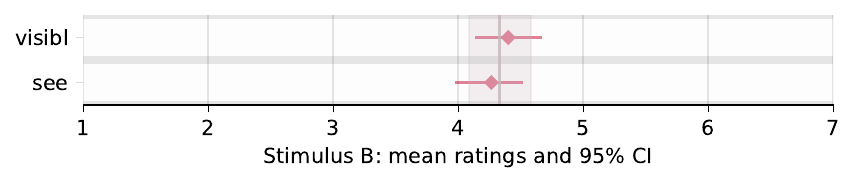}
    \caption{\SDF individual items scores and average score in \stimVB.}
    \label{app:fig:valid_B_SDF}
\end{figure}

\begin{figure} [h!]
    \centering
    \includegraphics[width=\linewidth]{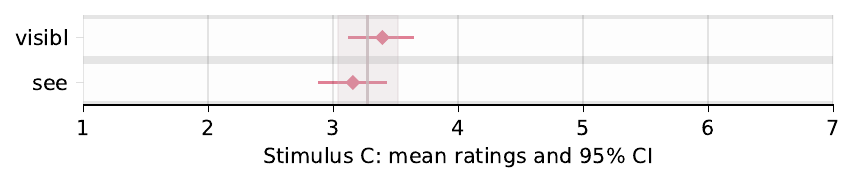}
    \caption{\SDF individual items scores and average score in \stimVC.}
    \label{app:fig:valid_C_SDF}
\end{figure}

\begin{figure} [h!]
    \centering
    \includegraphics[width=\linewidth]{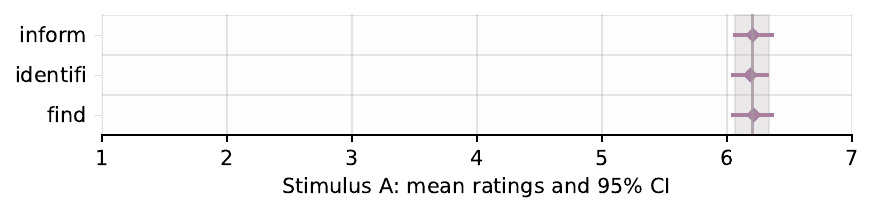}
    \caption{\SDR individual items scores and average score in \stimVA.}
    \label{app:fig:valid_A_SDR}
\end{figure}

\begin{figure} [h!]
    \centering
    \includegraphics[width=\linewidth]{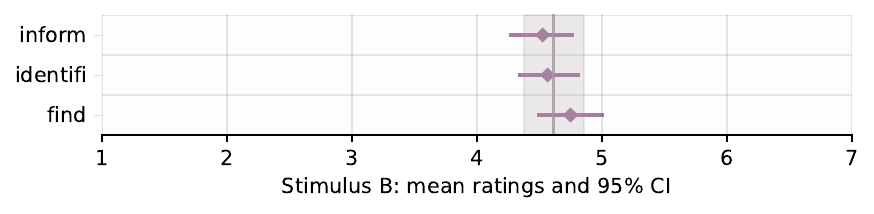}
    \caption{\SDR individual items scores and average score in \stimVB.}
    \label{app:fig:valid_B_SDR}
\end{figure}

\begin{figure} [h!]
    \centering
    \includegraphics[width=\linewidth]{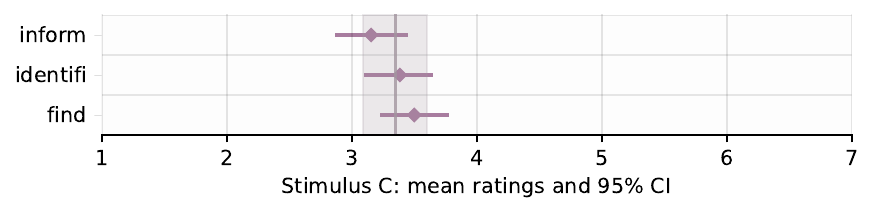}
    \caption{\SDR individual items scores and average score in \stimVC.}
    \label{app:fig:valid_C_SDR}
\end{figure}

We also plotted individual items ratings for each \PREVisColors subscale and stimulus in \autoref{app:fig:valid_A_SUn} to \autoref{app:fig:valid_C_SDR}.

\newpage
\section*{Figure credits and copyright}
\label{app:sec:figure_credits}
\autoref{app:fig:GeneaQuilts_Simpsons} is \textcopyright{} 2010 IEEE and we reused it (with permission) from Bezerianos \etal's paper \cite{bezerianos_2010_GeneaQuiltsSystem}.
All the remaining figures in this appendix are our own, and for them we retain the copyright but allow them to be used here. They are available under the \href{https://creativecommons.org/licenses/by/4.0/}{Creative Commons \ccLogo\,\ccAttribution\ \mbox{CC BY 4.0}} license and we share them in our supplemental material folder at \suppmat{osf.io/9cg8j}{9cg8j}.


%
%
%
%
%

\end{document}